\documentclass[11pt]{article}
\usepackage{fullpage}
\usepackage{amsmath,amsthm,amsfonts,amssymb,dsfont,bm}
\usepackage{enumerate,color,xcolor}
\usepackage{graphicx}
\usepackage{subfigure}
\usepackage[colorlinks,linkcolor=cyan,citecolor=blue]{hyperref}
\usepackage{url}

\numberwithin{equation}{section}
\theoremstyle{plain}

\newtheorem{theorem}{Theorem}
\newtheorem{corollary}[theorem]{Corollary}

\newtheorem{lemma}[theorem]{Lemma}
\newtheorem{proposition}[theorem]{Proposition}
\newtheorem{assumption}[theorem]{Assumption}

\newtheorem{remark}[theorem]{Remark}

\begin{document}
\begin{center}  
{\bf\Large Regime-Switching Langevin Monte Carlo Algorithms}
\end{center}

\author{}
\begin{center}
  {Xiaoyu Wang}\,\footnote{FinTech Thrust,
  Hong Kong University of Science and Technology (Guangzhou), Guangzhou, Guangdong, People's Republic of China; 
  xiaoyuwang@hkust-gz.edu.cn},
  Yingli Wang\,\footnote{School of Mathematics, Shanghai University of Finance and Economics, Shanghai, People's Republic of China; 2022310119@163.sufe.edu.cn},
  Lingjiong Zhu\,\footnote{Department of Mathematics, Florida State University, Tallahassee, Florida, United States of America; zhu@math.fsu.edu
 }
\end{center}

\begin{center}
 \today
\end{center}

\begin{abstract}
Langevin Monte Carlo (LMC) algorithms are popular Markov Chain Monte Carlo (MCMC) methods to sample a target probability distribution, which arises in many applications in machine learning. Inspired by regime-switching stochastic differential equations in the probability literature, we propose and study regime-switching Langevin dynamics (RS-LD) and regime-switching kinetic Langevin dynamics (RS-KLD). Based on their discretizations, we introduce regime-switching Langevin Monte Carlo (RS-LMC) 
and regime-switching kinetic Langevin Monte Carlo (RS-KLMC) algorithms, which can also be viewed as LMC and KLMC algorithms with random stepsizes. 
We also propose frictional-regime-switching kinetic Langevin dynamics (FRS-KLD)
and its associated algorithm frictional-regime-switching kinetic Langevin Monte Carlo (FRS-KLMC), 
which can also be viewed as the KLMC algorithm with random frictional coefficients.
We provide their 2-Wasserstein non-asymptotic convergence guarantees to the target distribution, and analyze the iteration complexities. Numerical experiments using both synthetic and real data are provided to illustrate the efficiency of our proposed algorithms.
\end{abstract}


\section{Introduction}

The problem of sampling a given target distribution of interest 
\begin{equation}\label{pi:eqn}
\pi(x)\propto e^{-f(x)},\qquad x\in\mathbb{R}^{d},
\end{equation}
is fundamental in many applications in machine learning, such as Bayesian learning. In Bayesian learning, 
one is interested in sampling a posterior distribution given in \eqref{pi:eqn}, with $f(x)=\sum_{i=1}^{n}f^{(i)}(x)$ where $f^{(i)}(x)$ is associated with the $i$-th data point and $n$ is the number of data points \cite{gelman1995bayesian,stuart2010inverse,andrieu2003introduction,teh2016consistency,gurbuzbalaban2021decentralized,EXTRALangevin}. Different choices of $f^{(i)}(x)$ functions correspond to different Bayesian problems, such as Bayesian statistical inference, Bayesian formulations of inverse problems, and Bayesian classification and regression tasks \cite{gelman1995bayesian,stuart2010inverse,andrieu2003introduction,teh2016consistency}.

One of the most widely used Markov Chain Monte Carlo methods for sampling in statistics are \emph{Langevin algorithms}, 
that allows one to sample from a given density of interest \eqref{pi:eqn}.
The classical Langevin algorithm is based on the
\emph{overdamped Langevin} stochastic differential equation (SDE); see e.g. \cite{Dalalyan,DM2016,DM2017,dalalyan2019user}:
\begin{equation}\label{eq:overdamped-2}
dX(t)=-\nabla f(X(t))dt+\sqrt{2}dB_{t},
\end{equation}
where $f:\mathbb{R}^{d}\rightarrow\mathbb{R}$
and $(B_{t})_{t\geq 0}$ is a standard $d$-dimensional Brownian motion that starts at zero
at time zero. Under some mild assumptions on $f$, the diffusion \eqref{eq:overdamped-2} admits a unique stationary distribution with the density $\pi(x) \propto e^{-f(x)}$,
also known as the \emph{Gibbs distribution} \cite{pavliotis2014stochastic}. For computational purposes, the diffusion \eqref{eq:overdamped-2} is simulated by considering its discretization. 
Among various proposed discretization schemes, Euler-Maruyama discretization is the simplest one and is known as the unadjusted Langevin algorithm in the literature \cite{DM2017,DM2016}:
\begin{equation}\label{discrete:overdamped}
x_{k+1}=x_{k}-\eta \nabla f(x_k)+\sqrt{2\eta}\xi_{k}\,,
\end{equation}
where $\eta>0$ is the stepsize parameter, and $\xi_k \in \mathbb{R}^d$ is a sequence of i.i.d. standard Gaussian random vectors $\mathcal{N}(0,I_{d})$. But then the discretized chain \eqref{discrete:overdamped} does not converge to the target $\pi$ and has a bias that needs to be properly characterized to provide performance guarantees \cite{dalalyan2019user}.
There has been growing interest in the non-asymptotic analysis of discretized Langevin diffusions, motivated by applications to large-scale data analysis and Bayesian inference. The discretized Langevin diffusions admit convergence guarantees to a stationary distribution in a variety of metrics and under various assumptions on $f$; see e.g. \cite{Dalalyan,DM2017,DM2016,CB2018,EH2020,dalalyan2019user,Barkhagen2021,Raginsky,xu2018global,Chau2019,Zhang2019}.

In this paper, we propose \textit{regime-switching Langevin Monte Carlo algorithm} (RS-LMC), 
which is based on the discretization of 
\textit{regime-switching Langevin dynamics} (RS-LD), a continuous-time regime-switching stochastic differential equation (SDE) that is introduced in the paper (Section~\ref{sec:RS-LMC}). 
There is a vast literature on regime-switching SDEs.
In terms of applications, regime-switching SDEs have been widely used in biology, control theory, 
mathematical finance, neuroscience, storage modeling and many other fields; 
in terms of theory, there have been extensive studies on 
ergodicity, recurrence, stochastic stability and numerical approximation schemes; see e.g. \cite{Ross1992,Basak1996,Shao2013,Shao2014,Cloez2015,Shao2015,ShaoEJP}, the books \cite{mao2006stochastic,Yin2010} and the references therein.
To the best of our knowledge, our work is the first one
that proposes and studies a Langevin SDE in the framework of regime-switching SDEs.

On the other hand, regime-switching Langevin Monte Carlo algorithm 
can also be viewed as the Langevin Monte Carlo algorithm with random stepsizes. 
There is a vast literature on optimization algorithms with 
deterministic and random stepsizes. 
It is argued that sometimes non-constant (and random) stepsizes can lead to better performance. Cyclic stepsizes where the stepsize changes in a cyclic fashion (between some lower and upper bounds) have been demonstrated to be numerically efficient in many problems; see e.g. \cite{Smith2017,smith2017exploring,Huang2017,ZhangCyclical2020,gulde2020deep,wang2022empirical}. 
Moreover, \cite{Kalousek2017} studies a steepest descent method with random stepsizes and shows that it can achieve faster asymptotic rate than gradient descent with constant stepsize without knowing the details of the Hessian information. 
\cite{musso2020stochastic} suggests that when the stepsizes are small, uniformly-distributed random stepsizes might yield better regularization without extra computational cost compared to constant stepsize.
Motivated by the literature that the heaviness of the tails (known as tail-index)
is linked to the generalization performance, \cite{cyclic2023} study the heavy-tail phenomenon in stochastic gradient descent with cyclic and random stepsizes, and provide a number of theoretical results that demonstrate how the tail-index varies on the stepsize scheduling. Their results bring a new understanding of the benefits of cyclic and randomized stepsizes compared to constant stepsize in terms of the tail behavior.
To the best of our knowledge, our work is the first one
that proposes and studies a Langevin Monte Carlo algorithm with random stepsizes in the context of sampling.

In the literature, there have been active studies of 
\textit{kinetic (underdamped) Langevin} diffusion
and its discretized algorithms \cite{EB1980,Bakry,Cheng,cheng-nonconvex,dalalyan2018kinetic,GGZ2,Ma2019,JianfengLu,mattingly2002ergodicity,Villani2009,CLW2020,Shen2019,Mangoubi-Smith-AAP,Mangoubi-Smith-AISTATS} based on the SDE:
\begin{align}
&dV(t)=-\gamma V(t)dt-\nabla f(X(t))dt+\sqrt{2\gamma}dB_{t},\nonumber
\\
&dX(t)=V(t)dt,\label{underdamped}
\end{align}
where $(B_{t})_{t\geq 0}$ is a standard $d$-dimensional Brownian motion, and $\gamma>0$
is the friction coefficient.
Under mild smoothness and growth assumptions on $f$, 
the diffusion process $(V(t),X(t))$ converges a unique stationary distribution known as the \textit{Gibbs
distribution}, 
whose probability
density function $\pi(v,x)\propto e^{-f(x)-\frac{1}{2}\Vert v\Vert^{2}}$
where the $x$-marginal coincides with that of the overdamped Langevin diffusion \cite{herau2004isotropic,pavliotis2014stochastic,mattingly2002ergodicity,Villani2009,DMS2015,RS2018,Eberle}.
Kinetic Langevin diffusion \eqref{underdamped} and its discretizations are known to 
converge to the stationary distribution faster than the overdamped Langevin diffusion \eqref{eq:overdamped-2} 
under some settings
\cite{Eberle,JianfengLu,Ma2019,GGZ}.
Inspired by kinetic Langevin Monte Carlo algorithms in the literature,
we introduce two variants of regime-switching kinetic Langevin Monte Carlo algorithms (Section~\ref{sec:RS-KLMC}).
We first introduce \textit{regime-switching kinetic Langevin dynamics} (RS-KLD) (Section~\ref{sec:RS-KLD}), 
and based on its discretization, \textit{regime-switching kinetic Langevin Monte Carlo} (RS-KLMC) algorithm, which can be viewed as the KLMC algorithm with random stepsizes (Section~\ref{sec:RS-KLMC}). 
Next, we propose \textit{frictional-regime-switching kinetic Langevin dynamics} (FRS-KLD) (Section~\ref{sec:FRS-KLD})
and its associated algorithm \textit{frictional-regime-switching kinetic Langevin Monte Carlo} (FRS-KLMC), 
which can also be viewed as KLMC algorithm with random frictional coefficients (Section~\ref{sec:FRS-KLMC}).

Our contributions can be summarized as follows.

\begin{itemize}
\item 
We propose regime-switching Langevin dynamics (RS-LD), a novel continuous-time regime-switching SDE in the context of Langevin sampling. We show
that its invariant distribution is the Gibbs distribution (Theorem~\ref{thm:invariant:1}).
We obtain non-asymptotic convergence rate for RS-LD (Theorem~\ref{thm:RS-LD}). 
Based on its discretization, we propose regime-switching Langevin Monte Carlo (RS-LMC) algorithm, which can also be viewed as LMC with randomized stepsize. 
We obtain non-asymptotic convergence guarantees
for RS-LMC (Theorem~\ref{thm:non:asymptotic}) and its iteration complexity (Corollary~\ref{cor:iteration:complexity}).
The proof technique is based on conditioning on the regime-switching process, which is a continuous-time Markov chain (CTMC), and then applying
the synchronous coupling approach as in \cite{dalalyan2019user} for the classical LMC. 
Then, we take expectations over the CTMC process, and analyze this expectation by employing the Perron-Frobenius theory, spectral analysis and a series of careful computations. 
\item 
We also propose regime-switching kinetic Langevin dynamics (RS-KLD) and frictional-regime-switching kinetic Langevin dynamics (FRS-KLD). We show the Gibbs distribution is their invariant distributions (Theorem~\ref{thm:invariant:underdamped:2}, Theorem~\ref{thm:invariant:underdamped:1}), and obtain non-asymptotic convergence rate 
(Theorem~\ref{thm:underdamped:alternative}, Theorem~\ref{thm:underdamped:continuous}). 
Based on their discretizations, we propose regime-switching kinetic Langevin Monte Carlo (RS-KLMC) and frictional-regime-switching kinetic Langevin Monte Carlo (RS-KLMC), which can also be viewed as KLMC with randomized stepsize and randomized friction coefficients respectively. We obtain non-asymptotic convergence guarantees (Theorem~\ref{thm:non:asymptotic:klmc}, Theorem~\ref{thm:non:asymptotic:frs-klmc}) and iteration complexities (Corollary~\ref{cor:iteration:complexity:klmc}, Corollary~\ref{cor:iteration:complexity:frs-klmc}). 
The proof technique is based on conditioning on the regime-switching process, which is a continuous-time Markov chain (CTMC), and then applying 
the synchronous coupling approach as in \cite{dalalyan2018kinetic} for the classical KLMC. 
Then, we take expectations over the CTMC process, and analyze
this expectation similarly as for RS-LMC.
\item
We conduct numerical experiments to demonstrate
the efficiency of the proposed algorithms. In a Baysesian linear regression problem, using synthetic data, we compare the performance of our proposed algorithms RS-LMC, RS-KLMC, FRS-KLMC with the classical LMC and KLMC algorithms using mean-squared error (MSE) (Section~\ref{subsec:linear}).
In a Bayesian logistic regression problem, using both synthetic and real data, we report the prediction accuracy of our proposed algorithms, and compare them with the classical methods (Section~\ref{subsec:logistic}).
Our numerical results show that in all the settings, our proposed algorithms can achieve a comparable or superior performance compared to the classical methods.
\end{itemize}

The rest of the paper is organized as follows. 
We first summarize the notations that will be used
in the rest of the paper. 
In Section~\ref{sec:RS-LMC}, we will introduce and study regime-switching Langevin Monte Carlo (RS-LMC) algorithm, based on the discretization of the regime-switching Langevin dynamics (RS-LD). 
In Section~\ref{sec:RS-KLMC}, we will introduce and study regime-switching kinetic Langevin Monte Carlo (RS-KLMC) algorithm, based on the discretization of the regime-switching kinetic Langevin dynamics (RS-KLD).
Numerical experiments will be presented in Section~\ref{sec:numerical}.
Finally, we conclude in Section~\ref{sec:conclusion}.
All the technical proofs will be provided in Appendix~\ref{sec:proofs}.

\paragraph{Notations.}
\begin{itemize}
\item 
For any $x\in\mathbb{R}^{d}$, define $\Vert x\Vert$ as its Euclidean norm.
For any $d$-dimensional random vector $X$,
define its $L^{2}$-norm as $\Vert X\Vert_{2}=\left(\mathbb{E}\Vert X\Vert^{2}\right)^{1/2}$. 
For any matrix $A\in\mathbb{R}^{m\times n}$, we define its Frobenius norm
as $\Vert A\Vert_{F}:=\sqrt{\sum_{i=1}^{m}\sum_{j=1}^{n}|a_{ij}|^{2}}$.
\item 
A differentiable function $f:\mathbb{R}^{d}\rightarrow\mathbb{R}$ is said to be $m$-strongly convex if 
    \[
        f(y) - f(x) - \langle \nabla f(x), y-x \rangle \ge \frac{m}{2}\|y-x\|^2, \quad \text{for any $x,y \in \mathbb{R}^d$},
    \]
and is said to be $M$-smooth if the gradient $\nabla f$ is $M$-Lipschitz continuous:
    \[
        \|\nabla f(y) - \nabla f(x)\| \le M\|y-x\|, \quad \text{for any $x,y \in \mathbb{R}^d$}.
    \]

\item 
Denote $\mathcal{P}_{2}(\mathbb{R}^{d})$
as the space consisting of all the Borel probability measures $\mu$
on $\mathbb{R}^{d}$ with the finite second moment
(based on the Euclidean norm).
For any $\nu_{1},\nu_{2}\in\mathcal{P}_{2}(\mathbb{R}^{d})$, 
the $2$-Wasserstein
distance $\mathcal{W}_{2}$ (see e.g. \cite{villani2008optimal}) between $\nu_{1}$ and $\nu_{2}$ is defined as:
\[
\mathcal{W}_{2}(\nu_{1},\nu_{2}):=\left(\inf\mathbb{E}\left[\Vert Y_{1}-Y_{2}\Vert^{2}\right]\right)^{1/2},
\]
where the infimum is taken over all joint distributions of the random variables $Y_{1},Y_{2}$ with marginal distributions
$\nu_{1},\nu_{2}$ respectively.
\end{itemize}

\section{Regime-Switching Langevin Monte Carlo Algorithms}\label{sec:RS-LMC}

\subsection{Regime-Switching Langevin Dynamics}\label{sec:RS-LD}

We introduce the \textit{regime-switching Langevin dynamics} (RS-LD):
\begin{equation}\label{eq:regime}
dX(t)=-\beta(t)\nabla f(X(t))dt+\sqrt{2\beta(t)}dB_{t},
\end{equation}
where $(B_{t})_{t\geq 0}$ is a standard $d$-dimensional Brownian motion
and $(\beta(t))_{t\geq 0}$ is a positive stochastic process, that is independent
of the Brownian motion $(B_{t})_{t\geq 0}$. In particular, we assume
that there are $N$ regimes $\{\bar{\beta}_{1},\bar{\beta}_{2},\ldots,\bar{\beta}_{N}\}$
and $(\beta(t))_{t\geq 0}$ is a continuous-time Markov process with the finite
state space $\{\bar{\beta}_{1},\bar{\beta}_{2},\ldots,\bar{\beta}_{N}\}$ 
with explicit transition matrix. We assume that $\beta(t)$ has the infinitesimal generator
\begin{equation}\label{eq:generator_mc}
\mathcal{L}_{\beta}g(\bar{\beta}_{i}):=\sum_{j\neq i}q_{ij}\left[g(\bar{\beta}_{j})-g(\bar{\beta}_{i})\right],
\end{equation}
for any $i=1,2,\ldots,N$.
Then the infinitesimal generator of the joint process $(\beta(t),X(t))$ is given by
\begin{align}\label{generator:eqn}
\mathcal{L}g(\bar{\beta}_{i},x)=-\bar{\beta}_{i}\sum_{j=1}^{d}\frac{\partial f}{\partial x_{j}}\frac{\partial g}{\partial x_{j}}
+\bar{\beta}_{i}\sum_{j=1}^{d}\frac{\partial^{2}g}{\partial x_{j}^{2}}
+\sum_{j\neq i}q_{ij}\left[g(\bar{\beta}_{j},x)-g(\bar{\beta}_{i},x)\right],
\end{align}
for any $i=1,2,\ldots,N$ and $x\in\mathbb{R}^{d}$.

\subsubsection{Assumptions}

Throughout our analysis, we impose the following conditions on the potential function $f:\mathbb{R}^d \to \mathbb{R}$. 

\begin{assumption}[Properties of the Potential Function]\label{assump:f}
For some positive constants $m<M$, the twice continuously differentiable function $f$ is $m$-strongly convex and $M$-smooth.
\end{assumption}

\begin{assumption}[Properties of the Regime-Switching Process]\label{assump:ctmc}
The continuous-time Markov chain $(\beta(t))_{t\ge 0}$ in the finite state space $\{\bar{\beta}_1, \dots, \bar{\beta}_N\}$ is irreducible. 
\end{assumption}

Assumption~\ref{assump:f} is often used in the literature of Langevin Monte Carlo sampling, see e.g. \cite{dalalyan2019user,dalalyan2018kinetic,gurbuzbalaban2021decentralized,EXTRALangevin}.
Assumption~\ref{assump:ctmc} is a standard condition for finite-state continuous-time Markov chains. It guarantees two crucial properties: first, the existence of a unique stationary distribution $\psi$, and second, that the generator matrix $\mathbf Q$ defined below has a strictly positive spectral gap. The existence of the spectral gap implies that the process is exponentially ergodic, meaning that the distribution of $\beta(t)$ converges to $\psi$ at an exponential rate in metrics such as the total variation distance (see, e.g., \cite{levin2017markov}).


We introduce the \textit{regime-switching Langevin Monte Carlo} (RSLMC) algorithm, which is a discrete-time approximation of the continuous-time regime-switching Langevin dynamics \eqref{eq:regime}. 
Under Assumption~\ref{assump:f}, the drift and diffusion coefficients of the SDE \eqref{eq:regime} are locally Lipschitz continuous and satisfy a linear growth condition. Therefore, there exists a unique strong solution to the SDE for all time $t \ge 0$ (see, e.g., \cite[Chapter 3]{mao2006stochastic}).

\paragraph{The Generator Matrix.}
The dynamics of the continuous-time Markov chain $(\beta(t))_{t\ge 0}$ is governed by the generator operator $\mathcal{L}_{\beta}$ defined in \eqref{eq:generator_mc}. For our finite state space, this operator has a unique matrix representation, the $N \times N$ generator matrix (or Q-matrix) $\mathbf{Q} = (q_{ij})$, where the off-diagonal entries $q_{ij}$ ($i\neq j$) are the transition rates from state $i$ to $j$, and the diagonal entries are $q_{ii} = -\sum_{j\neq i}q_{ij}$. The total exit rate from state $i$ is thus $q_i := -q_{ii} = \sum_{j \neq i} q_{ij}$.

\subsubsection{Invariant Distribution}

Under Assumptions~\ref{assump:f} and \ref{assump:ctmc}, the regime-switching process $(\beta(t), X(t))$ is known to be exponentially ergodic, which guarantees the existence of a unique stationary distribution \cite[Theorem 2.1]{Shao2015}. In this section, we explicitly identify this unique distribution. We show that it is given by the product measure $\pi=\psi\otimes\pi$, where $\pi\propto e^{-f(x)}$ is the Gibbs distribution and $\psi$ is the stationary distribution of the switching process. This also implies that the marginal stationary distribution for the process $X(t)$ in \eqref{eq:regime} is the Gibbs distribution $\pi$.

\begin{theorem}\label{thm:invariant:1}
Let $\psi=(\psi_{1},\ldots,\psi_{N})$ be the invariant distribution 
for $\beta(t)$, i.e. $\mathbb{P}(\beta(\infty)=\bar{\beta}_{i})=\psi_{i}$
for every $i=1,2,\ldots,N$. Then $\pi=\psi\otimes\pi$, where $\pi\propto e^{-f(x)}$,
is an invariant distribution of the joint process $(\beta(t),X(t))$.
In particular, the Gibbs distribution $\pi\propto e^{-f(x)}$ is an invariant distribution for 
the regime-switching Langevin dynamics $X(t)$.
\end{theorem}

\subsubsection{Convergence Analysis}

Next, we obtain the non-asymptotic 2-Wasserstein convergence guarantees for the continuous-time regime-switching Langevin dynamics $X(t)$ in \eqref{eq:regime} to the Gibbs distribution $\pi$.

\begin{theorem}\label{thm:RS-LD}
For any $t\geq 0$, 
\begin{equation}
\mathcal{W}_{2}(\mathrm{Law}(X(t)),\pi)
\leq\sqrt{\left\langle e^{(\mathbf{Q}-2m\Lambda)t}\mathbf{1},\psi\right\rangle}\mathcal{W}_{2}(\mathrm{Law}(X(0)),\pi),
\end{equation}
where $\Lambda$ is the diagonal matrix with diagonal entries $\bar{\beta}_{i}$, and $\psi$ is the stationary distribution for the process $(\beta(t))_{t\ge 0}$, from which the initial state $\beta(0)$ is drawn.
\end{theorem}


\subsection{Regime-Switching Langevin Monte Carlo Algorithms}\label{sec:RS:LMC}

In this section, we analyze the properties of a discrete-time implementation of the regime-switching Langevin dynamics \eqref{eq:regime}. For computational purposes, the continuous-time process must be discretized. We propose regime-switching Langevin Monte Carlo (RS-LMC) algorithm based on the Euler-Maruyama scheme and provide non-asymptotic guarantees on its sampling error, measured in the $2$-Wasserstein distance. Our analysis adapts the synchronous coupling method, a powerful technique used for analyzing standard Langevin Monte Carlo algorithms, to our regime-switching framework.

Let $\eta > 0$ be a fixed stepsize. Given the current state $(x_k, \beta_k)$ at step $k$, the next state $(x_{k+1}, \beta_{k+1})$ is generated as follows:

\begin{enumerate}
    \item \textbf{Regime Update:} The next regime, $\beta_{k+1}$, is sampled from the current regime, $\beta_k = \bar{\beta}_i$, using a first-order approximation of the true transition probabilities. The transition probabilities for the RS-LMC algorithm are defined as:
    \begin{equation}
        P_{ij}(\eta):= 
        \begin{cases} 
            q_{ij}\eta & \text{if } j \neq i, \\
            1 - q_i\eta & \text{if } j = i,
        \end{cases}
    \end{equation}
    where we assume the stepsize $\eta$ is sufficiently small such that $q_i\eta\le1$ for every $i$.

Our proof is constructed to explicitly handle the error introduced by this approximation. We are able to bound the discrepancy between the approximate discrete process and the true continuous one.

    \item \textbf{Position Update:} The position $x_{k+1}$ is updated using the current regime $\beta_k$:
    \begin{equation}
        x_{k+1} = x_k - \eta \beta_k \nabla f(x_k) + \sqrt{2\eta\beta_k} \xi_k,
    \end{equation}
    where $(\xi_k)_{k\ge0}$ is a sequence of i.i.d. standard Gaussian random vectors in $\mathbb{R}^d$.
\end{enumerate}

Let $\nu_k$ denote the distribution of $(x_n)_{n\ge0}$ at step $k$. Since the regime chain $(\beta_n)_{n\ge0}$ is independent of the position dynamics, we can consider the discretization algorithm in the following way.
Given the regime chain $(\beta_n)_{n\ge0}$, 
we define $\nu_{\beta,k}$ as the distribution 
of $(x_{n})_{n\geq 0}$ at step $k$ conditional on $(\beta_n)_{n\ge0}$.
This procedure defines a Markov chain $(x_n)_{n \ge 0}$ on the state space $\mathbb{R}^d$. Our goal is to bound the 2-Wasserstein distance between $\nu_k$ and the true invariant distribution $\pi$, where $\pi \propto e^{-f(x)}$.

\subsubsection{Convergence Analysis}\label{sec:overdamped_SynchronousCoupling}
To analyze the convergence of the distribution $\nu_k$ to $\pi$, we adapt the synchronous coupling methodology. 
The discrete-time process $(x_n)_{n\ge0}$ is constructed as a numerical approximation whose random components are directly coupled to those of the continuous process. This coupling is specified as follows. The standard Gaussian vector $\xi_k$ used to update the position $x_k$ is generated from the increment of the underlying Brownian motion $B_t$ as
  $\xi_k = \frac{B_{k\eta} - B_{(k-1)\eta}}{\sqrt{\eta}}$ for $k \ge 1$.
This ensures that the random noise in the discrete process is consistent with the continuous-time process, allowing us to analyze their convergence properties effectively.

\paragraph{Notation.}
Before stating the bound, we clarify the notation. Let $\mathbf{Q} = (q_{ij})$ be the $N \times N$ generator matrix of the regime-switching process. The eigenvalues of $\mathbf{Q}$, denoted by $\lambda_i(\mathbf{Q})$, may be complex in general but are known to have non-positive real parts. 
We also denote $\Lambda$ as the diagonal matrix with diagonal entries $\bar{\beta}_{i}$.

\begin{proposition}[Recursive Error Bound for RS-LMC]
\label{prop:recursive_error_bound}
    Let $\mathcal W_2(\nu_{k},\pi)$ denote the 2-Wasserstein distance between the law of $x_{k}$ and stationary distribution $\pi$. For 
    \[
        \eta\le\min\left(\frac{2}{\beta_{\max}(m+M)},\frac1{m\beta_{\max}},-\frac1{2\min_{1\leq i\leq N}\left\{ \operatorname{Re}\left(\lambda_i(\mathbf{Q} - m\Lambda)\right) \right\}}\right),
    \]
   the 2-Wasserstein distance $\mathcal W_2(\nu_{k},\pi)$ is bounded by the following recursion:
\begin{equation}\label{eq:scalar_recursion_new}
        \mathcal W_2^2(\nu_{k},\pi) \le 2\left(1 - \frac\alpha2\eta\right)^k \mathcal W_2^2(\nu_{0},\pi)+ C \eta,
    \end{equation}
    where 
    \begin{align*}
    C&:=2\left(1.65 M\sqrt{d}\frac{\beta_{\max}^{3/2}}{m\beta_{\min}}\right)^2,
    \\
        \alpha &:= -\max_{1\leq i\leq N} \left\{ \operatorname{Re}\left(\lambda_i(\mathbf{Q} - m\Lambda)\right) \right\}, \\
        C_M &:= \frac{1}{2} \max_{1\leq i\leq N}\left\{|\lambda_i(\mathbf{Q} - m\Lambda)|^2\right\} + \|\mathbf Q^2\|+2m\|\mathbf Q\Lambda\|+\frac12m^2\|\Lambda^2\|,
    \end{align*}
    where $\beta_{\max}:=\max_{1\leq i\leq N}\bar{\beta}_{i}$ and $\beta_{\min}:=\min_{1\leq i\leq N}\bar{\beta}_{i}$.
\end{proposition}

By unrolling the recursion from the proposition, we can establish an upper bound on $\mathcal{W}_{2}(\nu_{K}, \pi)$ after a total of $K$ iterations, which provides the non-asymptotic convergence guarantee of our RS-LMC algorithm to the target distribution.

\begin{theorem}[Non-Asymptotic Error Bound for RS-LMC]\label{thm:non:asymptotic}
Under the same conditions as in Proposition~\ref{prop:recursive_error_bound}, the distribution $\nu_K$ of the $K$-th iterate of the RS-LMC algorithm satisfies:
\begin{equation}
    \mathcal{W}_{2}(\nu_{K}, \pi) \le \left(1-\frac\alpha2\eta\right)^{K/2} \mathcal{W}_{2}(\nu_{0}, \pi) + \sqrt{\frac{2C\eta}{\alpha}},
\end{equation}
where the constants $C$ and $\alpha$ are explicitly defined in Proposition \ref{prop:recursive_error_bound}.
\end{theorem}

By using Theorem~\ref{thm:non:asymptotic}, we can obtain the iteration complexity of RS-LMC algorithm.

\begin{corollary}[Iteration Complexity for RS-LMC]\label{cor:iteration:complexity}
Under the assumptions in Theorem~\ref{thm:non:asymptotic}, for any given accuracy level $\epsilon>0$, 
we have $\mathcal{W}_{2}(\nu_{K}, \pi) \le \epsilon$ 
provided that
\[
\eta \le \frac{\epsilon^2 \alpha}{8C},
\]
and
\[
    K \ge \frac{4}{\alpha\eta}\log\left(\frac{2\mathcal{W}_{2}(\nu_{0}, \pi)}{\epsilon}\right).
\]
In particular, with $\eta = \frac{\epsilon^2 \alpha}{8C}$, the iteration complexity is given by
\[
    K= \mathcal O\left(\frac{1}{\epsilon^2}\log\left(\frac{1}{\epsilon}\right)\right).
\]
\end{corollary}

The iteration complexity derived in Corollary~\ref{cor:iteration:complexity}, $K=\tilde{\mathcal{O}}\left(\frac{32C}{\alpha^2\epsilon^2}\right)$ where the notation $\tilde{\mathcal{O}}$ ignores the logarithmic dependence, reveals the algorithm's dependence on the key problem parameters.

\begin{enumerate}
    \item \textbf{Dependence on dimension $d$:} The complexity $K$ is proportional to the constant $C$. From its definition, we can see that $C\propto d$. Therefore, the iteration complexity $K$ is linear with respect to the dimension $d$.

    \item \textbf{Dependence on $f$:} The dependence on function $f$ is captured by the strong convexity constant $m$ and the smoothness constant $M$. The constant $C$ depends on the square of the condition number, i.e., $C \propto (\frac{M}{m})^2$. The convergence rate $\alpha$ also depends on $m$. Consequently, the iteration complexity $K$ has a polynomial dependence on the condition number $M/m$.

    \item \textbf{Dependence on the CTMC dynamics:} The dynamics of the continuous-time Markov chain (CTMC) is determined by the generator matrix $\mathbf{Q}$ and the regime values in the diagonal matrix $\Lambda$. The iteration complexity $K$ is inversely proportional to $\alpha^2$, where $\alpha = -\max_{1\leq i\leq N} \{ \operatorname{Re}(\lambda_i(\mathbf{Q} - m\Lambda)) \}$. A process with a larger spectral gap or larger regime values $\{\bar\beta_i\}$ will result in a larger $\alpha$, leading to faster convergence and a smaller number of required iterations.
\end{enumerate}


\section{Regime-Switching Kinetic Langevin Monte Carlo Algorithms}\label{sec:RS-KLMC}

\subsection{Regime-Switching Kinetic Langevin Dynamics}\label{sec:RS-KLD}

In this section, we introduce the \textit{regime-switching kinetic Langevin dynamics} (RS-KLD):
\begin{align}
&dV(t)=-\gamma\beta(t)V(t)dt-\beta(t)\nabla f(X(t))dt+\sqrt{2\gamma\beta(t)}dB_{t},\nonumber
\\
&dX(t)=\beta(t)V(t)dt,\label{underdamped:regime:switching:2}
\end{align}
where $(B_{t})_{t\geq 0}$ is a standard $d$-dimensional Brownian motion, and $(\beta(t))_{t\geq 0}$
is a positive stochastic process, that is independent
of the Brownian motion $(B_{t})_{t\geq 0}$. 
In particular, we assume
that there are $N$ regimes $\{\bar{\beta}_{1},\bar{\beta}_{2},\ldots,\bar{\beta}_{N}\}$
and $(\beta(t))_{t\geq 0}$ is a continuous-time Markov process with the finite
state space $\{\bar{\beta}_{1},\bar{\beta}_{2},\ldots,\bar{\beta}_{N}\}$ 
with explicit transition matrix, and $(\beta(t))_{t\geq 0}$ is characterized by 
the infinitesimal generator given in \eqref{eq:generator_mc}. 

\subsubsection{Assumptions}

For the analysis of the RS-KLD and its discretization, we impose the same set of assumptions as in the overdamped case. Specifically, we require Assumption \ref{assump:f} on the potential function $f$, and Assumption \ref{assump:ctmc} on the continuous-time Markov chain $(\beta(t))_{t\ge 0}$.

\subsubsection{Invariant Distribution}

Under the same assumptions on $f$ and the irreducibility of the switching process $(\beta(t))_{t\ge 0}$, the joint process $(\beta(t), V(t), X(t))$ can be shown to be exponentially ergodic. This guarantees the existence of a unique stationary distribution \cite{Shao2015}. In what follows, we explicitly identify this distribution.
In particular, we will show that $\psi\otimes\mathcal{N}(0,I_{d})\otimes\pi$, where $\mathcal{N}(0,I_{d})\otimes\pi\propto e^{-f(x)-\frac{1}{2}\Vert v\Vert^{2}}$,
is an invariant distribution of the joint process $(\beta(t),V(t),X(t))$.
In particular, the Gibbs distribution $\propto e^{-f(x)}$ is an invariant distribution for 
the regime-switching kinetic Langevin dynamics $X(t)$ in \eqref{underdamped:regime:switching:2}.

\begin{theorem}\label{thm:invariant:underdamped:2}
Let $\psi=(\psi_{1},\ldots,\psi_{N})$ be the invariant distribution 
for $\beta(t)$, i.e. $\mathbb{P}(\beta_{\infty}=\bar{\beta}_{i})=\psi_{i}$
for every $i=1,2,\ldots,N$. Then $\psi\otimes\mathcal{N}(0,I_{d})\otimes\pi$, where $\mathcal{N}(0,I_{d})\otimes\pi\propto e^{-f(x)-\frac{1}{2}\Vert v\Vert^{2}}$,
is an invariant distribution of the joint process $(\beta(t),V(t),X(t))$.
In particular, the Gibbs distribution $\pi\propto e^{-f(x)}$ is an invariant distribution for 
the regime-switching kinetic Langevin dynamics $X(t)$.
\end{theorem}

\subsubsection{Convergence Analysis}

Next, we obtain the non-asymptotic 2-Wasserstein convergence guarantees for the continuous-time regime-switching kinetic Langevin dynamics $X(t)$ in \eqref{underdamped:regime:switching:2} to the Gibbs distribution $\pi$.

\begin{theorem}\label{thm:underdamped:alternative}
Let $V(0)\sim\mathcal{N}(0,I_{d})$ and $\beta(0)\sim\psi$. 
For any $t\geq 0$,
\begin{align}
&\mathcal{W}_{2}(\mathrm{Law}(X(t)),\pi)
\nonumber
\\
&\leq
\frac{\sqrt{2(\lambda_{+}^{2}+\lambda_{-}^{2})}}{\lambda_{+}-\lambda_{-}}
\left(\left\langle \exp\left\{\left(\mathbf{Q}+\frac{2(\lambda_{-}^{2}-m)\vee(M-\lambda_{+}^{2})}{\lambda_{+}-\lambda_{-}}\Lambda\right)t\right\}\mathbf{1},\psi\right\rangle\right)^{1/2}
\mathcal{W}_{2}(\mathrm{Law}(X(0)),\pi),
\end{align}
where $\lambda_{+}$ and $\lambda_{-}$ are two arbitrary positive numbers
such that $\lambda_{+}+\lambda_{-}=\gamma$ with $\lambda_{+}>\lambda_{-}$,
and $\Lambda$ is the diagonal matrix with diagonal entries $\bar{\beta}_{i}$, and $\psi$ is the stationary distribution for the process $(\beta(t))_{t\ge 0}$, from which the initial state $\beta(0)$ is drawn.
\end{theorem}

\begin{remark}
Assume $\gamma^{2}\geq 2(M+m)$. By taking
$\lambda_{-}=\frac{\gamma-\sqrt{\gamma^{2}-4m}}{2}\geq\frac{m}{\gamma}$ in Theorem~\ref{thm:underdamped:alternative}, we get
\begin{align}
\mathcal{W}_{2}(\mathrm{Law}(X(t)),\pi)
\leq
\left(\frac{2\gamma^{2}-4m}{\gamma^{2}-4m}\right)^{1/2}
\left(\left\langle \exp\left\{\left(\mathbf{Q}-\frac{2m}{\gamma}\Lambda\right)t\right\}\mathbf{1},\psi\right\rangle\right)^{1/2}
\mathcal{W}_{2}(\mathrm{Law}(X(0)),\pi).
\end{align}
\end{remark}

\subsection{Regime-Switching Kinetic Langevin Monte Carlo Algorithm}\label{sec:RS-KLMC:discrete}

In this section, we introduce the regime-switching kinetic Langevin Monte Carlo (RS-KLMC) algorithm, a discrete-time implementation of the regime-switching kinetic Langevin dynamics \eqref{underdamped:regime:switching:2}. Our analysis will aim to provide non-asymptotic guarantees on its sampling error, measured in the $2$-Wasserstein distance.

Let $\eta > 0$ be a fixed stepsize. Given the current state $(x_k, v_k, \beta_k)$ at step $k$, the next state $(x_{k+1}, v_{k+1}, \beta_{k+1})$ is generated as follows:

\begin{enumerate}
    \item \textbf{Regime Update:} The next regime, $\beta_{k+1}$, is sampled from the current regime, $\beta_k = \bar{\beta}_i$, using a first-order approximation of the true transition probabilities. The transition probabilities for the RS-KLMC algorithm are defined as:
    \begin{equation}
        P_{ij}(\eta) := 
        \begin{cases} 
            q_{ij}\eta & \text{if } j \neq i, \\
            1 - q_i\eta & \text{if } j = i,
        \end{cases}
    \end{equation}
    where we assume the stepsize $\eta$ is sufficiently small such that $q_i\eta \le 1$ for every $i$.

    \item \textbf{Position and Velocity Update:} The position $x_{k+1}$ and velocity $v_{k+1}$ are updated using the current regime $\beta_k$. We build upon the discretization scheme introduced for KLMC in \cite{dalalyan2018kinetic}. 

    We update the position and velocity as a single block:
    \begin{align}
        v_{k+1} &= \psi_0(\beta_k\eta) v_k - \psi_1(\beta_k\eta) \nabla f(x_k) + \sqrt{2\gamma}\xi_{k+1}^{(v)}, \nonumber\\
        x_{k+1} &= x_k + \psi_1(\beta_k\eta) v_k - \psi_2(\beta_k\eta) \nabla f(x_k) + \sqrt{2\gamma}\xi_{k+1}^{(x)},\label{discrete:scheme:beta:underdamped}
    \end{align}
    where for any $t\geq 0$,
     \begin{align*}
        \psi_0(t) = e^{-\gamma t}, \quad
        \psi_1(t) = \int_0^{t} \psi_0(s) ds = \frac{1 - e^{-\gamma t}}{\gamma}, \quad
        \psi_2(t) = \int_0^{t} \psi_1(s) ds = \frac{t - \psi_1(t)}{\gamma},
    \end{align*}
   and $\left(\xi_{k+1}^{(v)}, \xi_{k+1}^{(x)}\right)$ is a $2d$-dimensional centered Gaussian random vector, and its covariance matrix is given by $\int_{0}^{\beta_{k}\eta}[\psi_{0}(t),\psi_{1}(t)]^{\top}[\psi_{0}(t),\psi_{1}(t)]dt$; see \cite[p. 1961-1962]{dalalyan2018kinetic}. 
\end{enumerate}

Note that the discretization scheme \eqref{discrete:scheme:beta:underdamped} is 
finer than the Euler-Maruyama discretization scheme
and it is equivalent to the following formulation. 
For any $k$, $(v_{k},x_{k})$ has the same distribution as $({V}(k\eta),{X}(k\eta))$, where for any $k\eta\leq t<(k+1)\eta$, $({V}(t),{X}(t))$ satisfies the SDE:
\begin{align}
&d{V}(t)=-\gamma\beta_{\lfloor t/\eta \rfloor}{V}(t)dt-\beta_{\lfloor t/\eta \rfloor}\nabla f(X(t))dt+\sqrt{2\gamma\beta_{\lfloor t/\eta \rfloor}}dB_{t},
\\
&d{X}(t)=\beta_{\lfloor t/\eta \rfloor}{V}(t)dt.
\end{align}

Let $\nu_k$ and $\mu_k$ denote the marginal distributions of the position $x_k$ and velocity $v_k$ at step $k$, respectively. Since the dynamics of position and velocity depend on the realization of the regime chain $(\beta_n)_{n\ge0}$, we can first analyze the algorithm conditional on this path.
Given a realization of the regime chain $(\beta_n)_{n\ge0}$, we define $(\nu_{\beta,k}, \mu_{\beta,k})$ as the joint distribution of $(x_k, v_k)$ at step $k$. This procedure defines a Markov chain $(x_n, v_n)_{n\ge0}$ in the state space $\mathbb{R}^d \times \mathbb{R}^d$. Our ultimate goal remains to bound the distance between the marginal position distribution $\nu_k$ and the true invariant distribution $\pi$, where $\pi \propto e^{-f(x)}$.

\subsubsection{Convergence Analysis}\label{sec:kinetic_SynchronousCoupling}


\begin{proposition}[Recursive Error Bound for RS-KLMC]
\label{prop:recursive_bound_RS_KLMC}
Let $\nu_k$ be the marginal distribution of the position $x_k$ after $k$ iterations of the RS-KLMC algorithm. Under Assumptions~\ref{assump:f} and \ref{assump:ctmc}, and for a sufficiently small stepsize $\eta$ satisfying
\[
    \eta \le \min\left\{\frac{m}{4\beta_{\max}\gamma M}, \frac{m\gamma}{(m^2+1.5M\gamma^2)\beta_{\max}}, \frac{2\gamma}{m\beta_{\min}}\right\},
\]
the squared 2-Wasserstein distance is bounded by:
\[
    \mathcal W_2^2(\nu_k,\pi) \le 4\left(1-\frac{\alpha}{2}\eta\right)^k\mathcal W_2^2(\nu_0,\pi) + \frac{2C}{\gamma^2}\eta^2,
\]
where $\alpha$ is the spectral decay rate and $C$ is a constant that are defined as:
\begin{align*}
    \alpha := -\max_{1\leq i\leq N}\left\{ \operatorname{Re}\left(\lambda_i\left(\mathbf{Q} - \frac{m}{\gamma}\Lambda\right)\right) \right\},\qquad 
    C := \frac{18 M^2\beta_{\max}^4 d}{m^2\beta_{\min}^2},
\end{align*}
where $\beta_{\max}:=\max_{1\leq i\leq N}\bar{\beta}_{i}$ and $\beta_{\min}:=\min_{1\leq i\leq N}\bar{\beta}_{i}$.
\end{proposition}

By unrolling the recursion from the proposition, we can establish an upper bound on $\mathcal{W}_{2}(\nu_{k}, \pi)$ after a total of $k$ iterations, which provides the non-asymptotic convergence guarantee of our RS-KLMC algorithm to the target distribution.

\begin{theorem}[Non-Asymptotic Error Bound for RS-KLMC]
\label{thm:non:asymptotic:klmc}
Under the same conditions as in Proposition~\ref{prop:recursive_bound_RS_KLMC}, the marginal distribution $\nu_K$ of the $K$-th iterate of the RS-KLMC algorithm satisfies:
\begin{equation}
    \mathcal{W}_2(\nu_K, \pi) \le 2\left(1-\frac{\alpha}{2}\eta\right)^{K/2} \mathcal{W}_2(\nu_0, \pi) + \sqrt{\frac{2C}{\gamma^2}}\eta,
\end{equation}
where the constants $C$ and $\alpha$ are explicitly defined in Proposition~\ref{prop:recursive_bound_RS_KLMC}.
\end{theorem}

By using Theorem~\ref{thm:non:asymptotic:klmc}, we can obtain the iteration complexity of RS-KLMC algorithm.

\begin{corollary}[Iteration Complexity for RS-KLMC]
\label{cor:iteration:complexity:klmc}
Under the assumptions in Theorem~\ref{thm:non:asymptotic:klmc}, for any given accuracy level $\epsilon>0$, we have $\mathcal{W}_{2}(\nu_{K}, \pi) \le \epsilon$ provided that
\[
\eta \le \frac{\epsilon\gamma}{2\sqrt{2C}},
\]
and
\[
    K \ge \frac{4}{\alpha\eta}\log\left(\frac{4\mathcal{W}_{2}(\nu_{0}, \pi)}{\epsilon}\right).
\]
In particular, with $\eta = \frac{\epsilon\gamma}{2\sqrt{2C}}$, the iteration complexity is given by
\[
    K = \mathcal O\left(\frac{1}{\epsilon}\log\left(\frac{1}{\epsilon}\right)\right).
\]
\end{corollary}

The iteration complexity derived in Corollary~\ref{cor:iteration:complexity:klmc}, $K= \tilde{\mathcal{O}}\left(\frac{1}{\alpha\epsilon}\sqrt{\frac{2C}{\gamma^2}}\right)$ where the notation $\tilde{\mathcal{O}}$ ignores the logarithmic dependence, reveals the algorithm's dependence on the key problem parameters, and shows an improvement over the overdamped case.

\begin{enumerate}
    \item \textbf{Dependence on dimension $d$:} The complexity $K$ is proportional to $\sqrt{\frac{2C}{\gamma^2}}$, which is proportional to $\sqrt{d}$. Therefore, the iteration complexity $K$ has a square root dependence on the dimension $d$, an improvement over the linear dependence in the overdamped case.

    \item \textbf{Dependence on $f$:} The dependence on $f$ is captured by the condition number $\kappa = M/m$. The constant $\sqrt{\frac{2C}{\gamma^2}}$ is proportional to $\kappa$. The complexity $K$ is therefore proportional to $\kappa$, which is an improvement over the $\kappa^2$ dependence in the overdamped case.

    \item \textbf{Dependence on the CTMC dynamics:} This dependence is structurally similar to the overdamped case. The complexity $K$ is inversely proportional to the spectral decay rate $\alpha$, where $\alpha = -\max_{1\leq i\leq N} \{ \operatorname{Re}(\lambda_i(\mathbf{Q} - \frac{m}{\gamma}\Lambda)) \}$. A process with a larger spectral gap (a more negative real part of the eigenvalues of $\mathbf{Q}$) or larger regime values $\{\bar\beta_i\}$ will result in a larger $\alpha$, leading to faster convergence.
\end{enumerate}

\subsection{Frictional-Regime-Switching Kinetic Langevin Dynamics}\label{sec:FRS-KLD}

In this section, we introduce a variant of the \textit{regime-switching kinetic Langevin dynamics} (RS-KLD), where the friction coefficient $\gamma(t)$ follows a regime-switching process, and we name this variant \textit{frictional-regime-switching kinetic Langevin dynamics} (FRS-KLD):
\begin{align}
&dV(t)=-\gamma(t)V(t)dt-\nabla f(X(t))dt+\sqrt{2\gamma(t)}dB_{t},\nonumber
\\
&dX(t)=V(t)dt,\label{underdamped:regime:switching}
\end{align}
where $(B_{t})_{t\geq 0}$ is a standard $d$-dimensional Brownian motion, $(\gamma(t))_{t\geq 0}$
is a positive stochastic process, that is independent
of the Brownian motion $(B_{t})_{t\geq 0}$. 
In particular, we assume
that there are $N$ regimes $\{\bar{\gamma}_{1},\bar{\gamma}_{2},\ldots,\bar{\gamma}_{N}\}$
and $(\gamma(t))_{t\geq 0}$ is a continuous-time Markov process with the finite
state space $\{\bar{\gamma}_{1},\bar{\gamma}_{2},\ldots,\bar{\gamma}_{N}\}$ 
with explicit transition matrix. We assume that the diffusion part has the infinitesimal generator
$\mathcal{L}_1$,
\[
\mathcal{L}_1g(\bar{\gamma}_{i},v,x)
=-\bar{\gamma}_{i}\sum_{j=1}^{d}v_{j}\frac{\partial g}{\partial v_{j}}
-\sum_{j=1}^{d}\frac{\partial f}{\partial x_{j}}\frac{\partial g}{\partial v_{j}}
+\bar{\gamma}_{i}\sum_{j=1}^{d}\frac{\partial^{2}g}{\partial v_{j}^{2}}
+\sum_{j=1}^{d}v_{j}\frac{\partial g}{\partial x_{j}},
\] 
and $\gamma(t)$ has the infinitesimal generator
\begin{align}\label{gamma:infinitesimal}
\mathcal{L}_2g(\bar{\gamma}_{i})=\sum_{j\neq i}q_{ij}\left[g(\bar{\gamma}_{j})-g(\bar{\gamma}_{i})\right],
\end{align}
for any $i=1,2,\ldots,N$.
Then the infinitesimal generator of the joint process $(\gamma(t),V(t),X(t))$ is given by
\begin{equation}\label{eq:full_generator_kinetic}
\begin{aligned}
\mathcal{L}g(\bar{\gamma}_{i},v,x)
&=(\mathcal L_1+\mathcal L_2)g(\bar{\gamma}_{i},v,x)=
-\bar{\gamma}_{i}\sum_{j=1}^{d}v_{j}\frac{\partial g}{\partial v_{j}}
-\sum_{j=1}^{d}\frac{\partial f}{\partial x_{j}}\frac{\partial g}{\partial v_{j}}
+\bar{\gamma}_{i}\sum_{j=1}^{d}\frac{\partial^{2}g}{\partial v_{j}^{2}}
+\sum_{j=1}^{d}v_{j}\frac{\partial g}{\partial x_{j}}
\nonumber
\\
&\qquad\qquad\qquad
+\sum_{j\neq i}q_{ij}\left[g(\bar{\gamma}_{j},v,x)-g(\bar{\gamma}_{i},v,x)\right],
\end{aligned}
\end{equation}
for any $i=1,2,\ldots,N$ and $v,x\in\mathbb{R}^{d}$.

\subsubsection{Assumptions}

For the analysis of the FRS-KLD and its discretization, we impose the same set of assumptions as in the overdamped case. Specifically, we require Assumption~\ref{assump:f} on the potential function $f$, and Assumption~\ref{assump:ctmc} on the continuous-time Markov chain $(\gamma(t))_{t\ge 0}$ (with $\beta(t)$ replaced by $\gamma(t)$).

\subsubsection{Invariant Distribution}

Under the same assumptions on $f$ and the irreducibility of the switching process $(\gamma(t))_{t\ge 0}$, the joint process $(\gamma(t), V(t), X(t))$ can be shown to be exponentially ergodic. This guarantees the existence of a unique stationary distribution \cite{Shao2015}. In what follows, we explicitly identify this distribution.
In particular, we will show that $\psi\otimes\mathcal{N}(0,I_{d})\otimes\pi$, where $\mathcal{N}(0,I_{d})\otimes\pi\propto e^{-f(x)-\frac{1}{2}\Vert v\Vert^{2}}$,
is an invariant distribution of the joint process $(\gamma(t),V(t),X(t))$.
In particular, the Gibbs distribution $\pi\propto e^{-f(x)}$ is an invariant distribution for 
the frictional-regime-switching kinetic Langevin dynamics $X(t)$ in \eqref{underdamped:regime:switching}.

\begin{theorem}\label{thm:invariant:underdamped:1}
Let $\psi=(\psi_{1},\ldots,\psi_{N})$ be the invariant distribution 
for $\gamma(t)$, i.e. $\mathbb{P}(\gamma_{\infty}=\bar{\gamma}_{i})=\psi_{i}$
for every $i=1,2,\ldots,N$. Then $\psi\otimes\mathcal{N}(0,I_{d})\otimes\pi$, where $\mathcal{N}(0,I_{d})\otimes\pi\propto e^{-f(x)-\frac{1}{2}\Vert v\Vert^{2}}$,
is an invariant distribution of the joint process $(\gamma(t),V(t),X(t))$.
In particular, the Gibbs distribution $\propto e^{-f(x)}$ is an invariant distribution for 
the frictional-regime-switching kinetic Langevin dynamics $X(t)$.
\end{theorem}

\subsubsection{Convergence Analysis}

Next, we obtain the non-asymptotic 2-Wasserstein convergence guarantees for the continuous-time frictional-regime-switching kinetic Langevin dynamics $X(t)$ in \eqref{underdamped:regime:switching} to the Gibbs distribution $\pi$.

\begin{theorem}\label{thm:underdamped:continuous}
Assume $\min_{1\leq i\leq N}\bar{\gamma}_{i}\geq\max(\sqrt{2},\sqrt{m+M})$.
Let $V(0)\sim\mathcal{N}(0,I_{d})$ and $\gamma(0)\sim\psi$. 
For any $t\geq 0$,
\begin{equation}
\mathcal{W}_{2}(\mathrm{Law}(X(t)),\pi)
\leq\sqrt{\left\langle e^{(\mathbf{Q}-2m\Lambda_{\gamma}^{-1})t}\mathbf{1},\psi\right\rangle}\mathcal{W}_{2}(\mathrm{Law}(X(0)),\pi),
\end{equation}
where $\Lambda_{\gamma}^{-1}$ is the diagonal matrix with diagonal entries $1/\bar{\gamma}_{i}$, and $\psi$ is the stationary distribution for the process $(\gamma(t))_{t\ge 0}$, from which the initial state $\gamma(0)$ is drawn.
\end{theorem}

\subsection{Frictional-Regime-Switching Kinetic Langevin Monte Carlo Algorithm}\label{sec:FRS-KLMC}

In this section, we propose \textit{frictional-regime-switching kinetic Langevin Monte Carlo} (FRS-KLMC) algorithm, based on discretization of the frictional-regime-switching kinetic Langevin dynamics, as introduced in Section~\ref{sec:RS-KLD}. We adopt the discretization scheme 
of the kinetic Langevin Monte Carlo (KLMC) algorithm from \cite[Eqn. (8)]{dalalyan2018kinetic} to our setting where the friction coefficient is a time-varying process.
Let $\eta > 0$ be a fixed stepsize. 

\begin{enumerate}
    \item \textbf{Regime Update:} The friction regime $\gamma_{k+1}$ is sampled from the current regime $\gamma_k$ using the first-order approximation of the transition probabilities, $P_{ij}(\eta)$, derived from the generator matrix $\mathbf{Q}$. This step is identical to the regime update in the RS-LMC algorithm for the overdamped case.

    \item \textbf{Position and Velocity Update:} Given the regime chain $(\gamma_n)_{n\ge0}$ and the state $(x_{\gamma,k}, v_{\gamma,k})$, the next state $(x_{\gamma,k+1}, v_{\gamma,k+1})$ is generated as follows: Let $\gamma_k$ be the current friction coefficient. 
    
    We update the position and velocity as a single block:
    \begin{equation}\label{discretization:underdamped}
        \begin{pmatrix} v_{\gamma,k+1} \\ x_{\gamma,k+1} \end{pmatrix}
        =
        \begin{pmatrix}
            \psi_0(\eta, \gamma_k)v_{\gamma,k} - \psi_1(\eta, \gamma_k)\nabla f(x_{\gamma,k}) \\
            x_{\gamma,k} + \psi_1(\eta, \gamma_k)v_{\gamma,k} - \psi_2(\eta, \gamma_k)\nabla f(x_{\gamma,k})
        \end{pmatrix}
        + \sqrt{2\gamma_k} \begin{pmatrix} \xi_{k+1}^{(v)} \\ \xi_{k+1}^{(x)} \end{pmatrix},
    \end{equation}
    where for any $t\geq 0$ and $\gamma>0$,
     \begin{align*}
        \psi_0(t,\gamma) = e^{-\gamma t}, \,
        \psi_1(t,\gamma) = \int_0^{t} \psi_0(s,\gamma) ds = \frac{1 - e^{-\gamma t}}{\gamma},\,
        \psi_2(t,\gamma) = \int_0^{t} \psi_1(s,\gamma) ds = \frac{t - \psi_1(t)}{\gamma},
    \end{align*}
   and $\left(\xi_{k+1}^{(v)}, \xi_{k+1}^{(x)}\right)$ is a $2d$-dimensional centered Gaussian random vector, and its covariance matrix is given by $\int_{0}^{\eta}[\psi_{0}(t,\gamma_{k}),\psi_{1}(t,\gamma_{k})]^{\top}[\psi_{0}(t,\gamma_{k}),\psi_{1}(t,\gamma_{k})]dt$.
\end{enumerate}

Note that the discretization scheme \eqref{discretization:underdamped} is 
finer than the Euler-Maruyama discretization scheme
and is equivalent to the following formulation. 
For any $k$, $(v_{\gamma,k},x_{\gamma,k})$ has the same distribution as $({V}(k\eta),{X}(k\eta))$, where for any $k\eta\leq t<(k+1)\eta$, $({V}(t),{X}(t))$ satisfies the SDE:
\begin{align}
&d{V}(t)=-\gamma_{\lfloor t/\eta \rfloor}{V}(t)dt-\nabla f(X(t))dt+\sqrt{2\gamma_{\lfloor t/\eta \rfloor}}dB_{t},
\\
&d{X}(t)={V}(t)dt.
\end{align}

The discretization procedure \eqref{discretization:underdamped} defines the FRS-KLMC algorithm. The subsequent analysis will aim to prove a non-asymptotic bound on the 2-Wasserstein distance for the law of the iterates $(x_k, v_k, \gamma_k)$ generated by this algorithm.

\subsubsection{Convergence Analysis}

\begin{proposition}[Recursive Error Bound for FRS-KLMC]
\label{prop:frs_klmc_final}
Let $\nu_K$ be the marginal distribution of the position $x_K$ after $K$ iterations of the FRS-KLMC algorithm. Under Assumptions~\ref{assump:f} and \ref{assump:ctmc}, and assuming that for $\min_{1\leq i\leq N} \bar\gamma_i \ge \max(\sqrt{2}, \sqrt{M+m})$, for 
\[
    \eta\le\min\left(\sqrt{\frac{m}{1.5M\gamma_{\max}}},\frac{m\gamma_{\min}}{m^2+1.5M\gamma_{\max}^2},\frac{m}{4\gamma_{\max}M}\right),
\]
the squared 2-Wasserstein distance is bounded by:
\begin{align*}
    \mathcal W_2^2(\nu_{K},\pi) 
     \le & 2\left(1-\frac\alpha2\eta\right)^K \mathcal W_2^2(\nu_0,\pi) 
     + \frac{2\gamma_{\max}^2M^2\eta^4}{9m^2}\left(2\sqrt d+\sqrt{\sum_{i=1}^N\psi_i\bar\gamma_i^2}\mathcal W_2(\nu_0,\pi)\right)^2,
\end{align*}
where $\gamma_{\max}:=\max_{1\leq i\leq N}\bar{\gamma}_{i}$, $\gamma_{\min}:=\min_{1\leq i\leq N}\bar{\gamma}_{i}$ and 
\[
    \alpha := -\max_{1\leq i\leq N} \left\{ \operatorname{Re}\left(\lambda_i(\mathbf{Q} - 2m\Lambda_{\gamma}^{-1})\right) \right\},\quad
        \Lambda_{\gamma}^{-1}:=\text{diag}\left(\frac1{\bar\gamma_i},\ldots,\frac1{\bar\gamma_N}\right).
\]
\end{proposition}

By unrolling the recursion from the proposition, we can establish an upper bound on $\mathcal{W}_{2}(\nu_{k}, \pi)$ after a total of $k$ iterations, which provides the non-asymptotic convergence guarantee of our FRS-KLMC algorithm to the target distribution.

\begin{theorem}[Non-Asymptotic Error Bound for FRS-KLMC]\label{thm:non:asymptotic:frs-klmc}
Under the same conditions as in Proposition~\ref{prop:frs_klmc_final}, the marginal distribution $\nu_K$ of the $K$-th iterate of the FRS-KLMC algorithm satisfies:
\begin{equation}
    \mathcal{W}_2(\nu_K, \pi) \le \sqrt{2} \left(1-\frac{\alpha}{2}\eta\right)^{K/2} \mathcal{W}_2(\nu_0,\pi) + C_B \eta^2,
\end{equation}
where the constant $C_B$ is given by
\[
    C_B := \frac{\sqrt{2}\gamma_{\max}M}{3m}\left(2\sqrt d+\sqrt{\sum_{i=1}^N\psi_i\bar\gamma_i^2}\mathcal{W}_2(\nu_0,\pi)\right),
\]
and $\alpha$ is defined as in Proposition~\ref{prop:frs_klmc_final}.
\end{theorem}

By using Theorem~\ref{thm:non:asymptotic:frs-klmc}, we can obtain the iteration complexity of FRS-KLMC algorithm.

\begin{corollary}[Iteration Complexity for FRS-KLMC]\label{cor:iteration:complexity:frs-klmc}
Under the assumptions in Theorem~\ref{thm:non:asymptotic:frs-klmc}, for any given accuracy level $\epsilon > 0$, we can achieve $\mathcal{W}_2(\nu_K, \pi) \le \epsilon$ by choosing the stepsize $\eta$ and the number of iterations $K$ appropriately. Specifically, if we choose the stepsize $\eta$ such that
\[
    \eta \le \sqrt{\frac{\epsilon}{2C_B}},
\]
then the required number of iterations $K$ is
\[
    K \ge \frac{4}{\alpha\eta} \log\left(\frac{2\sqrt{2}\mathcal{W}_2(\nu_0,\pi)}{\epsilon}\right).
\]
In particular, by choosing $\eta = \mathcal{O}(\sqrt{\epsilon})$, the iteration complexity is given by
\[
    K = \mathcal{O}\left(\frac{1}{\sqrt{\epsilon}}\log\left(\frac{1}{\epsilon}\right)\right).
\]
\end{corollary}

The iteration complexity derived in Corollary~\ref{cor:iteration:complexity:frs-klmc}, which is $K=\tilde{\mathcal{O}}(\frac{\sqrt{C_B}}{\alpha\sqrt{\epsilon}})$ where the $\tilde{\mathcal{O}}$ notation ignores logarithmic factors, reveals the algorithm's dependence on key problem parameters and demonstrates a notable acceleration compared to the other proposed algorithms.

\begin{enumerate}
    \item \textbf{Dependence on dimension $d$ and $f$}: The complexity $K$ is proportional to $\sqrt{C_B}$, which in turn depends polynomially on the dimension $d$ and the condition number $\kappa=M/m$. The constant $C_B$ is proportional to $\kappa$ and contains a $\sqrt{d}$ term. Consequently, the iteration complexity $K$ has a dependence of roughly $\mathcal{O}(\sqrt{\kappa}d^{1/4})$, which is a significant improvement over the $\mathcal{O}(\kappa\sqrt{d})$ dependence of the RS-KLMC algorithm and the $\mathcal{O}(\kappa^2 d)$ dependence of the RS-LMC algorithm.

    \item \textbf{Dependence on the CTMC dynamics}: The complexity $K$ is inversely proportional to the spectral decay rate $\alpha$, where $\alpha = -\max_{1\leq i\leq N} \{ \operatorname{Re}(\lambda_i(\mathbf{Q} - 2m\Lambda_{\gamma}^{-1})) \}$. This means that the dynamics of the continuous-time Markov chain (CTMC), determined by the generator matrix $\mathbf{Q}$ and the friction regimes $\{\bar\gamma_i\}$, are crucial for convergence. A process with a larger spectral gap (a more negative real part for the eigenvalues of $\mathbf{Q} - 2m\Lambda_{\gamma}^{-1}$) will result in a larger $\alpha$, leading to faster convergence and a smaller number of required iterations.

    \item \textbf{Dependence on friction coefficients $\{\bar\gamma_i\}$}: The friction coefficients affect the complexity in two ways. They directly influence the spectral decay rate $\alpha$ through the matrix $\Lambda_{\gamma}^{-1}$, and they also impact the constant term $C_B$ via $\gamma_{\max}$. Therefore, the entire set of friction values, not just the minimum, plays a role in determining the algorithm's overall efficiency.
\end{enumerate}


\section{Numerical Experiments}\label{sec:numerical}

This section provides numerical experiments to demonstrate the efficiency of our proposed algorithms. First, in Section~\ref{subsec:linear}, we study a Bayesian linear regression problem with synthetic data; compare the \emph{mean squared error} (MSE) of our proposed regime-switching Langevin Monte Carlo (RS-LMC) algorithm (Section~\ref{sec:RS:LMC}) with the classical LMC, and compare the regime-switching kinetic Langevin Monte Carlo (RS-KLMC) algorithm (Section~\ref{sec:RS-KLMC:discrete}), and frictional-regime-switching kinetic Langevin Monte Carlo (FRS-KLMC) algorithm (Section~\ref{sec:FRS-KLMC}) with the classical KLMC. Subsequently, in Section~\ref{subsec:logistic}, we demonstrate the performance of our methods on a Bayesian logistic regression problem. We report the prediction \emph{accuracy} calculated as the proportion of correct labels in the entire dataset using both synthetic and real-world data.


\subsection{Bayesian Linear Regression}
\label{subsec:linear}

\begin{figure}[htbp]
\centering
\includegraphics[scale=0.4]{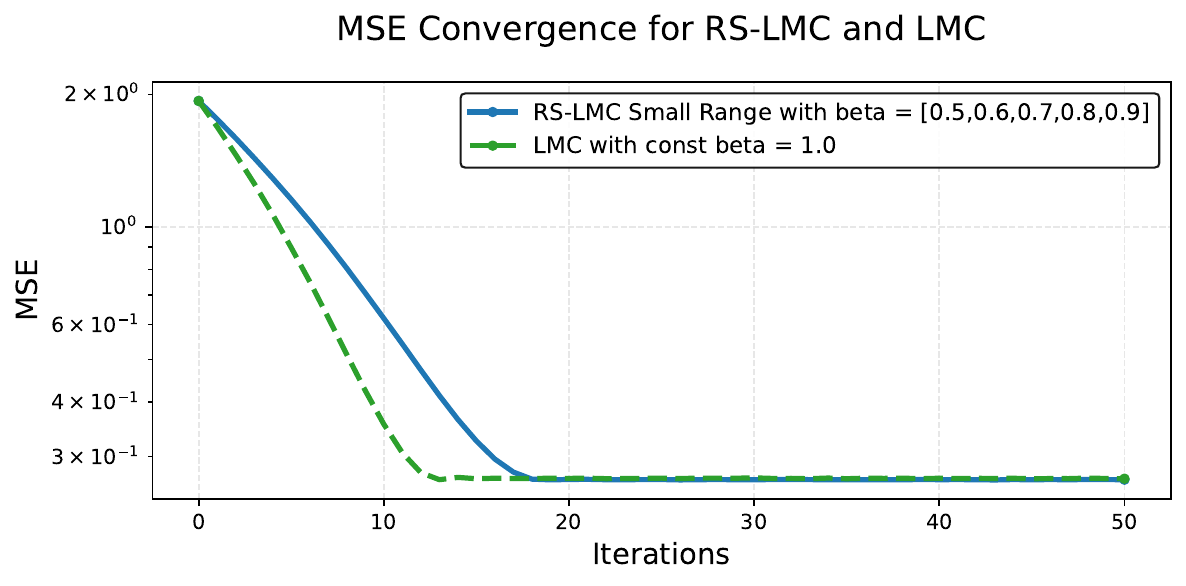} 
\includegraphics[scale=0.4]{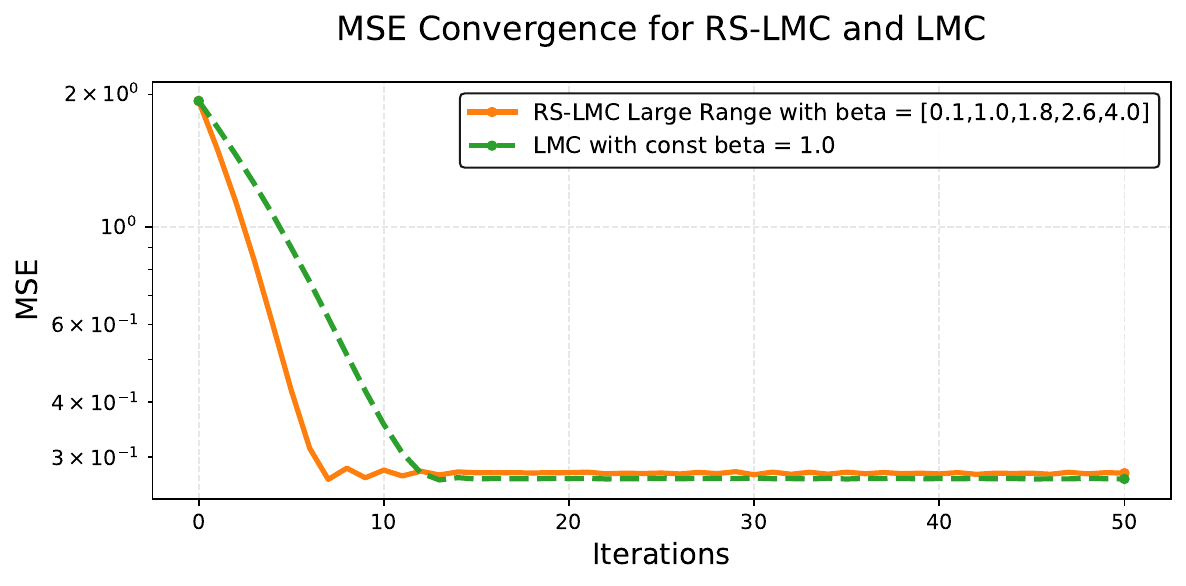} 
\caption{MSE for RS-LMC and LMC.}
\label{fig:mse:lmc}
\end{figure}

In this section, we consider the Bayesian linear regression model as follows:
\begin{equation}
y_j=x_*^{\top} a_j+\delta_j,\quad \delta_j \sim \mathcal{N}(0,0.25), \quad a_j \sim \mathcal{N}\left(0, 0.5 I_3\right), \quad x_*=[1,-0.7,0.5]^{\top}, \quad j = 1,\ldots,n,
\end{equation}
where $\mathbf{1}$ denotes an all-one vector, and the prior distribution of $a_j\sim \mathcal{N}\left(0, \lambda I_3\right)$ is Gaussian, with $I_3$ being the $3 \times 3$ identity matrix. 
Our goal is to sample the posterior distribution given by
\begin{equation}
\pi(a) \propto \exp \left\{-\frac{1}{2} \sum_{j=1}^n\left(y_j-x^{\top} a_j\right)^2-\frac{1}{2\lambda}\Vert a \Vert^2\right\},
\end{equation}
where $n$ is the total number of data points in the training set. In order to present the performance of convergence, we compute the MSE at the $k$-th iterate defined by the following formula:
\begin{equation}
\mathrm{MSE}_k:=\frac{1}{n} \sum_{j=1}^n\left(y_j-\left(x_k\right)^{\top} a_j\right)^2.
\end{equation}
In this experiment, we design switching regimes with small and large values of $\bar{\beta}_{i}$, $i=1,\ldots,N$, and two generator matrices for RS-LMC. In particular, we take the state space $\{\bar{\beta}_{i}:i=1,\ldots,N\}$ of the regime process $(\beta(t))_{t\geq 0}$ as:
\begin{equation}\label{beta:small:large}
\beta_{\mathrm{small}} := \{0.5,0.6,0.7,0.8,0.9\}, \quad \beta_{\mathrm{large}} := \{0.1,1.0,1.8,2.6,4.0\},
\end{equation}
and the generator matrices  $\mathbf{Q}_1$ and $\mathbf{Q}_2$ as follows:
\begin{equation}\label{Q:eqn}
\mathbf{Q}_{1} = {\footnotesize
\begin{bmatrix}
 -0.6 & 0.2 & 0.2 & 0.1 & 0.1 \\
 0.1 & -0.5 & 0.2 & 0.1 & 0.1 \\
0.1 & 0.1 & -0.5 & 0.2 & 0.1 \\
0.1 & 0.1 & 0.2 & -0.6 & 0.2 \\
0.1 & 0.1 & 0.2 & 0.2 & -0.6 
\end{bmatrix}  
},
\quad 
\mathbf{Q}_{2}  = {\footnotesize
\begin{bmatrix}
-0.5 & 0.2 & 0.1 & 0.1 & 0.1 \\
0.1 & -0.5 & 0.2 & 0.1 & 0.1 \\
0.1 & 0.1 & -0.6 & 0.2 & 0.2 \\
0.1 & 0.1 & 0.2 & -0.7 & 0.3 \\
0.1 & 0.1 & 0.2 & 0.3 & -0.7
\end{bmatrix}
}.
\end{equation}
We implement the RS-LMC and LMC algorithms using the state space and generator matrices described above in \eqref{beta:small:large}-\eqref{Q:eqn} and summarize our numerical results for RS-LMC and LMC in Figure~\ref{fig:mse:lmc}.

We observe from Figure~\ref{fig:mse:lmc} that if we choose the range of the regimes of $(\beta(t))_{t\geq 0}$, i.e. the range of its state space $\{\bar{\beta}_{i}:i=1,\ldots,N\}$, to be wide (orange line), RS-LMC can achieve faster convergence compared to LMC; however, the convergence rate of RS-LMC is worse than that of LMC by choosing the values of the regime parameters, i.e. the values in the state space of $(\beta(t))_{t\geq 0}$ (blue line), to be small. It confirms our theoretical result in Corollary~\ref{cor:iteration:complexity} that larger $\bar{\beta}_{i}$'s induce a larger $\alpha$ in Corollary~\ref{cor:iteration:complexity}, which can lead to faster convergence. 

\begin{figure}[htbp]
\centering
\includegraphics[scale=0.33]{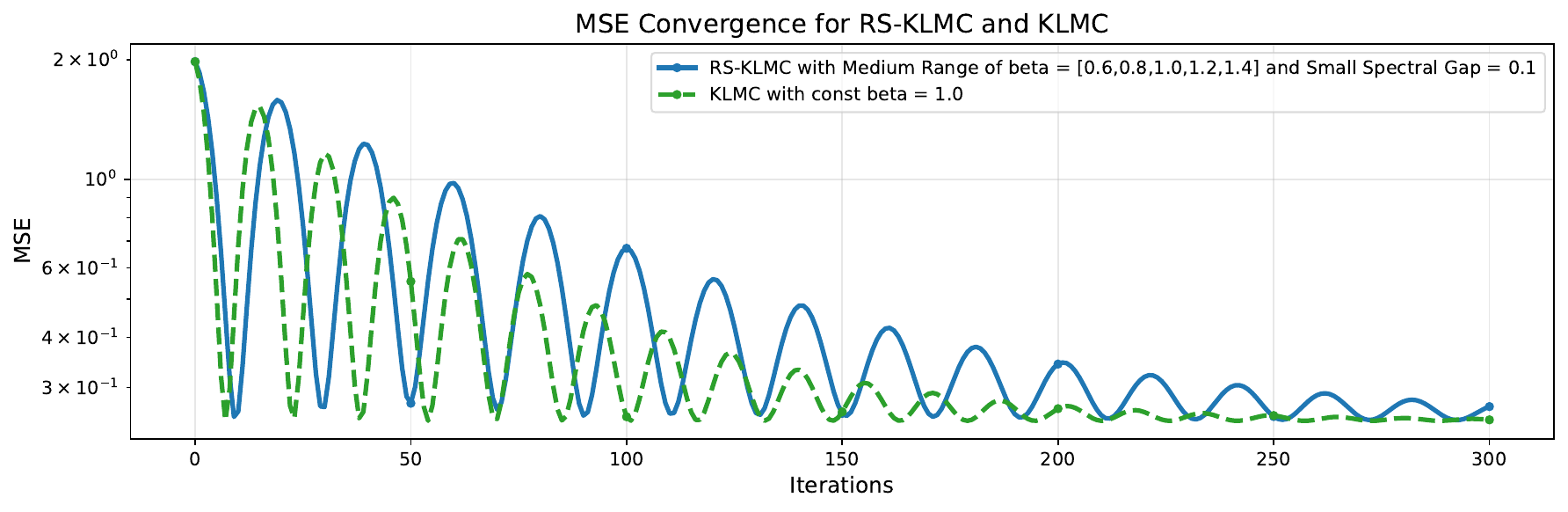} 
\includegraphics[scale=0.33]{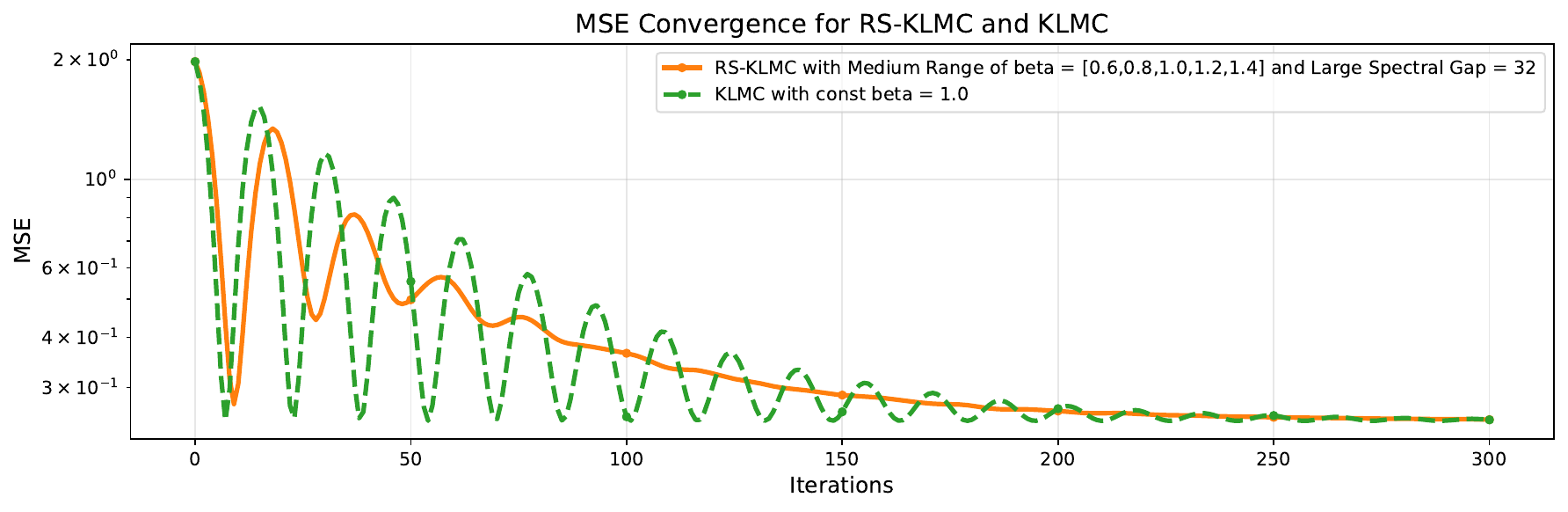} 
\caption{MSE for RS-KLMC and KLMC.}
\label{fig:mse:klmc}
\end{figure}

In the next experiment, we compare RS-KLMC to KLMC. We fix the state space of the regime process $(\beta(t))_{t\geq 0}$ to concentrate around $1.0$ as the average such that the regime-switching range (state space) is $\{0.6, 0.8, 1.0, 1.2, 1.4\}$. We investigate the impact of the spectrum of the generator matrices $\mathbf{Q}$ on the performance of the proposed algorithms.
In particular, we choose
$$
\begin{aligned}
\mathbf{Q}_{\mathrm{small}} = {\footnotesize
\begin{bmatrix}
-0.6 & 0.2 & 0.2 & 0.1 & 0.1 
\\
0.1 & -0.5 & 0.2 & 0.1 & 0.1
\\ 
0.1 & 0.1 & -0.5 & 0.2 & 0.1
\\
0.1 & 0.1 & 0.2 & -0.6 & 0.2
\\
0.1 & 0.1 & 0.2 & 0.2 & -0.6
\end{bmatrix}
}, \quad 
\mathbf{Q}_{\mathrm{large}} = 
{\footnotesize
\begin{bmatrix}
-32.0 & 8.0 & 8.0 & 8.0 & 8.0
\\
8.0 & -32.0 & 8.0 & 8.0 & 8.0
\\ 
8.0 & 8.0 & -32.0 & 8.0 & 8.0
\\
8.0 & 8.0 & 8.0 & -32.0 & 8.0
\\
8.0 & 8.0 & 8.0 & 8.0 & -32.0
\end{bmatrix},
}
\end{aligned}
$$
where the matrix $\mathbf{Q}_{\mathrm{small}}$ is chosen such that it has a relatively small spectral gap $\lambda_{\mathrm{small}} = 0.1$ and the matrix $\mathbf{Q}_{\mathrm{large}}$ is chosen such that it has a relatively large spectral gap $\lambda_{\mathrm{large}} = 32$. Moreover, we fix the friction coefficient $\gamma = 1.5$ in the experiment to freeze its effect on the convergence. We summarize the MSE convergence results in Figure~\ref{fig:mse:klmc} for RS-KLMC and KLMC.

We observe in Figure~\ref{fig:mse:klmc} that RS-KLMC with the generator matrix $\mathbf{Q}_{\mathrm{large}}$ whose spectral gap is large can accelerate convergence compared to KLMC. Moreover, RS-LMC with the generator matrix $\mathbf{Q}_{\mathrm{small}}$ obtains a comparable performance. This numerical observation validates our theoretical results in Corollary~\ref{cor:iteration:complexity:klmc} that the algorithm with the generator matrix equipped with a larger spectral gap induces a larger $\alpha$ in Corollary~\ref{cor:iteration:complexity:klmc} which can lead to faster convergence. 

In the third experiment, we explore the convergence of FRS-KLMC when the friction coefficient $(\gamma(t))_{t\geq 0}$ is regime-switching and compare it with KLMC without regime-switching. 
For comparison, we design small and large friction regime ranges, i.e. the state space $\{\bar{\gamma}_{i}:i=1,\ldots,N\}$ of $(\gamma(t))_{t\geq 0}$, as the following:
$$
\gamma_{\mathrm{small}} := \{0.05, 0.08, 0.1, 0.12\}, \quad \gamma_{\mathrm{large}} := \{8.0, 10.0, 12.0, 16.0\}.
$$
In addition, we fix the generator matrix as 
$$
\mathbf{Q}_{\mathrm{large}}=
\begin{bmatrix}
-36.0 & 12.0 & 12.0 & 12.0
\\
12.0 & -36.0 & 12.0 & 12.0
\\
12.0 & 12.0 & -36.0 & 12.0
\\
12.0 & 12.0 & 12.0 & -36.0
\end{bmatrix},
$$
such that it has a large spectral gap $\lambda_{\mathrm{large}} = 48$.

\begin{figure}[htbp]
\centering
\includegraphics[scale=0.33]{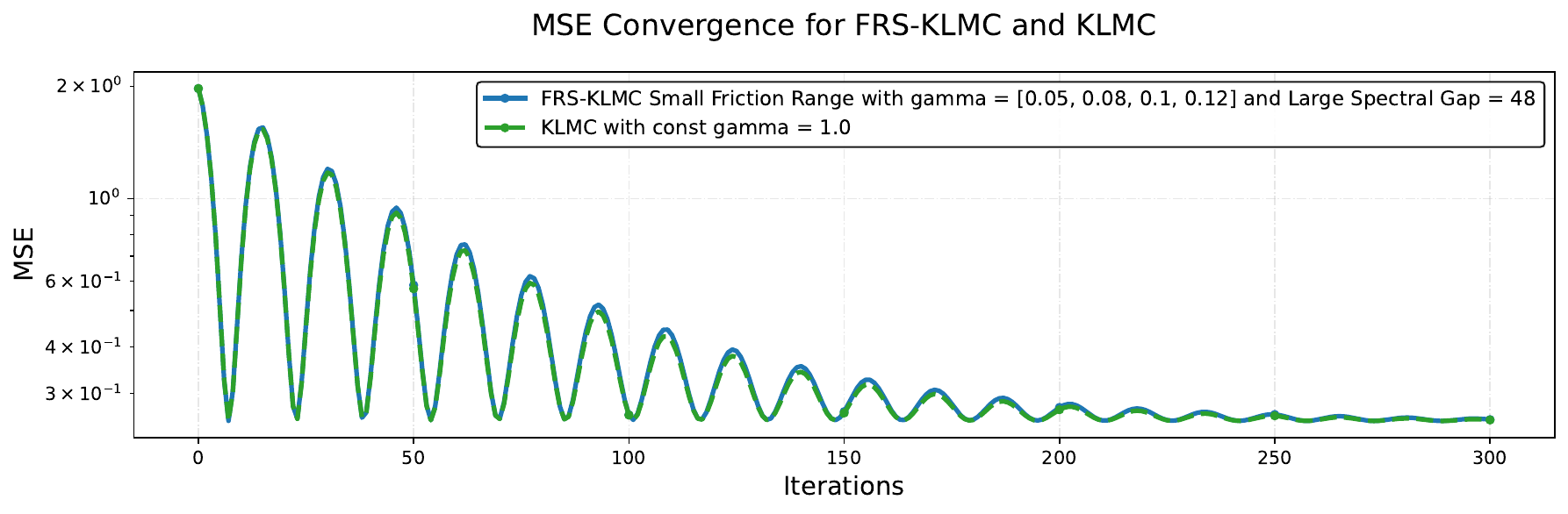} 
\includegraphics[scale=0.33]{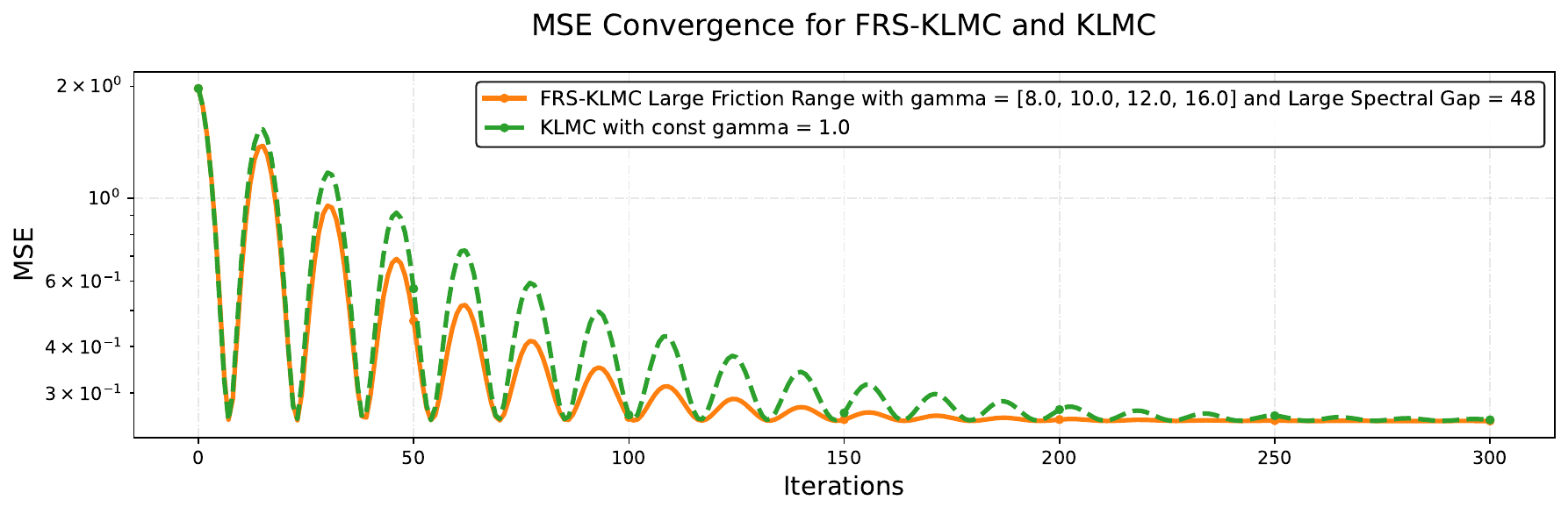} 
\caption{MSE for FRS-KLMC and KLMC.}
\label{fig:mse:flmc}
\end{figure}

We observe from the plots in Figure~\ref{fig:mse:flmc} that even if the algorithm with the generator metrix is equipped with a large spectral gap, 
it is unable to provide acceleration 
when the friction regimes have a narrower range.

On the other hand, if the friction regime spans over some relatively larger values, FRS-KLMC can accelerate the convergence in this Bayesian linear regression task. This also confirms our theoretical conclusion from Corollary~\ref{cor:iteration:complexity:frs-klmc} that the set of friction values plays an important role in the algorithm's performance.


\subsection{Bayesian Logistic Regression}
\label{subsec:logistic}

In this section, we aim to test the performance of our algorithms in binary classification problems by considering the Bayesian logistic regression model on both synthetic and real data (Iris\footnote{Iris - UCI Machine Learning Repository, \url{ https://archive.ics.uci.edu/dataset/53/iris}} and MAGIC Gamma Telescope\footnote{MAGIC Gamma Telescope - UCI Machine Learning Repository, \url{ http://archive.ics.uci.edu/dataset/159/magic+gamma+telescope}}).

Suppose we have access to a dataset $Z=\left\{z_j\right\}_{j=1}^n$ where $z_j=\left(X_j, y_j\right), X_j \in \mathbb{R}^d$ are the features and $y_j \in\{0,1\}$ are the labels with the assumption that $X_j$ are independent and the probability distribution of $y_j$ given $X_j$ and the regression coefficients $c \in \mathbb{R}^d$ are given by
\begin{equation}
\mathbb{P}\left(y_j=1 \mid X_j, c\right)=\frac{1}{1+e^{-c^{\top} X_j}},
\end{equation}
where the prior distribution is Gaussian $p(c) \sim \mathcal{N}\left(0, \lambda I_3\right)$ for some $\lambda>0$, where $I_3$ is the $3 \times 3$ identity matrix. Our goal for the Bayesian logistic regression problem is to sample from $\pi(c) \propto e^{-f(c)}$, where the negative log likelihood $f(c)$ is defined as:
\begin{equation}
f(c):=-\sum_{j=1}^n \log p\left(y_j \mid X_j, c\right)-\log p(c)=\sum_{j=1}^n \log \left(1+e^{-c^{\top} X_j}\right)+ \frac{1}{2 \lambda}\|c\|^2.
\end{equation}
In the experiment with synthetic data, we use $20,000$ samples.
In the experiments using real data, 
the dataset MAGIC Gamma Telescope has $19,020$ samples and $10$ features, and the dataset Iris has $150$ samples and $4$ features. 
To efficiently implement our algorithms, instead of using the full gradient, 
we employ a stochastic gradient using mini-batches with batch-size $b\ll n$ in our experiments; see e.g. \cite{Raginsky,GGZ}. 
As the classical LMC with stochastic gradients and the classical KLMC with stochastic gradients
are known as \textit{stochastic gradient Langevin dynamics} (SGLD) and \textit{stochastic gradient Hamiltonian Monte Carlo} (SGHMC), respectively, in the literature, see e.g. \cite{Raginsky,GGZ},
we name our proposed regime-switching algorithms with stochastic gradient as
\textit{regime-switching stochastic gradient Langevin dynamics} (RS-SGLD), 
\textit{regime-switching stochastic gradient Hamiltonian Monte Carlo} (RS-SGHMC),
and 
\textit{frictional-regime-switching stochastic gradient Hamiltonian Monte Carlo} (FRS-SGHMC).
In the following experiments, we use a stepsize $\eta = 10^{-4}$, a batch-size $b = 100$ for synthetic data and dataset MAGIC Gamma Telescope, which have a larger sample set, and a batch-size $b = 50$ for the dataset Iris.

We provide two comparisons: one between RS-SGHMC, FRS-SGHMC, and RS-SGLD; and another between RS-SGHMC, FRS-SGHMC, and SGHMC, or between RS-SGLD and SGLD. To demonstrate the efficiency of the regime-switching mechanism, we choose the state space $\{\overline{\beta}_i: i=1,\ldots,N\}$ such that its entries concentrate around the constant $\bar{\beta} := 1$ for comparison with SGHMC. Likewise, the state space $\{\overline{\gamma}_i: i=1,\ldots,N\}$ for FRS-SGHMC is selected such that its entries concentrate around the constant friction $\bar{\gamma} := 0.65$ used in our RS-SGHMC setting. Moreover, the generator matrices are chosen to be
$$
\mathbf{Q}_{\overline{\beta}}={\footnotesize
\begin{bmatrix}
0.6 & 0.2 & 0.2 & 0.1 & 0.1
\\
0.1 & -0.5 & 0.2 & 0.1 & 0.1
\\
0.1 & -0.5 & 0.2 & 0.1 & 0.1     
\\
0.1 & 0.1 & 0.2 & 0.2 & -0.6
\end{bmatrix}
}, \quad
\mathbf{Q}_{\overline{\gamma}}= {\footnotesize
\begin{bmatrix}
-0.6 & 0.2 & 0.2 & 0.2
\\
0.1 & -0.5 & 0.2 & 0.2
\\
0.1 & 0.1 & -0.5 & 0.3   
\\
0.1 & 0.1 & 0.3 & -0.5
\end{bmatrix}
}.
$$
This setup ensures that any performance differences are mainly attributable to the regime-switching mechanism itself. 

\paragraph{Synthetic Data.} In this example with $d=3$, we first generate $n = 20,000$ synthetic data by the following model
$$
X_j \sim \mathcal{N}\left(0,2 I_3\right), \quad p_j \sim \mathcal{U}(0,1), \quad y_j
= \begin{cases}1 & \text { if } p_j \leq \frac{1}{1+e^{-c^{\top} X_j}} \\ 0 & \text { otherwise }
\end{cases},
$$
where $\mathcal{U}(0,1)$ is the uniform distribution on $[0,1]$ and the prior distribution of $c \in \mathbb{R}^3$ is Gaussian $c \sim \mathcal{N}(0,\lambda I_3)$ with $\lambda = 2$. We execute the algorithms with a stepsize of $\eta = 10^{-4}$, a batch-size of $b = 20$, and for $2000$ iterations. 

\begin{figure}[htbp]
\centering
\includegraphics[scale=0.38]{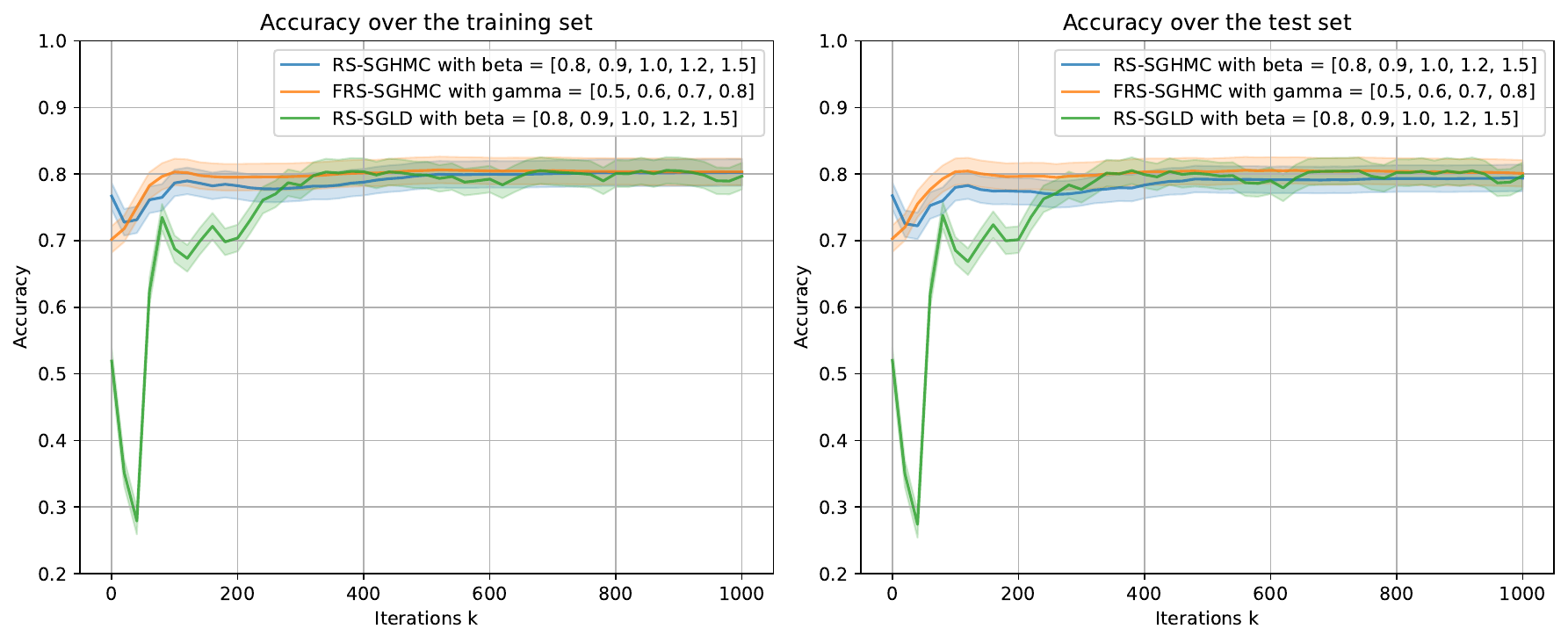}  
\caption{Comparisons within regime-switching algorithms over the synthetic data.}
\label{fig:syn:flmc_lmc:1}
\end{figure}

The results of the comparison between RS-SGHMC, FRS-SGHMC, and RS-SGLD are presented in Figure~\ref{fig:syn:flmc_lmc:1}.

\begin{figure}[htbp]
\centering
\includegraphics[scale=0.38]
{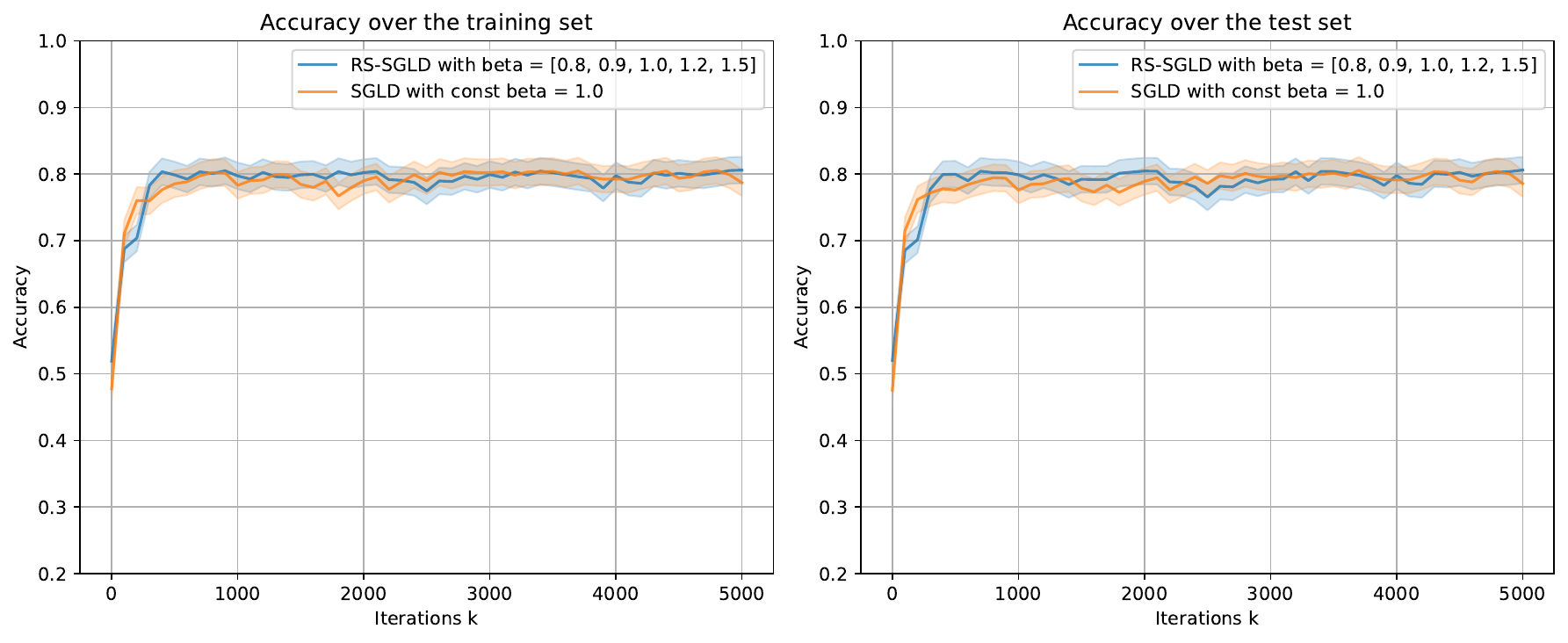} 
\includegraphics[scale=0.38]{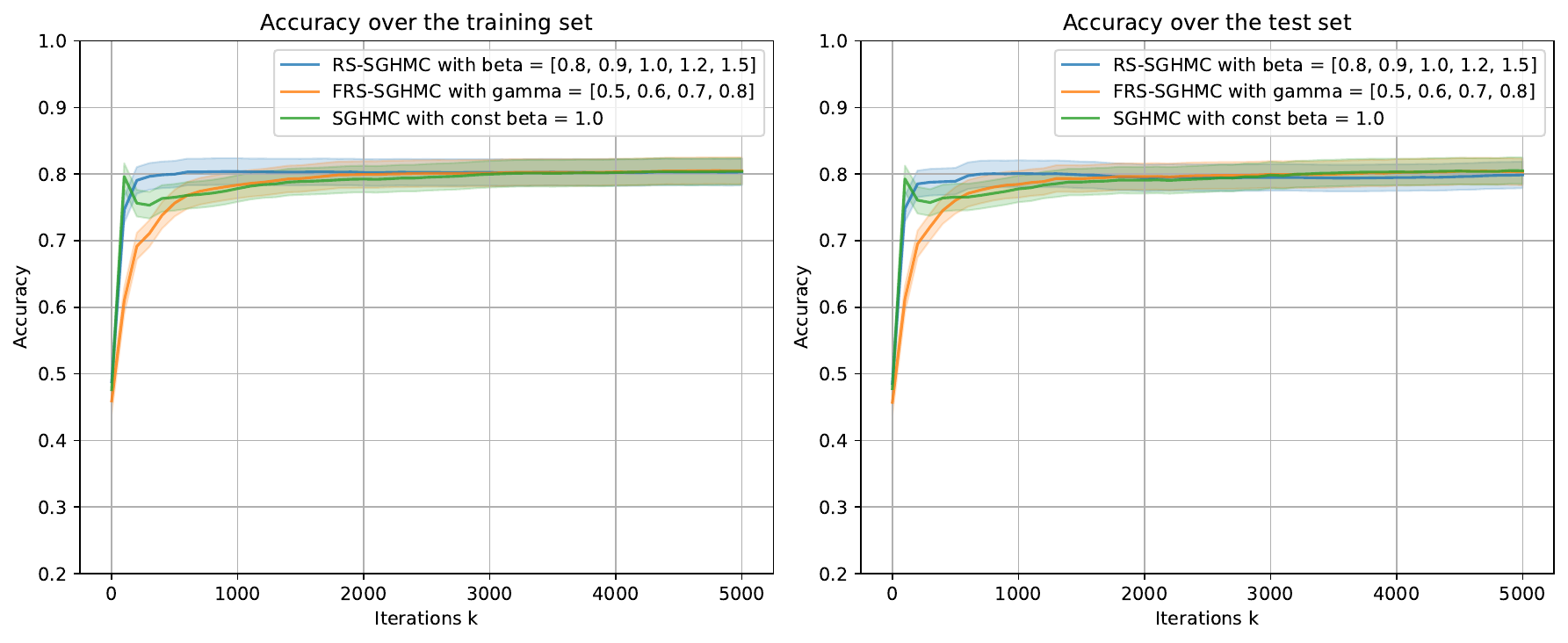}
\caption{Comparisons between regime-switching (RS-SGLD, RS-SGHMC, FRS-SGHMC) and non-regime-swithcing (SGLD, SGHMC) algorithms over the synthetic data.}
\label{fig:syn:flmc_lmc:2}
\end{figure}

We also present the comparison between RS-SGHMC, FRS-SGHMC, and SGHMC, as well as between RS-SGLD and SGLD in Figure~\ref{fig:syn:flmc_lmc:2}.

Two key observations can be made from Figure~\ref{fig:syn:flmc_lmc:1} and Figure~\ref{fig:syn:flmc_lmc:2}. First, the superior performance of RS-SGHMC and FRS-SGHMC over RS-SGLD (Figure~\ref{fig:syn:flmc_lmc:1}) demonstrates that momentum-based and non-reversible RS-SGHMC and FRS-SGHMC can achieve acceleration, as in the case of classical KMLC discussed in~\cite{Ma2019,GGZ,GGZ2}. Second, even with conservatively chosen parameters, such that the state space $\{\overline{\beta}_i: i=1,\ldots,N\}$ narrowly concentrates around $\bar{\beta} := 1$ and the friction state space $\{\overline{\gamma}_i: i=1,\ldots,N\}$ for FRS-SGHMC narrowly concentrates around $\bar{\gamma} := 0.65$, both RS-SGHMC and FRS-SGHMC achieve higher accuracy than SGHMC (Figure~\ref{fig:syn:flmc_lmc:2}). However, RS-SGLD and SGLD have comparable performance in this experiment. This indicates that the both regime-switching and frictional-regime-switching SGHMC algorithms can provide a distinct performance advantage under this conservative setting.

\begin{figure}[htbp]
\centering
\includegraphics[scale=0.45]{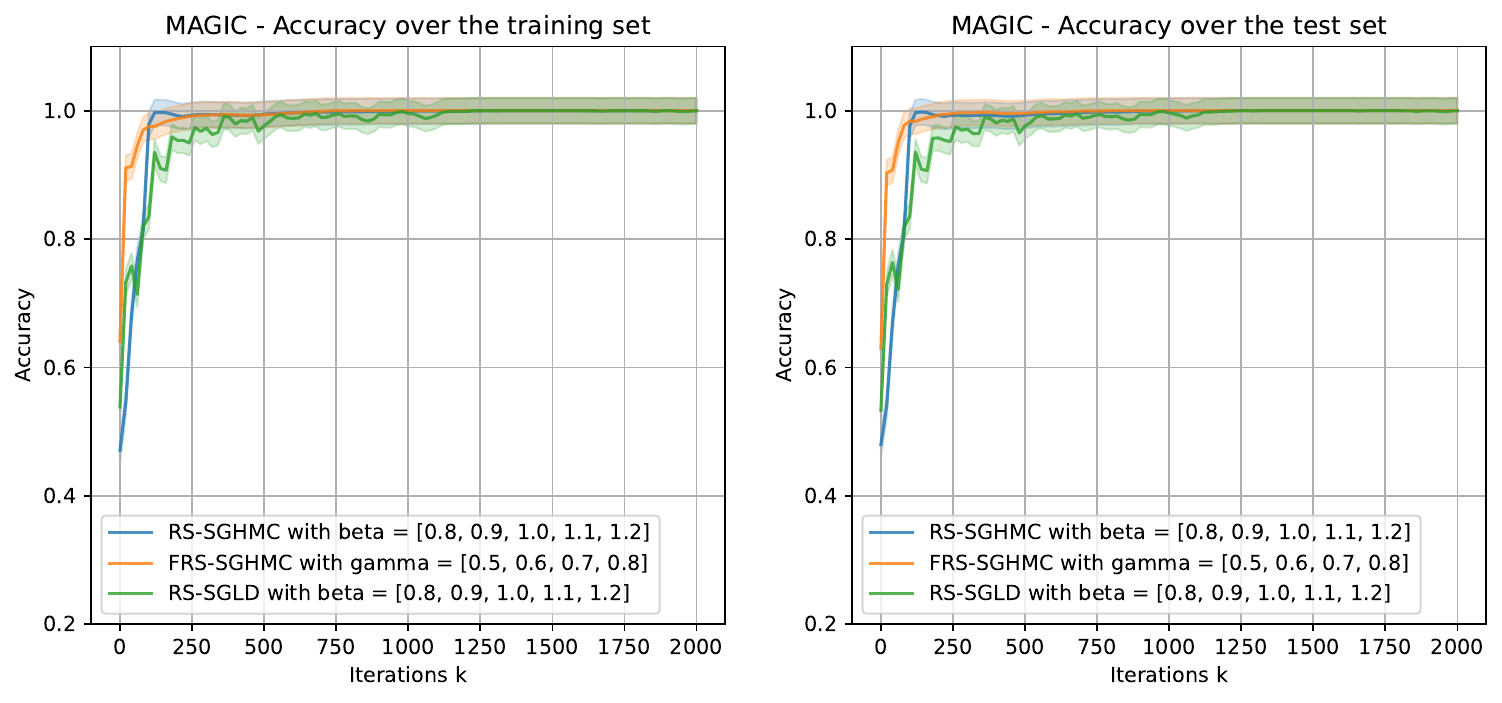}
\includegraphics[scale=0.45]{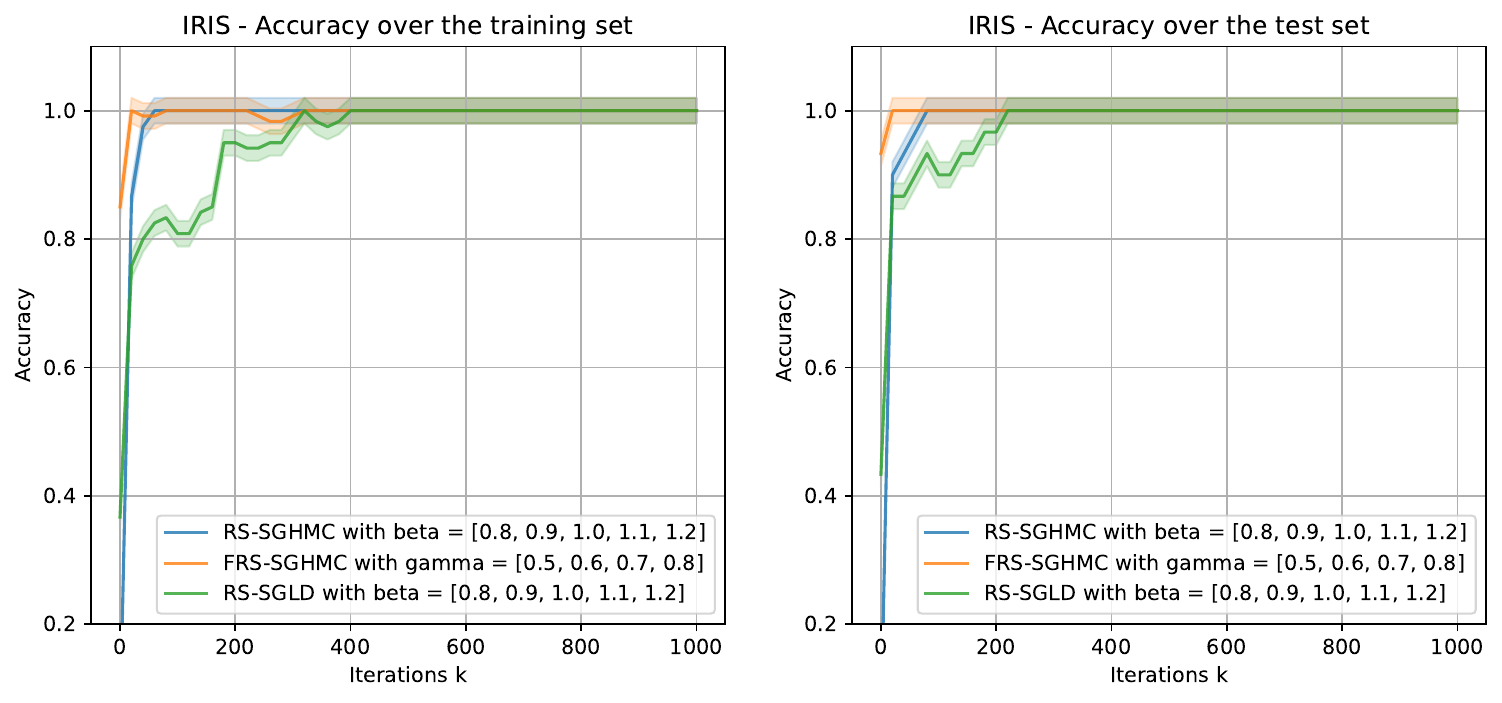} 
\caption{Comparisons within regime-switching algorithms over the real data.}
\label{fig:real:flmc_lmc:1}
\end{figure}

\paragraph{Real Data.} We use real datasets in the following experiments under the same setting as the one with synthetic data. We implement either $1000$ iterations (Iris dataset) or 2000 iterations (MAGIC dataset) to get Figure~\ref{fig:real:flmc_lmc:1} to compare various regime-swithcing algorithms.

We observe from these figures that RS-SGHMC and FRS-SGHMC consistently outperform RS-SGLD. This superiority is most pronounced on the Iris dataset, which has a small sample size of $150$, where the difference in accuracy is substantial. Moreover, even on the larger MAGIC dataset ($19,020$ samples), RS-SGHMC and FRS-SGHMC still show a measurable performance improvement.

\begin{figure}[htbp]
\centering
\includegraphics[scale=0.41]{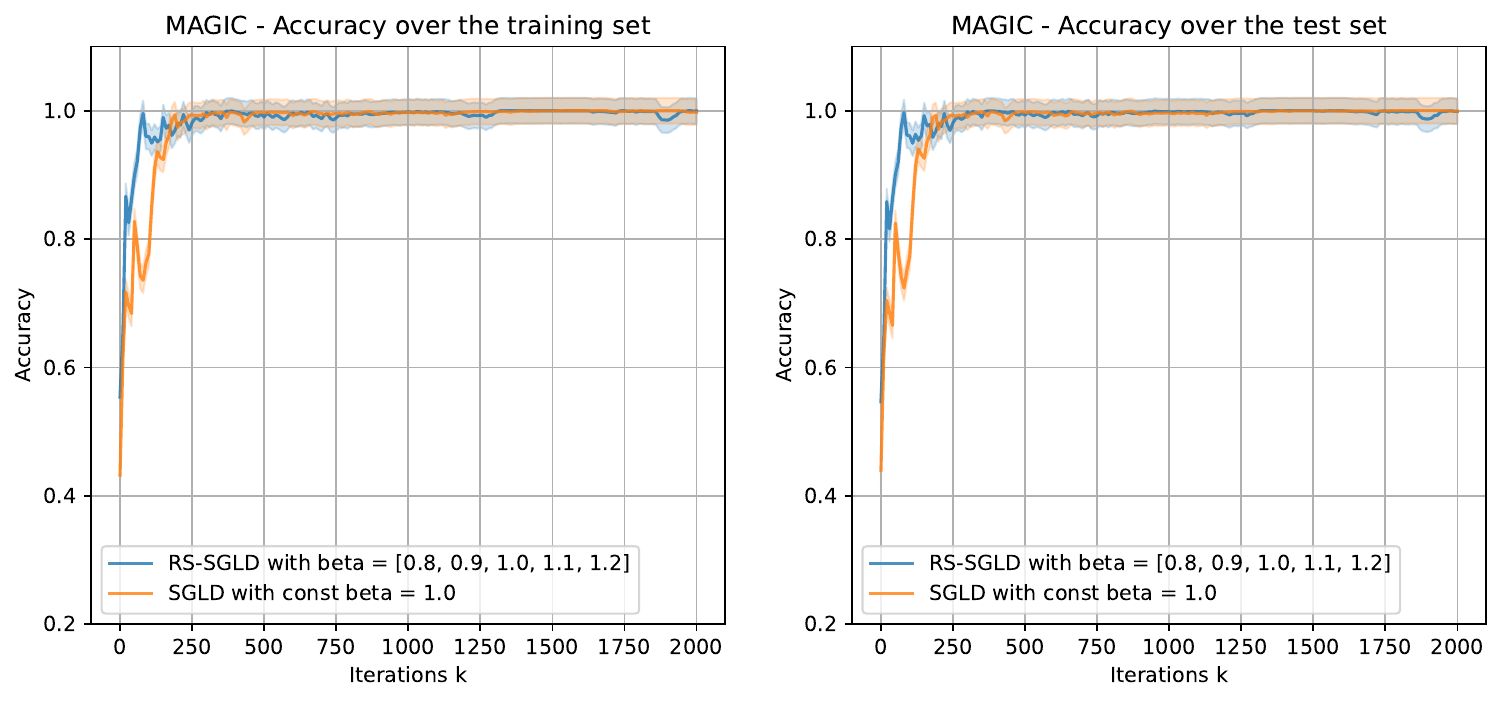} 
\includegraphics[scale=0.41]
{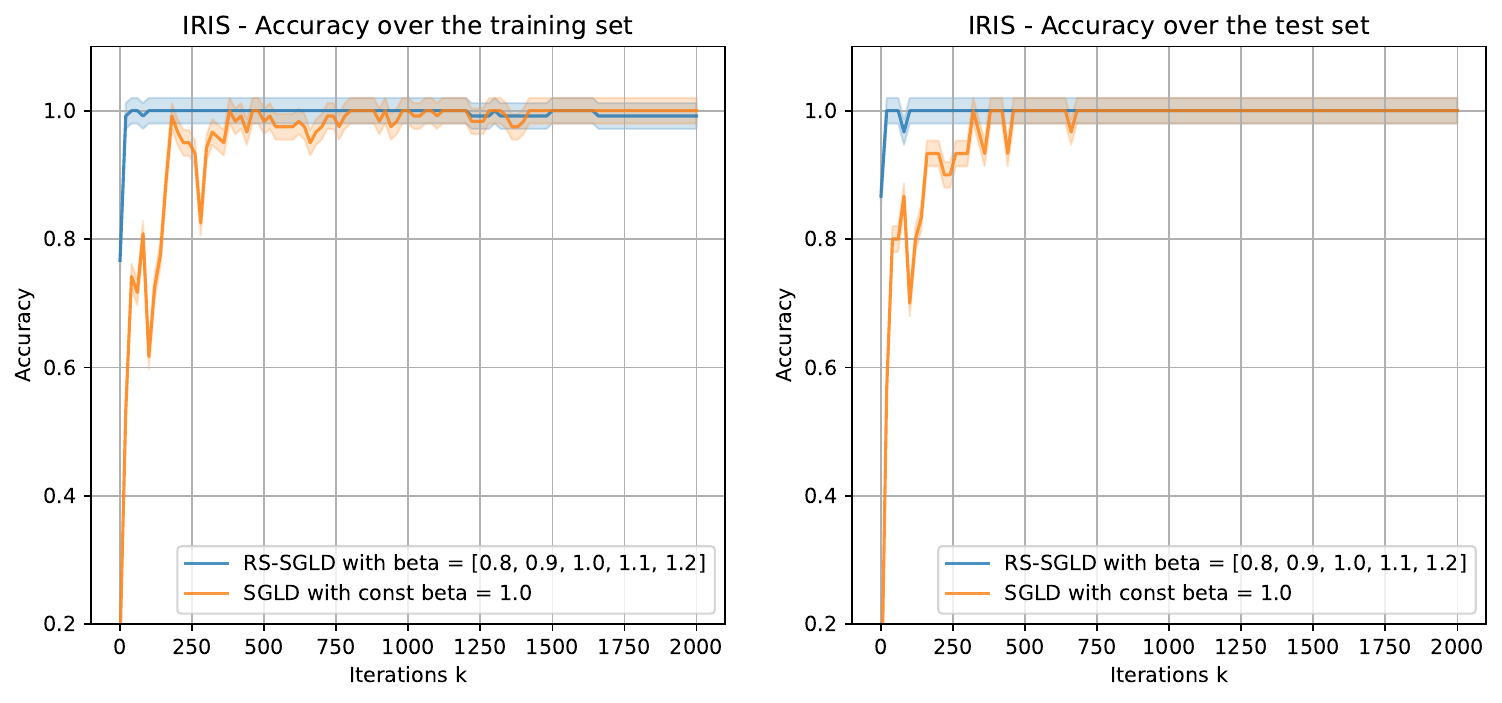}    
\caption{
Comparing RS-SGLD to SGLD.}
\label{fig:real:flmc_lmc:2}
\end{figure}

In the next experiment, we compare RS-SGLD to SGLD by iterating algorithms $2000$ iterations, and we summarize our results in Figure~\ref{fig:real:flmc_lmc:2}.

These plots demonstrate that RS-SGLD is more stable than SGLD for achieving the same accuracy, even with a small batch-size on both large and small sample sets. In the example using the Iris dataset ($150$ samples), SGLD exhibits unstable changes between iterations $500$ and $2000$ over the training set. In contrast, RS-SGLD maintains consistent convergence performance throughout.

\begin{figure}[htbp]
\centering
\includegraphics[scale=0.41]{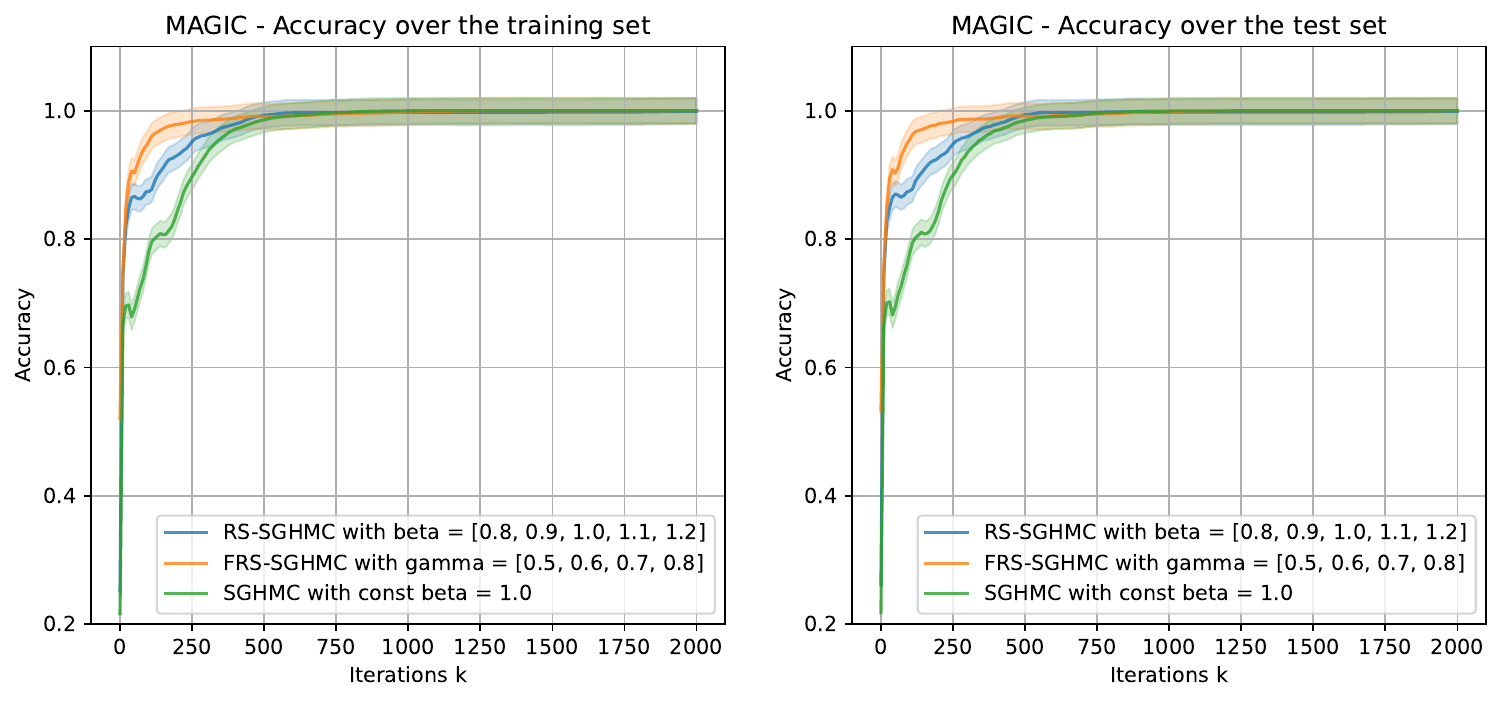} 
\includegraphics[scale=0.41]
{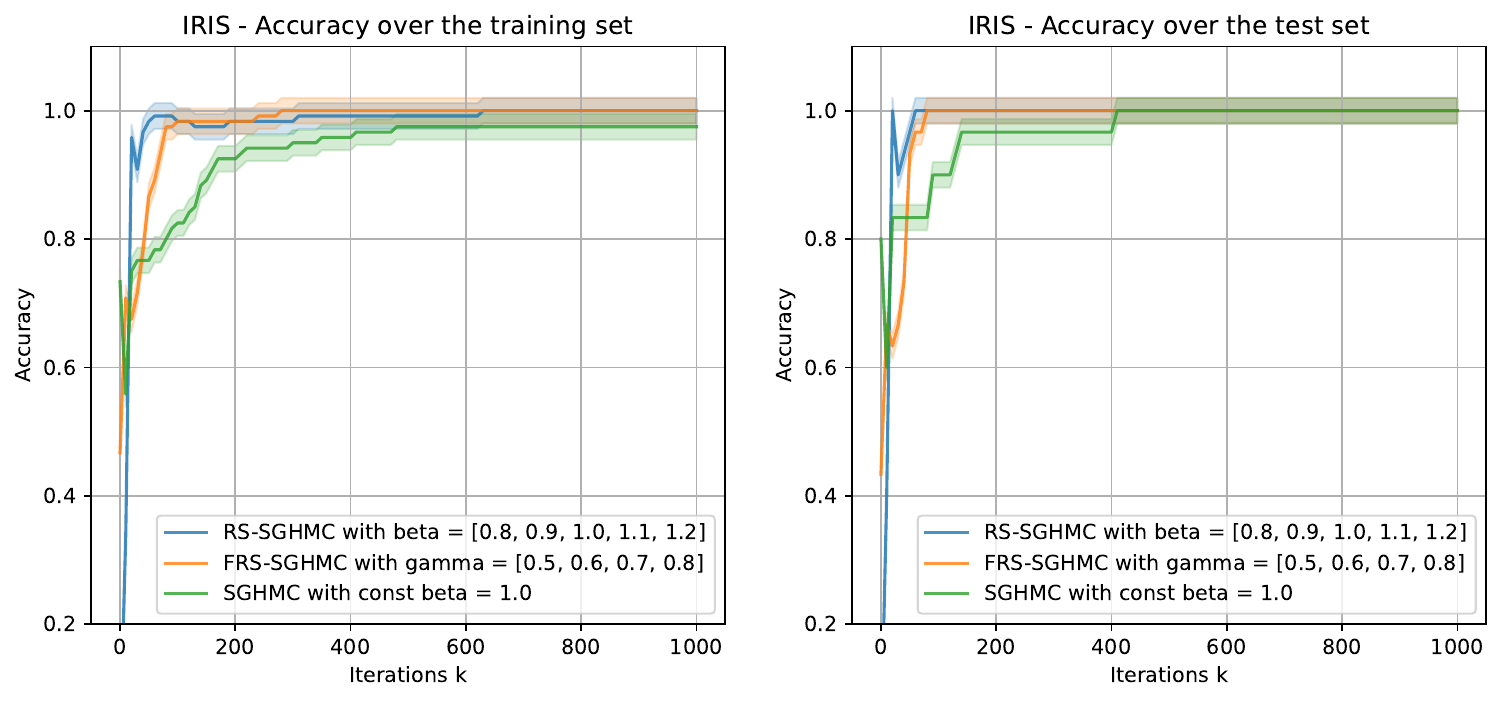}  
\caption{Comparing RS-SGHMC, FRS-SGHMC to SGHMC.}
\label{fig:real:flmc_lmc:3}
\end{figure}

In the following experiment, we compare RS-SGHMC, FRS-SGHMC to SGHMC by iterating algorithms $1000$ iterations over Iris dataset and $2000$ iterations over MAGIC dataset. We summarize our results in Figure~\ref{fig:real:flmc_lmc:3}.

From these plots, we conclude that RS-SGHMC and FRS-SGHMC outperform SGHMC by achieving the same high accuracy in fewer iterations. In particular, both regime-switching and frictional-regime-switching algorithms converge to high accuracy much faster than SGHMC over the dataset (Iris dataset, 150 samples) has limited samples. These results indicate that the regime-switching mechanism improves performance by accelerating convergence and preserving stability.

\section{Conclusion}\label{sec:conclusion}

In this paper, we proposed and studied regime-switching Langevin dynamics (RS-LD) and regime-switching kinetic Langevin dynamics (RS-KLD). 
These continuous-time stochastic differential equations (SDE) belong
to the class of regime-switching SDEs in the probability literature.
We also introduced regime-switching Langevin Monte Carlo (RS-LMC) algorithm
and regime-switching kinetic Langevin Monte Carlo (RS-KLMC) algorithm, 
based on the discretizations of RS-LD
and RS-KLD respectively.
From another perspective, the RS-LMC and RS-KLMC algorithms can also be viewed as the LMC and KLMC algorithms with random stepsizes. 
We also proposed frictional-regime-switching kinetic Langevin dynamics (FRS-KLD)
and its associated algorithm frictional-regime-switching kinetic Langevin Monte Carlo (FRS-KLMC), 
which can also be viewed as the KLMC algorithm with random frictional coefficients.
We provided their 2-Wasserstein non-asymptotic convergence guarantees to the target distribution, and analyzed the iteration complexities. Numerical experiments were provided for Bayesian linear regression and Bayesian logistic regression problems using synthetic and real data, and our proposed algorithms achieved a comparable or superior performance compared to the classical methods.

\section*{Acknowledgments}
Xiaoyu Wang is supported by the Guangzhou-HKUST(GZ) Joint Funding Program (No.2025A03J3556) and 
Guangzhou Municipal Key Laboratory of Financial Technology Cutting-Edge Research \\(No.2024A03J0630).
Lingjiong Zhu is partially supported by the grants NSF DMS-2053454, NSF DMS-2208303.

\bibliographystyle{alpha}
\bibliography{regime}


\appendix

\section{Technical Proofs}\label{sec:proofs}

\subsection{Proof of Theorem~\ref{thm:invariant:1}}

\begin{proof}
Recall from \eqref{eq:generator_mc} that the infinitesimal generator of the $\beta(t)$ is given by
\begin{align}
\mathcal{L}_{\beta}g(\bar{\beta}_{i})=\sum_{j\neq i}q_{ij}\left[g(\bar{\beta}_{j})-g(\bar{\beta}_{i})\right],
\end{align}
for any $i=1,2,\ldots,N$. One can compute that its adjoint operator is given by:
\begin{align}
\mathcal{L}_{\beta}^{\ast}g(\bar{\beta}_{i})=\sum_{j\neq i}\left[q_{ji}g(\bar{\beta}_{j})-q_{ij}g(\bar{\beta}_{i})\right],
\end{align}
for any $i=1,2,\ldots,N$.
Since $\psi=(\psi_{1},\psi_{2},\ldots,\psi_{N})$ is the invariant distribution of $\beta(t)$, 
by abusing the notation and defining $\psi(\beta):=\psi_{i}$ for any $\beta=\beta_{i}$, we have
\begin{align}
\mathcal{L}_{\beta}^{\ast}\psi(\bar{\beta}_{i})=\sum_{j\neq i}\left[q_{ji}\psi(\bar{\beta}_{j})-q_{ij}\psi(\bar{\beta}_{i})\right]
=\sum_{j\neq i}\left[q_{ji}\psi_{j}-q_{ij}\psi_{i}\right]=0,
\end{align}
for any $i=1,2,\ldots,N$.
Moreover, the standard overdamped Langevin SDE:
\begin{equation}
dX(t)=-\nabla f(X(t))dt+\sqrt{2}dB_{t},
\end{equation}
has the infinitesimal generator given by
\begin{align}
\mathcal{L}_{o}g(x):=-\sum_{j=1}^{d}\frac{\partial f}{\partial x_{j}}\frac{\partial g}{\partial x_{j}}
+\sum_{j=1}^{d}\frac{\partial^{2}g}{\partial x_{j}^{2}},
\end{align}
for any $x\in\mathbb{R}^{d}$ and its adjoint operator is given by:
\begin{align}
\mathcal{L}_{o}^{\ast}g(x):=\sum_{j=1}^{d}\frac{\partial}{\partial x_{j}}\left[\frac{\partial f}{\partial x_{j}}g(x)\right]
+\sum_{j=1}^{d}\frac{\partial^{2}g}{\partial x_{j}^{2}},
\end{align}
for any $x\in\mathbb{R}^{d}$. Since $\pi\propto e^{-f(x)}$ is the invariant
distribution for the standard overdamped Langevin SDE, we have
\begin{align}
\mathcal{L}_{o}^{\ast}e^{-f(x)}=\sum_{j=1}^{d}\frac{\partial}{\partial x_{j}}\left[\frac{\partial f}{\partial x_{j}}e^{-f(x)}\right]
+\sum_{j=1}^{d}\frac{\partial^{2}e^{-f(x)}}{\partial x_{j}^{2}}=0.
\end{align}
Finally, one can compute that the adjoint operator of the infinitesimal generator of
of the joint process $(\beta(t),X(t))$ is given by:
\begin{align}
\mathcal{L}^{\ast}g(\bar{\beta}_{i},x)=\bar{\beta}_{i}\sum_{j=1}^{d}\frac{\partial}{\partial x_{j}}\left[\frac{\partial f}{\partial x_{j}}g(\beta,x)\right]
+\bar{\beta}_{i}\sum_{j=1}^{d}\frac{\partial^{2}g}{\partial x_{j}^{2}}
+\sum_{j\neq i}\left[q_{ji}g(\bar{\beta}_{j},x)-q_{ij}g(\bar{\beta}_{i},x)\right],
\end{align}
for any $i=1,2,\ldots,N$ and $x\in\mathbb{R}^{d}$ and
\begin{align}
\mathcal{L}^{\ast}\psi(\bar{\beta}_{i})e^{-f(x)}
&=\bar{\beta}_{i}\sum_{j=1}^{d}\frac{\partial}{\partial x_{j}}\left[\frac{\partial f}{\partial x_{j}}\psi(\bar{\beta}_{i})e^{-f(x)}\right]
+\bar{\beta}_{i}\sum_{j=1}^{d}\frac{\partial^{2}\psi(\bar{\beta}_{i})e^{-f(x)}}{\partial x_{j}^{2}}
\nonumber
\\
&\qquad\qquad\qquad
+\sum_{j\neq i}\left[q_{ji}\psi(\bar{\beta}_{j})e^{-f(x)}-q_{ij}\psi(\bar{\beta}_{i})e^{-f(x)}\right]
\nonumber
\\
&=\bar{\beta}_{i}\psi_{i}\left(\sum_{j=1}^{d}\frac{\partial}{\partial x_{j}}\left[\frac{\partial f}{\partial x_{j}}e^{-f(x)}\right]
+\sum_{j=1}^{d}\frac{\partial^{2}e^{-f(x)}}{\partial x_{j}^{2}}\right)
+e^{-f(x)}\sum_{j\neq i}\left[q_{ji}\psi_{j}-q_{ij}\psi_{i}\right]=0,
\end{align}
for any $i=1,2,\ldots,N$ and $x\in\mathbb{R}^{d}$.
Hence, we conclude that $\pi=\psi\otimes\pi$
is an invariant distribution of the joint process $(\beta(t),X(t))$.
In particular, the Gibbs distribution $\pi$ is an invariant distribution for 
the regime-switching Langevin dynamics $X(t)$ in \eqref{eq:regime}.
This completes the proof.
\end{proof}


\subsection{Proof of Theorem~\ref{thm:RS-LD}}

\begin{proof}
We adopt the synchronous coupling method.
Let $X(t),\tilde{X}(t)$ be driven by the same $(\beta(t),B_{t})$ starting at $X(0)$ and $\tilde{X}(0)$ respectively:
\begin{align}
&dX(t)=-\beta(t)\nabla f(X(t))dt+\sqrt{2\beta(t)}dB_{t},
\\
&d\tilde{X}(t)=-\beta(t)\nabla f(\tilde{X}(t))dt+\sqrt{2\beta(t)}dB_{t}.
\end{align}
By It\^{o}'s formula, we can compute that
\begin{align}
&e^{2m\int_{0}^{t}\beta(s)ds}\Vert X(t)-\tilde{X}(t)\Vert^{2}
\nonumber
\\
&=\Vert X(0)-\tilde{X}(0)\Vert^{2}
-2\int_{0}^{t}\beta(s)e^{2m\int_{0}^{s}\beta(u)du}\left\langle X(s)-\tilde{X}(s),\nabla f(X(s))-\nabla f(\tilde{X}(s))\right\rangle ds
\nonumber
\\
&\qquad
+\int_{0}^{t}2m\beta(s)e^{2m\int_{0}^{s}\beta(u)du}\Vert X(s)-\tilde{X}(s)\Vert^{2}ds
\nonumber
\\
&\leq
\Vert X(0)-\tilde{X}(0)\Vert^{2},
\end{align}
where we used the $m$-strong convexity of $f$.
Therefore, we get
\begin{equation}
\Vert X(t)-\tilde{X}(t)\Vert^{2}
\leq e^{-2m\int_{0}^{t}\beta(s)ds}\Vert X(0)-\tilde{X}(0)\Vert^{2}.
\end{equation}
By letting $(\beta(0),\tilde{X}(0))$ follow
the invariant distribution $\psi\otimes\pi$ 
such that $\mathbb{E}\Vert X(0)-\tilde{X}(0)\Vert^{2}=\mathcal{W}_{2}^{2}(\mathrm{Law}(X(0)),\pi)$, we obtain
\begin{align}
\mathcal{W}_{2}^{2}(\mathrm{Law}(X(t)),\pi)
&\leq 
\mathbb{E}_{(\beta(0),\tilde{X}(0))\sim\psi\otimes\pi}\left[e^{-2m\int_{0}^{t}\beta(s)ds}\Vert X(0)-\tilde{X}(0)\Vert^{2}\right]
\nonumber
\\
&=\mathbb{E}_{\beta(0)\sim\psi}\left[e^{-2m\int_{0}^{t}\beta(s)ds}\right]\mathcal{W}_{2}^{2}(\mathrm{Law}(X(0)),\pi).
\end{align}
Let $u(t):=(u_{1}(t),\ldots,u_{N}(t))$, where $u_{i}(t):=\mathbb{E}_{\beta(0)=\bar{\beta}_{i}}\left[e^{-2m\int_{0}^{t}\beta(s)ds}\right]$. 
By Feynman-Kac formula, 
\begin{equation}
\frac{\partial u}{\partial t}=\mathbf{Q}u-2m\Lambda u,
\end{equation}
where $\Lambda$ is the diagonal matrix with diagonal entries $\bar{\beta}_{i}$, which implies that
\begin{equation}
u(t)=e^{(\mathbf{Q}-2m\Lambda)t}\mathbf{1},
\end{equation}
where $\mathbf{1}$ is an all-one vector.
This implies that 
\begin{equation}
\mathbb{E}_{\beta(0)\sim\psi}\left[e^{-2m\int_{0}^{t}\beta(s)ds}\right]=\left\langle e^{(\mathbf{Q}-2m\Lambda)t}\mathbf{1},\psi\right\rangle.
\end{equation}
This completes the proof.
\end{proof}


\subsection{Proof of Proposition~\ref{prop:recursive_error_bound}}
\label{appendix:overdamped_recursive_error_bound}
This proof treats the regime-switching parameter $\beta_k$ as a source of structured randomness for the stepsize. 

\begin{proof}
The proof proceeds in two main steps.
 
\paragraph{Step 1: Establishing a Conditional One-Step Error Bound.}

Given the regime chain $(\beta_n)_{n\ge0}$, let $(\mathcal F_{\beta,n})_{n\ge0}$ be the $\sigma$-algebra generated by $(x_{\beta,n})_{n\ge0}$. 
We define the \textbf{continuous-time process} $(L_\beta(t))_{t\ge0}$ as follows:

\begin{equation}\label{L:t:SDE}
        dL_\beta(t) = -\beta_{\lfloor t/\eta\rfloor} \nabla f(L_\beta(t))dt + \sqrt{2\beta_{\lfloor t/\eta\rfloor}}dB_t,
    \end{equation}
where $(B_t)_{t\ge0}$ is a standard $d$-dimensional Brownian motion. Let $(L_\beta(t))_{t\ge0}$ start from the stationary distribution $\pi$. We analyze in the time interval $[k\eta,(k+1)\eta]$ for $(L_\beta(t))_{t\ge0}$, and step $k$ to step $k+1$ for $(x_{\beta,n})_{n\ge0}$. 

The first step in this proof is to establish a rigorous, non-asymptotic bound on the conditional expectation of the squared error. \eqref{L:t:SDE} can be understood as ``piecewise'' overdamped Langevin dynamics. Hence, $L_\beta(0)\sim \pi$ implies $L_\beta(k\eta)\sim \pi$. Define
\[
    W_{\beta,k}^2:=\mathcal{W}_{2}^{2}(\mathrm{Law}(x_{\beta,k}),\mathrm{Law}(L_\beta(k\eta)))=\mathcal{W}_{2}^{2}(\nu_{\beta,k},\pi), \qquad k\ge1.
\]

Recall in \cite[p. 5282-5284]{dalalyan2019user}, for classic overdamped Langevin algorithm given stepsize $h_{k+1}$ from step $k$ to step $k+1$: 
\begin{equation}\label{discrete:classical}
x_{k+1}=x_{k}-h_{k+1}\nabla f(x_{k})+\sqrt{2h_{k+1}}\xi_{k},
\end{equation}
their 2-Wasserstein distance has the relationship
\begin{equation}\label{eq:dalalyan_result}
    \mathcal W_2(\nu_{h,k+1},\pi)\le (1-mh_{k+1})\mathcal W_2(\nu_{h,k},\pi)+1.65M\sqrt{d}h_{k+1}^{3/2},
\end{equation}
provided that $h_{k+1}\le \frac2{m+M}$, the condition in Theorem 1 in \cite{dalalyan2019user},
where $\nu_{h,k}$ denotes the law of $x_{k}$ in \eqref{discrete:classical}, 
$d$ is the dimension and $M$ the smoothness of the potential $f$.
Since for $t\in[k\eta,(k+1)\eta]$, $(\beta_{\lfloor t/\eta\rfloor})_{t\ge0}$ remains constant, we can use the classic result in \cite{dalalyan2019user}.

Applying \eqref{eq:dalalyan_result}, for $\eta\leq\frac{2}{\beta_{\max}(m+M)}$, we have
\[
    W_{\beta,k+1} \le (1-m\eta\beta_{k} )W_{\beta,k}+1.65M\sqrt{d}(\beta_{k}\eta)^{3/2}.
\]

By iterating, define $\beta_{\max}=\max_{1\le k\le N}\bar\beta_k$ and $\beta_{\min}=\min_{1\le k\le N}\bar\beta_k$,
\begin{align*}
    W_{\beta,K}
    & \le \left( \prod_{k=0}^{K-1}\left(1-m\eta\beta_k\right) \right)W_{\beta,0} + 1.65M\sqrt{d}\eta^{3/2}\sum_{j=0}^{K-1}\left( \prod_{k=j+1}^{K-1} \left(1-m\eta\beta_k\right)\right)\beta_j^{3/2} \\
    & \le \left( \prod_{k=0}^{K-1}\left(1-m\eta\beta_k\right) \right)W_{\beta,0} + 1.65M\sqrt{d}\eta^{3/2} \cdot \beta_{\max}^{3/2} \sum_{j=0}^{K-1} (1-m\eta\beta_{\min})^{K-j-1} \\
    & \le \left( \prod_{k=0}^{K-1}\left(1-m\eta\beta_k\right) \right)W_{\beta,0} + 1.65M\sqrt{d}\eta^{3/2} \cdot \beta_{\max}^{3/2} \cdot \frac{1}{m\eta\beta_{\min}} \\
    & = \left( \prod_{k=0}^{K-1}\left(1-m\eta\beta_k\right) \right)W_{\beta,0} + 1.65 M\sqrt{d}\frac{\beta_{\max}^{3/2}}{m\beta_{\min}}\eta^{1/2},
\end{align*}
which implies
\[
    W_{\beta,K}^2
    \le 2\left( \prod_{k=0}^{K-1}\left(1-m\eta\beta_k\right) \right)^2W_{\beta,0}^2+2\left(1.65 M\sqrt{d}\frac{\beta_{\max}^{3/2}}{m\beta_{\min}}\right)^2\eta.
\]
Taking expectation on both sides w.r.t. $(\beta_{k})_{k=0}^{K-1}$ and use inequality $\mathcal{W}_{2}^{2}(\nu_{K},\pi)
\leq\mathbb{E}\mathcal{W}_{2}^{2}(\nu_{\beta,k},\pi)$, we have
\[
    \mathcal W_2^2(\nu_{K},\pi)
    \le 2\mathbb E\left[\left( \prod_{k=0}^{K-1}\left(1-m\eta\beta_k\right) \right)^2\right]\mathbb E[W_{\beta,0}^2]+2\left(1.65 M\sqrt{d}\frac{\beta_{\max}^{3/2}}{m\beta_{\min}}\right)^2\eta.
\]
Since $x_{0}$ is independent of $(\beta_{n})_{n\geq 0}$, $\nu_{\beta,0}=\nu_{0}$.
Hence, the above inequality can be further written as:
\begin{align*}
    \mathcal W_2^2(\nu_{K},\pi)
    \le &2\mathbb E\left[\left( \prod_{k=0}^{K-1}\left(1-m\eta\beta_k\right) \right)^2\right]\mathcal W_2^2(\nu_{0},\pi)+2\left(1.65 M\sqrt{d}\frac{\beta_{\max}^{3/2}}{m\beta_{\min}}\right)^2\eta.
\end{align*}
For $\eta\le\frac{1}{m\beta_{\max}}$, the RHS is smaller than
\begin{align*}
    2\mathbb E\left[\prod_{k=0}^{K-1}\left(1-m\eta\beta_k\right) \right]\mathcal W_2^2(\nu_{0},\pi)+2\left(1.65 M\sqrt{d}\frac{\beta_{\max}^{3/2}}{m\beta_{\min}}\right)^2\eta.
\end{align*}

The main technical challenge is to bound the expectation of the product of correlated random variables. We use the standard inequality $1-x \le e^{-x}$ for $x \ge 0$:
\begin{align}
    \mathbb{E}\left[ \prod_{k=0}^{K-1} (1 - m\eta\beta_k) \right] &\le \mathbb{E}\left[ \prod_{k=0}^{K-1} \exp(-m\eta\beta_k) \right] \nonumber = \mathbb{E}\left[ \exp\left(-m\eta \sum_{k=0}^{K-1} \beta_k\right) \right].
\end{align}
This transforms the difficult problem of analyzing an expected product into the more standard problem of analyzing the moment generating function of the integrated Markov chain.

\paragraph{Step 2: Non-Asymptotic Analysis of the Exponential Term}
The goal of this step is to derive a rigorous upper bound for the exponential decay of the term $\mathbb{E}\left[ \exp\left(-\theta \sum_{k=0}^{K-1} \beta_k\right) \right]$, where $\theta = m\eta$. This will establish the exponential convergence of the leading error term.

\noindent\textbf{1. The Tilted Transition Operator and Perron-Frobenius Theory.}
As established previously using the Law of Total Expectation, the conditional expectation vector $\mathbf{u}_K$, with components $u_K(i) := \mathbb{E}\left[ \exp\left(-\theta \sum_{k=0}^{K-1} \beta_k\right) \bigg| \beta_0 = \bar{\beta}_i \right]$, satisfies the exact linear recursion:
\begin{equation}
    \mathbf{u}_K = \mathbf{T}_\theta \mathbf{u}_{K-1},\qquad K\ge1,
\end{equation}
where the tilted matrix is given by $(\mathbf{T}_\theta)_{ij} = P_{ij}(\eta)e^{-\theta\bar{\beta}_j}$. By induction, this means $\mathbf{u}_K = (\mathbf{T}_\theta)^K \mathbf{u}_0$. The vector $\mathbf{u}_0$ represents the initial state; for this expectation, we can consider $\mathbf{u}_0 = \mathbf{1}$ (the all-ones vector), corresponding to an expectation of 1 at $K=0$.

The matrix $\mathbf{P}(\eta)$ has strictly positive entries on its diagonal (for small $\eta$) and non-negative off-diagonal entries. Assuming the chain is irreducible (Assumption 2), $\mathbf{P}(\eta)$ is an irreducible non-negative matrix. The diagonal matrix $\Lambda_\theta$ has strictly positive entries. Therefore, the tilted matrix $\mathbf{T}_\theta = \mathbf{P}(\eta)\Lambda_\theta$ is also a non-negative and irreducible matrix.

By the Perron-Frobenius theorem for non-negative irreducible matrices, $\mathbf{T}_\theta$ has a simple, positive eigenvalue equal to its spectral radius, which we denote by $\rho(\mathbf{T}_\theta)$. Furthermore, there exists a corresponding right eigenvector, $\mathbf{v}$, with all components strictly positive, satisfying:
\[
    \mathbf{T}_\theta \mathbf{v} = \rho(\mathbf{T}_\theta) \mathbf{v}, \quad \text{where } v(i) > 0 \text{ for all } 1\leq i\leq N.
\]

\noindent\textbf{2. Deriving the Inequality Bound.}
Since $\mathbf{v}$ is a vector with strictly positive components, we can find a finite, positive constant $C_v$ such that our initial vector $\mathbf{u}_0 = \mathbf{1}$ is bounded component-wise by a multiple of $\mathbf{v}$:
\[
    u_0(i) = 1 \le C_v \cdot v(i) \quad \text{for all } 1\leq i\leq N, \quad \text{where } C_v = \frac{1}{\min_{1\leq i\leq N}v(i)}.
\]
We now prove by induction that $\mathbf{u}_K \le C_v \left(\rho(\mathbf{T}_\theta)\right)^K \mathbf{v}$ for all $K \ge 0$.
The base case $K=0$ holds by construction. Assume the inequality holds for $K-1$. For step $K$, we have:
\begin{align*}
    \mathbf{u}_K = \mathbf{T}_\theta \mathbf{u}_{K-1} &\le \mathbf{T}_\theta \left( C_v \left(\rho(\mathbf{T}_\theta)\right)^{K-1} \mathbf{v} \right) && \text{(since } \mathbf{T}_\theta \text{ is non-negative)} \\
    &= C_v \left(\rho(\mathbf{T}_\theta)\right)^{K-1} (\mathbf{T}_\theta \mathbf{v}) && \text{(linearity)} \\
    &= C_v \left(\rho(\mathbf{T}_\theta)\right)^{K-1} (\rho(\mathbf{T}_\theta) \mathbf{v}) && \text{(by eigenvector property)} \\
    &= C_v \left(\rho(\mathbf{T}_\theta)\right)^K \mathbf{v}.
\end{align*}
The induction holds. Now, if we assume the process starts from a distribution $\boldsymbol{\psi}_0$, the total expectation is $\boldsymbol{\psi}_0^\top\mathbf{u}_K$:
\begin{align*}
     \mathbb{E}_{\boldsymbol{\psi}_0}\left[ \exp\left(-\theta \sum_{k=0}^{K-1} \beta_k\right) \right] = \boldsymbol{\psi}_0^\top\mathbf{u}_K &\le \boldsymbol{\psi}_0^\top \left( C_v \left(\rho(\mathbf{T}_\theta)\right)^K \mathbf{v} \right) \\
     &= \left( C_v \boldsymbol{\psi}_0^\top \mathbf{v} \right) \left(\rho(\mathbf{T}_\theta)\right)^K.
\end{align*}
The term $(C_v \boldsymbol{\psi}_0^\top \mathbf{v})$ is a finite constant. This gives the rigorous inequality:
\begin{equation}
    \mathbb{E}\left[ \exp\left(-\theta \sum_{k=0}^{K-1} \beta_k\right) \right] \le C \cdot \left( \rho(\mathbf{T}_\theta) \right)^K.
\end{equation}

\noindent\textbf{3. Spectral Analysis and Non-Asymptotic Decay Rate.}
The goal is to find a rigorous, non-asymptotic upper bound for the spectral radius $\rho(\mathbf{T}_\theta)$ with $\theta = m\eta$. This is the key to determining the exponential decay rate of the leading error term in the random recursion approach.

First, we analyze the structure of the tilted matrix $\mathbf{T}_{m\eta}$. By expanding its definition, we can express it as a first-order perturbation of the identity matrix:
\begin{align*}
    \mathbf{T}_{m\eta} &= \mathbf{P}(\eta) \Lambda_{m\eta} = \left(\mathbf{I} + \eta\mathbf{Q} + \frac12\mathbf Q^2\eta^2+o(\eta^2)\right) 
    \left(\mathbf{I} - m\eta \Lambda +\frac14m^2\Lambda^2\eta^2+ o(\eta^2)\right) \\
    &=\mathbf I+\eta(\mathbf Q-m\Lambda)+\left( \frac12\mathbf Q^2-m\mathbf Q\Lambda+\frac14m^2\Lambda^2 \right)\eta^2+o(\eta^2)\\
    &\le \mathbf{I} + \eta \underbrace{\left( \mathbf{Q} - m\Lambda \right)}_{\mathbf{M}'} + \mathbf{R}_\eta,
\end{align*}
where $\mathbf{R}_\eta$ is a remainder matrix whose norm can be bounded by $\|\mathbf{R}_\eta\| \le K_R\eta^2$ with
\[
    K_R=\|\mathbf Q^2\|+2m\|\mathbf Q\Lambda\|+\frac12m^2\|\Lambda^2\|.
\]
For each eigenvalue $\lambda_i(\mathbf{T}_{m\eta})$, we can write:
\[
    \lambda_i(\mathbf{T}_{m\eta}) = 1 + \eta \lambda_i(\mathbf{M}') + r_i(\eta),
\]
where the remainder term $r_i(\eta)$ is of order $\mathcal{O}(\eta^2)$, i.e., $|r_i(\eta)| \le K_R\eta^2$.
Now, we derive a non-asymptotic bound for the first term. Let $\lambda_i(\mathbf{M}') = a_i + ib_i$. The squared modulus is given exactly by:
\[
    \left|1 + \eta \lambda_i(\mathbf{M}')\right|^2 = (1+\eta a_i)^2 + (\eta b_i)^2 = 1 + 2\eta a_i + \eta^2(a_i^2+b_i^2) = 1 + 2\eta \operatorname{Re}(\lambda_i) + \eta^2 |\lambda_i|^2.
\]
Using the inequality $\sqrt{1+x} \le 1 + x/2$ (valid for $x \ge -1$), 
for $\eta\leq-\frac1{2\min_{1\leq i\leq N}\left\{ \operatorname{Re}\left(\lambda_i(\mathbf{Q} - m\Lambda)\right) \right\}}$ (We need to let $1 + 2\eta \operatorname{Re}(\lambda_i(\mathbf{M}'))>0$ for all $i=1,2,\ldots,N$. Gershgorin Circle Theorem guarantees for all $i=1,2,\ldots,N$, $\operatorname{Re}(\lambda_i(\mathbf{Q} - m\Lambda))<0$, so we take mininum here), we can bound the modulus:
\begin{align*}
    \left|1 + \eta \lambda_i(\mathbf{M}')\right| &= \sqrt{1 + 2\eta \operatorname{Re}(\lambda_i(\mathbf{M}')) + \eta^2 |\lambda_i(\mathbf{M}')|^2} \\
    &\le 1 + \frac{1}{2}\left(2\eta \operatorname{Re}(\lambda_i(\mathbf{M}')) + \eta^2 |\lambda_i(\mathbf{M}')|^2\right) \\
    &= 1 + \eta \operatorname{Re}(\lambda_i(\mathbf{M}')) + \frac{\eta^2}{2} \left|\lambda_i(\mathbf{M}')\right|^2.
\end{align*}

Combining these bounds, we get a fully non-asymptotic inequality for each eigenvalue's modulus:
\[
    |\lambda_i(\mathbf{T}_{m\eta})| \le 1 + \eta \operatorname{Re}(\lambda_i(\mathbf{M}')) + \frac{\eta^2}{2} |\lambda_i(\mathbf{M}')|^2 + K_R\eta^2.
\]
The spectral radius $\rho(\mathbf{T}_{m\eta})$ is the maximum of these moduli. Taking the maximum over all $i$:
\[
    \rho(\mathbf{T}_{m\eta}) \le 1 + \eta \max_{1\leq i\leq N}\left\{ \operatorname{Re}(\lambda_i(\mathbf{M}')) \right\} + \eta^2 \left( \frac{1}{2} \max_{1\leq i\leq N}\left\{|\lambda_i(\mathbf{M}')|^2\right\} + K_R \right).
\]
We now define the rate $\alpha$ and the constant $C_M$ based on the spectrum of $\mathbf{M}'$:
\begin{align*}
    \alpha &= -\max_{1\leq i\leq N}\left\{ \operatorname{Re}\left(\lambda_i(\mathbf{Q} - m\Lambda)\right) \right\}, \\
    C_M &= \frac{1}{2} \max_{1\leq i\leq N}\left\{|\lambda_i(\mathbf{Q} - m\Lambda)|^2\right\} + K_R.
\end{align*}
With these definitions, we arrive at the desired rigorous and non-asymptotic bound for the spectral radius:
\begin{equation}\label{eq:nonasymptotic_rho_bound_final}
    \rho(\mathbf{T}_{m\eta}) \le 1 - \alpha\eta + C_M\eta^2.
\end{equation}
To obtain a purely linear decay factor, we can absorb the higher-order term by imposing a condition on $\eta$. Our goal is to find a new effective rate $\alpha'$ such that $1 - \alpha\eta + C_M\eta^2 \le 1 - \alpha'\eta$.

Let us choose, for instance, $\alpha' = \alpha/2$. We seek the condition on $\eta$ under which the following holds:
\[
    1 - \alpha\eta + C_M\eta^2 \le 1 - \frac{\alpha}{2}\eta.
\]
Rearranging the terms, this is equivalent to:
\begin{align*}
    C_M\eta^2 \le \alpha\eta - \frac{\alpha}{2}\eta=\frac{\alpha}{2}\eta,
\end{align*}
which is equivalent to
\[
\eta \le \frac{\alpha}{2C_M},
\]
since $\eta>0$.
This provides an explicit and computable upper bound on the stepsize $\eta$. Therefore, by restricting $\eta$ to this range, we can absorb the quadratic term.

This leads to the final, rigorous, and non-asymptotic bound on the spectral radius. Provided that $\eta \le \frac{\alpha}{2C_M}$, we have:
\begin{equation}
    \rho(\mathbf{T}_{m\eta}) \le 1 - \frac{\alpha}{2}\eta.
\end{equation}
The proof is complete.
\end{proof}

\subsection{Proof of Theorem~\ref{thm:non:asymptotic}}

\begin{proof}
The proof follows by unrolling the recursion for the squared error established in the proof of Proposition~\ref{prop:recursive_error_bound}. Let $W_k = \mathcal{W}_{2}^2(\nu_k, \pi)$.
We start with the inequality $W_{k+1} \le (1-\frac{\alpha}{2}\eta)W_k + C\eta^2$. Unrolling this for $K$ steps yields:
\begin{align*}
    W_K &\le \left(1-\frac{\alpha}{2}\eta\right)^K W_{0} + C\eta^2 \sum_{j=0}^{K-1}\left(1-\frac{\alpha}{2}\eta\right)^j 
    \\
    &\le \left(1-\frac{\alpha}{2}\eta\right)^K W_0 + \frac{C\eta^2}{1-(1-\frac{\alpha}{2}\eta)} = \left(1-\frac{\alpha}{2}\eta\right)^K W_0 + \frac{2C\eta}{\alpha}.
\end{align*}
The result is obtained by taking the square root of both sides and using the inequality $\sqrt{a+b} \le \sqrt{a} + \sqrt{b}$. This completes the proof.
\end{proof}

\subsection{Proof of Corollary~\ref{cor:iteration:complexity}}
\begin{proof}
It follows from Theorem~\ref{thm:non:asymptotic} that
\begin{equation}
    \mathcal{W}_{2}(\nu_{K}, \pi) \le \left(1-\frac{\alpha}{2}\eta\right)^{K/2} \mathcal{W}_{2}(\nu_{0}, \pi) + \sqrt{\frac{2C\eta}{\alpha}}.
\end{equation}
First, we choose $\eta$ to ensure the asymptotic bias is at most $\epsilon/2$, 
that is $\sqrt{\frac{2C}{\alpha}}\sqrt{\eta} \le \frac{\epsilon}{2}$, which is equivalent to
\[
\eta \le \frac{\epsilon^2 \alpha}{8C}.
\]
Given $\eta$, we choose $K$ such that the contraction term is smaller than $\epsilon/2$
:\[
    \left(1-\frac{\alpha}{2}\eta\right)^{K/2} \mathcal{W}_{2}(\nu_{0}, \pi) \le e^{-K\alpha\eta/4} \mathcal{W}_{2}(\nu_{0}, \pi) \le \frac{\epsilon}{2},
\]
which implies that the number of iterations $K$ must satisfy:
\[
    K \ge \frac{4}{\alpha\eta}\log\left(\frac{2\mathcal{W}_{2}(\nu_{0}, \pi)}{\epsilon}\right).
\]
Substituting the value of $\eta = \frac{\epsilon^2 \alpha}{8C}$, the total number of iterations required is:
\[
    K \ge \frac{32C}{\alpha^2\epsilon^2}\log\left(\frac{2\mathcal{W}_{2}(\nu_{\beta,0}, \pi)}{\epsilon}\right) = O\left(\frac{1}{\epsilon^2}\log\left(\frac{1}{\epsilon}\right)\right).
\]
This completes the proof.
\end{proof}


\subsection{Proof of Theorem~\ref{thm:invariant:underdamped:2}}

\begin{proof}
Recall from \eqref{eq:generator_mc} that the infinitesimal generator of the $\beta(t)$ is given by
\begin{align}
\mathcal{L}_{\beta}g(\bar{\beta}_{i})=\sum_{j\neq i}q_{ij}\left[g(\bar{\beta}_{j})-g(\bar{\beta}_{i})\right],
\end{align}
for any $i=1,2,\ldots,N$. One can compute that its adjoint operator is given by:
\begin{align}
\mathcal{L}_{\beta}^{\ast}g(\bar{\beta}_{i})=\sum_{j\neq i}\left[q_{ji}g(\bar{\beta}_{j})-q_{ij}g(\bar{\beta}_{i})\right],
\end{align}
for any $i=1,2,\ldots,N$.
Since $\psi=(\psi_{1},\psi_{2},\ldots,\psi_{N})$ is the invariant distribution of $\beta(t)$, 
by abusing the notation and defining $\psi(\beta):=\psi_{i}$ for any $\beta=\beta_{i}$, we have
\begin{align}
\mathcal{L}_{\beta}^{\ast}\psi(\bar{\beta}_{i})=\sum_{j\neq i}\left[q_{ji}\psi(\bar{\beta}_{j})-q_{ij}\psi(\bar{\beta}_{i})\right]
=\sum_{j\neq i}\left[q_{ji}\psi_{j}-q_{ij}\psi_{i}\right]=0,
\end{align}
for any $i=1,2,\ldots,N$.
Next, one can compute that the adjoint operator of the infinitesimal generator of
the joint process $(\beta(t),V(t),X(t))$ is given by:
\begin{align}
\mathcal{L}^{\ast}g(\bar{\beta}_{i},v,x)
&=\gamma\bar{\beta}_{i}\sum_{j=1}^{d}\frac{\partial}{\partial v_{j}}\left[v_{j}g\right]
+\sum_{j=1}^{d}\frac{\partial f}{\partial x_{j}}\frac{\partial g}{\partial v_{j}}
+\gamma\bar{\beta}_{i}\sum_{j=1}^{d}\frac{\partial^{2}g}{\partial v_{j}^{2}}
-\sum_{j=1}^{d}v_{j}\frac{\partial g}{\partial x_{j}}
\nonumber
\\
&\qquad\qquad\qquad
+\sum_{j\neq i}\left[q_{ji}g(\bar{\beta}_{j},v,x)-q_{ij}g(\bar{\beta}_{i},v,x)\right],
\end{align}
for any $i=1,2,\ldots,N$ and $x\in\mathbb{R}^{d}$ and finally, we can compute that
\begin{align}
&\mathcal{L}^{\ast}\psi(\bar{\beta}_{i})e^{-f(x)-\frac{1}{2}\Vert v\Vert^{2}}
\nonumber
\\
&=\gamma\bar{\beta}_{i}\sum_{j=1}^{d}\frac{\partial}{\partial v_{j}}\left[v_{j}\psi(\bar{\beta}_{i})e^{-f(x)-\frac{1}{2}\Vert v\Vert^{2}}\right]
+\bar{\beta}_{i}\sum_{j=1}^{d}\frac{\partial f}{\partial x_{j}}\frac{\partial \psi(\bar{\beta}_{i})e^{-f(x)-\frac{1}{2}\Vert v\Vert^{2}}}{\partial v_{j}}
\nonumber
\\
&\qquad\qquad\qquad\qquad\qquad\qquad
+\gamma\bar{\beta}_{i}\sum_{j=1}^{d}\frac{\partial^{2}\psi(\bar{\beta}_{i})e^{-f(x)-\frac{1}{2}\Vert v\Vert^{2}}}{\partial v_{j}^{2}}-\bar{\beta}_{i}\sum_{j=1}^{d}v_{j}\frac{\partial \psi(\bar{\beta}_{i})e^{-f(x)-\frac{1}{2}\Vert v\Vert^{2}}}{\partial x_{j}}
\nonumber
\\
&\qquad\qquad\qquad
+\sum_{j\neq i}\left[q_{ji}\psi(\bar{\beta}_{j})e^{-f(x)-\frac{1}{2}\Vert v\Vert^{2}}-q_{ij}\psi(\bar{\beta}_{i})e^{-f(x)-\frac{1}{2}\Vert v\Vert^{2}}\right].
\end{align}
We can compute that
\begin{align}
&\gamma\bar{\beta}_{i}\sum_{j=1}^{d}\frac{\partial}{\partial v_{j}}\left[v_{j}\psi(\bar{\beta}_{i})e^{-f(x)-\frac{1}{2}\Vert v\Vert^{2}}\right]
+\gamma\bar{\beta}_{i}\sum_{j=1}^{d}\frac{\partial^{2}\psi(\bar{\beta}_{i})e^{-f(x)-\frac{1}{2}\Vert v\Vert^{2}}}{\partial v_{j}^{2}}
\nonumber
\\
&=\gamma\bar{\beta}_{i}\psi(\bar{\beta}_{i})e^{-f(x)}
\sum_{j=1}^{d}\left(\frac{\partial}{\partial v_{j}}\left[v_{j}e^{-\frac{1}{2}\Vert v\Vert^{2}}\right]
+\frac{\partial^{2}e^{-\frac{1}{2}\Vert v\Vert^{2}}}{\partial v_{j}^{2}}\right)
=0,
\end{align}
and moreover
\begin{align}
&\bar{\beta}_{i}\sum_{j=1}^{d}\frac{\partial f}{\partial x_{j}}\frac{\partial \psi(\bar{\beta}_{i})e^{-f(x)-\frac{1}{2}\Vert v\Vert^{2}}}{\partial v_{j}}
-\bar{\beta}_{i}\sum_{j=1}^{d}v_{j}\frac{\partial \psi(\bar{\beta}_{i})e^{-f(x)-\frac{1}{2}\Vert v\Vert^{2}}}{\partial x_{j}}\nonumber
\\
&=\bar{\beta}_{i}\psi(\bar{\beta}_{i})\sum_{j=1}^{d}\left(\frac{\partial f}{\partial x_{j}}\frac{\partial e^{-f(x)-\frac{1}{2}\Vert v\Vert^{2}}}{\partial v_{j}}
-v_{j}\frac{\partial e^{-f(x)-\frac{1}{2}\Vert v\Vert^{2}}}{\partial x_{j}}\right)=0,
\end{align}
and finally
\begin{align}
\sum_{j\neq i}\left[q_{ji}\psi(\bar{\beta}_{j})e^{-f(x)-\frac{1}{2}\Vert v\Vert^{2}}-q_{ij}\psi(\bar{\beta}_{i})e^{-f(x)-\frac{1}{2}\Vert v\Vert^{2}}\right]
=e^{-f(x)-\frac{1}{2}\Vert v\Vert^{2}}\sum_{j\neq i}\left[q_{ji}\psi_{j}-q_{ij}\psi_{i}\right]=0,
\end{align}
for any $i=1,2,\ldots,N$ and $v,x\in\mathbb{R}^{d}$.
Hence, we conclude that
\begin{equation}
\mathcal{L}^{\ast}\psi(\bar{\beta}_{i})e^{-f(x)-\frac{1}{2}\Vert v\Vert^{2}}=0,
\end{equation}
for any $i=1,2,\ldots,N$ and $v,x\in\mathbb{R}^{d}$, 
and therefore $\psi\otimes\mathcal{N}(0,I_{d})\otimes\pi$
is an invariant distribution of the joint process $(\beta(t),V(t),X(t))$.
In particular, the Gibbs distribution $\pi\propto e^{-f(x)}$ is an invariant distribution for 
the regime-switching kinetic Langevin dynamics $X(t)$ in \eqref{underdamped:regime:switching:2}.
This completes the proof.
\end{proof}


\subsection{Proof of Theorem~\ref{thm:underdamped:alternative}}

\begin{proof}
Let $X(0),\tilde{X}(0)$ and $V(0)$ be three $d$-dimensional random vectors defined
in the same probability space such that $V(0)$ is independent of $(X(0),\tilde{X}(0))$,
$V(0)\sim\mu_{1}:=\mathcal{N}(0,I_{d})$, $X(0)\sim\mu_{2}$ and $\tilde{X}(0)\sim\tilde{\mu}_{2}$,
and finally $\mathcal{W}_{2}^{2}(\mu_{2},\tilde{\mu}_{2})=\mathbb{E}\left[\Vert X(0)-\tilde{X}(0)\Vert^{2}\right]$.

Let $(B_{t})_{t\geq 0}$ be a standard $d$-dimensional Brownian motion and $(\beta(t))_{t\geq )}$ be the CTMC process defined in the same probability space. 
We define $(X(t),V(t))_{t\geq 0}$ and $(\tilde{X}(t),\tilde{V}(t))_{t\geq 0}$
as two SDEs driven by the same Brownian motion $(B_{t})_{t\geq 0}$ and CTMC process $(\beta(t))_{t\geq 0}$:
\begin{align}
&dV(t)=(-\beta(t)\nabla f(X(t))-\gamma \beta(t)V(t))dt+\sqrt{2\gamma\beta(t)}dB_{t},
\nonumber
\\
&dX(t)=\beta(t)V(t)dt,
  \label{eqn:LangevinSDE:1}
\end{align}
and
\begin{align}
&d\tilde{V}_t=(-\beta(t)\nabla f(\tilde{X}(t))-\gamma \beta(t)\tilde{V}(t))dt+\sqrt{2\gamma\beta(t)}dB_{t},
\nonumber
\\
&d\tilde{X}(t)=\beta(t)\tilde{V}(t)dt,
  \label{eqn:LangevinSDE:2}
\end{align}
that start from $(X(0),V(0))$ and $(\tilde{X}(0),\tilde{V}(0))$ 
with $\tilde{V}(0)=V(0)$ and $\tilde{X}(0)\neq X(0)$.

Define:
\begin{align}
&\psi_{t}:=(V(t)+\lambda_{+}X(t))-(\tilde{V}(t)+\lambda_{+}\tilde{X}(t)),
\\
&z_{t}:=(-V(t)-\lambda_{-}X(t))+(\tilde{V}(t)+\lambda_{-}\tilde{X}(t)),
\end{align}
where $\lambda_{+}$ and $\lambda_{-}$ are two arbitrary positive numbers
such that $\lambda_{+}+\lambda_{-}=\gamma$ with $\lambda_{+}>\lambda_{-}$.

Note that it follows from Taylor's theorem that
\begin{equation}\label{eqn:Hessian}
\nabla f(X(t))-\nabla f(\tilde{X}(t))=H_{t}(X(t)-\tilde{X}(t)),
\end{equation}
where
\begin{equation}
H_{t}:=\int_{0}^{1}\nabla^{2}f\left(X(t)-y\left(X(t)-\tilde{X}(t)\right)\right)dy.
\end{equation}
Thus, it follows from \eqref{eqn:LangevinSDE:1}, \eqref{eqn:LangevinSDE:2} and \eqref{eqn:Hessian} that
\begin{align}
d\psi_{t}
&=\beta(t)\left[-\gamma(V_{t}-\tilde{V}_{t})-(\nabla f(X_{t})-\nabla f(\tilde{X}_{t}))+\lambda_{+}(V_{t}-\tilde{V}_{t})\right]dt
\nonumber
\\
&=\beta(t)\left[\frac{(\lambda_{+}-\gamma)(\lambda_{-}\psi_{t}+\lambda_{+}z_{t})}{\lambda_{-}-\lambda_{+}}-\frac{H_{t}(\psi_{t}+z_{t})}{\lambda_{+}-\lambda_{-}}\right]dt
\nonumber
\\
&=\beta(t)\frac{(\lambda_{-}^{2}I_{d}-H_{t})\psi_{t}+(\lambda_{-}\lambda_{+}I_{d}-H_{t})z_{t}}{\lambda_{+}-\lambda_{-}}dt,
\end{align}
where we used the identity $\lambda_{+}+\lambda_{-}=\gamma$. 
Similarly, one can compute that
\begin{align}
dz_{t}
&=\beta(t)\left[\gamma(V_{t}-\tilde{V}_{t})+(\nabla f(X_{t})-\nabla f(\tilde{X}_{t}))-\lambda_{-}(V_{t}-\tilde{V}_{t})\right]dt
\nonumber
\\
&=\beta(t)\left[\frac{(\gamma-\lambda_{-})(\lambda_{-}\psi_{t}+\lambda_{+}z_{t})}{\lambda_{-}-\lambda_{+}}+\frac{H_{t}(\psi_{t}+z_{t})}{\lambda_{+}-\lambda_{-}}\right]dt
\nonumber
\\
&=\beta(t)\frac{(H_{t}-\lambda_{-}\lambda_{+}I_{d})\psi_{t}+(H_{t}-\lambda_{+}^{2}I_{d})z_{t}}{\lambda_{+}-\lambda_{-}}dt.
\end{align}
Thus, we have
\begin{align}
d\left\Vert\left(\psi_{t}^{\top},z_{t}^{\top}\right)^{\top}\right\Vert^{2}
&=2\psi_{t}^{\top}d\psi_{t}+2z_{t}^{\top}dz_{t}
\nonumber\\
&=\frac{2\beta(t)}{\lambda_{+}-\lambda_{-}}
\left[\psi_{t}^{\top}(\lambda_{-}^{2}I_{d}-H_{t})\psi_{t}+z_{t}^{\top}(H_{t}-\lambda_{+}^{2}I_{d})z_{t}\right]dt.
\end{align}
Under our assumption, 
$mI_{d}\preceq H_{t}\preceq MI_{d}$.
Therefore, we get
\begin{align}
\frac{d}{dt}\left\Vert\left(\psi_{t}^{\top},z_{t}^{\top}\right)^{\top}\right\Vert^{2}
&\leq
\frac{2\beta(t)}{\lambda_{+}-\lambda_{-}}
\left[(\lambda_{-}^{2}-m)\Vert\psi_{t}\Vert^{2}+(M-\lambda_{+}^{2})\Vert z_{t}\Vert^{2}\right]
\nonumber
\\
&\leq
\frac{2\beta(t)[(\lambda_{-}^{2}-m)\vee(M-\lambda_{+}^{2})]}{\lambda_{+}-\lambda_{-}}\left\Vert\left(\psi_{t}^{\top},z_{t}^{\top}\right)^{\top}\right\Vert^{2}.
\end{align}
By Gronwall's inequality, we get that
for any $t\geq 0$,
\begin{align}
\left\Vert\left(\psi_{t}^{\top},z_{t}^{\top}\right)^{\top}\right\Vert^{2}
\leq
\exp\left\{\frac{2(\lambda_{-}^{2}-m)\vee(M-\lambda_{+}^{2})}{\lambda_{+}-\lambda_{-}}\int_{0}^{t}\beta(s)ds\right\}\left\Vert\left(\psi_{0}^{\top},z_{0}^{\top}\right)^{\top}\right\Vert^{2}.
\end{align}
Note that $X(t)-\tilde{X}(t)=\frac{\psi_{t}+z_{t}}{\lambda_{+}-\lambda_{-}}$
and $V(0)=\tilde{V}(0)$, we conclude that
\begin{align}
\Vert X(t)-\tilde{X}(t)\Vert
&\leq
\frac{\sqrt{2}}{\lambda_{+}-\lambda_{-}}
\left\Vert\left(\psi_{t}^{\top},z_{t}^{\top}\right)^{\top}\right\Vert
\nonumber
\\
&\leq
\frac{\sqrt{2(\lambda_{+}^{2}+\lambda_{-}^{2})}}{\lambda_{+}-\lambda_{-}}
\exp\left\{\frac{(\lambda_{-}^{2}-m)\vee(M-\lambda_{+}^{2})}{\lambda_{+}-\lambda_{-}}\int_{0}^{t}\beta(s)ds\right\}
\Vert X(0)-\tilde{X}(0)\Vert.
\end{align}
Therefore, we have
\begin{align}
&\mathcal{W}_{2}^{2}(\mathrm{Law}(X(t)),\mathrm{Law}(\tilde{X}(t)))
\nonumber
\\
&\leq
\frac{2(\lambda_{+}^{2}+\lambda_{-}^{2})}{(\lambda_{+}-\lambda_{-})^{2}}
\mathbb{E}\left[\exp\left\{\frac{2(\lambda_{-}^{2}-m)\vee(M-\lambda_{+}^{2})}{\lambda_{+}-\lambda_{-}}\int_{0}^{t}\beta(s)ds\right\}\right]
\mathbb{E}\Vert X(0)-\tilde{X}(0)\Vert^{2}.
\end{align}
By letting $\tilde{X}(0)\sim\pi$, $\tilde{V}(0)=V(0)\sim\mathcal{N}(0,I_{d})$ and $\beta(0)\sim\psi$, we have $\tilde{X}(t)\sim\pi$
for every $t$, and we conclude that
\begin{align}
&\mathcal{W}_{2}(\mathrm{Law}(X(t)),\pi)
\nonumber
\\
&\leq
\frac{\sqrt{2(\lambda_{+}^{2}+\lambda_{-}^{2})}}{\lambda_{+}-\lambda_{-}}
\left(\mathbb{E}_{\beta(0)\sim\psi}\left[\exp\left\{\frac{2(\lambda_{-}^{2}-m)\vee(M-\lambda_{+}^{2})}{\lambda_{+}-\lambda_{-}}\int_{0}^{t}\beta(s)ds\right\}\right]\right)^{1/2}
\mathcal{W}_{2}(\mathrm{Law}(X(0)),\pi).
\end{align}
Let $u(t):=(u_{1}(t),\ldots,u_{N}(t))$, where $u_{i}(t):=\mathbb{E}_{\beta(0)=\bar{\beta}_{i}}\left[\exp\left\{\frac{2(\lambda_{-}^{2}-m)\vee(M-\lambda_{+}^{2})}{\lambda_{+}-\lambda_{-}}\int_{0}^{t}\beta(s)ds\right\}\right]$. 
By Feynman-Kac formula, 
\begin{equation}
\frac{\partial u}{\partial t}=\mathbf{Q}u+\frac{2(\lambda_{-}^{2}-m)\vee(M-\lambda_{+}^{2})}{\lambda_{+}-\lambda_{-}}\Lambda u,
\end{equation}
where $\Lambda$ is the diagonal matrix with diagonal entries $\bar{\beta}_{i}$, which implies that
\begin{equation}
u(t)=\exp\left\{\left(\mathbf{Q}+\frac{2(\lambda_{-}^{2}-m)\vee(M-\lambda_{+}^{2})}{\lambda_{+}-\lambda_{-}}\Lambda\right)t\right\}\mathbf{1},
\end{equation}
where $\mathbf{1}$ is an all-one vector.
This implies that 
\begin{equation}
\mathbb{E}_{\beta(0)\sim\psi}\left[e^{\frac{2(\lambda_{-}^{2}-m)\vee(M-\lambda_{+}^{2})}{\lambda_{+}-\lambda_{-}}\int_{0}^{t}\beta(s)ds}\right]=\left\langle \exp\left\{\left(\mathbf{Q}+\frac{2(\lambda_{-}^{2}-m)\vee(M-\lambda_{+}^{2})}{\lambda_{+}-\lambda_{-}}\Lambda\right)t\right\}\mathbf{1},\psi\right\rangle.
\end{equation}
This completes the proof.
\end{proof}


\subsection{Proof of Proposition~\ref{prop:recursive_bound_RS_KLMC}}
\begin{proof}
    We couple the discrete algorithm process with a stationary continuous-time process and define the error in a transformed space.
    \begin{itemize}
    \item Let $\{(x_k, v_k, \beta_k)\}_{k\ge 0}$ be the state of the RS-KLMC algorithm.
    \item Let $\{(X_\beta(t), V_\beta(t))\}_{t\ge 0}$ be the stationary continuous RS-KLD process defined as
        \begin{align*}
            &dV_\beta(t)=-\gamma \beta_{\lfloor t/\eta \rfloor}V_\beta(t)dt-\beta_{\lfloor t/\eta \rfloor}\nabla f(X_\beta(t))dt+\sqrt{2\gamma\beta_{\lfloor t/\eta \rfloor}}dB_t,\\
            &dX_\beta(t)=\beta_{\lfloor t/\eta \rfloor}V_\beta(t)dt,
        \end{align*}
        where $(B_t)_{t\ge0}$ is a standard $d$-dimensional Brownian motion.
    \item We introduce the invertible transformation matrix $\mathbf P$ given in \cite{dalalyan2018kinetic}. 
    \begin{equation}
        \mathbf P = \frac{1}{\gamma}
        \begin{pmatrix}
            0 & -\gamma \mathbf I_d \\
            \mathbf I_d & \mathbf I_d
        \end{pmatrix}
        \quad \text{and} \quad
        \mathbf P^{-1} = 
        \begin{pmatrix}
            \mathbf I_d & \gamma \mathbf I_d \\
            -\mathbf I_d & 0
        \end{pmatrix}.
    \end{equation}
    \item The transformed error norm $A_{\beta,k}$ between the algorithm state $(x_k, v_k)$ and the stationary process state $(X_\beta(k\eta), V_\beta(k\eta))$ is defined the matrix $\mathbf P$:
    \begin{equation}
        A_{\beta,k} := \left\| \mathbf P^{-1} 
        \begin{pmatrix}
            v_{\beta,k} - V_\beta(k\eta) \\
            x_{\beta,k} - X_\beta(k\eta)
        \end{pmatrix} 
        \right\|_2,
    \end{equation}
    where $\|\cdot\|_2$ denotes $L^{2}$-norm, i.e. $\|\cdot\|_2:=(\mathbb E\|\cdot\|^2)^{1/2}$.
    
\end{itemize}
    Bounding $A_{\beta,k}$ provides a bound on the error of both position and velocity. Like the strategy we used to analyze the overdamped case, a single step of the RS-KLMC algorithm with physical stepsize $\eta$ under a fixed regime $\beta_k$ is mathematically equivalent to analyzing a standard KLMC algorithm (with constant friction $\gamma$) that takes a single step of effective size $h_k = \beta_k \eta$.

    Recall in \cite[p. 1972]{dalalyan2018kinetic}, for classic kinetic Langevin algorithm given stepsize $h\leq m/(4\gamma M)$ from step $k$ to step $k+1$, $A_{k+1}$ and $A_j$ have the relationship
    \begin{equation}\label{eq:dalalyan_result_kinetic}
    A_{k+1} \le 0.75Mh^{2}\sqrt{d} + (e^{-hm/\gamma}+0.75Mh^{2})A_{k},
    \end{equation}
    where $d$ is the dimension and $M$ the smoothness of the potential $f$.

    In our case, applying \eqref{eq:dalalyan_result_kinetic}, for $\eta\le\frac{m}{4\beta_{\max}\gamma M}$, we have
    \[
        A_{\beta,k+1}=0.75M(\beta_{k}\eta)^2\sqrt d+(e^{-\beta_k\eta m/\gamma}+0.75M(\beta_k\eta)^2)A_{\beta,k}.
    \]
    By iterating, we have
    \begin{align*}
        A_{\beta,K} 
        = & \left( \prod_{k=0}^{K-1} \left[ e^{-\eta m \beta_k/\gamma} + 0.75M(\eta\beta_k)^2 \right] \right) A_{\beta,0} \\
        & + \sum_{j=0}^{K-1} \left( \prod_{k=j+1}^{K-1} \left[ e^{-\eta m \beta_k/\gamma} + 0.75M(\eta\beta_k)^2 \right] \right) \left( 0.75M(\eta\beta_j)^2\sqrt{d} \right)\\
        \le & \left( \prod_{k=0}^{K-1} \left[ 1-\eta m \beta_k/\gamma+\frac12(\eta m \beta_k/\gamma)^2 + 0.75M(\eta\beta_k)^2 \right] \right) A_{\beta,0} \\
        & + \sum_{j=0}^{K-1} \left( \prod_{k=j+1}^{K-1} \left[ 1-\eta m \beta_k/\gamma+\frac12(\eta m \beta_k/\gamma)^2 + 0.75M(\eta\beta_k)^2 \right] \right) \left( 0.75M(\eta\beta_j)^2\sqrt{d} \right).
    \end{align*}
    For $\eta\le\frac{m\gamma}{(m^2+1.5M\gamma^2)\beta_{\max}}$, we have
    \[
       1-\frac{\eta m \beta_k}{\gamma}+\frac12\left(\frac{\eta m \beta_k}{\gamma}\right)^2 + 0.75M(\eta\beta_k)^2
       \leq 1-\frac{\eta m\beta_k}{2\gamma},
    \]
    which implies
    \begin{align*}
        A_{\beta,K} 
        \le & \left( \prod_{k=0}^{K-1} \left[ 1-\frac{\eta m \beta_k}{2\gamma} \right] \right) A_{\beta,0}  + \sum_{j=0}^{K-1} \left( \prod_{k=j+1}^{K-1} \left[ 1-\frac{\eta m \beta_k}{2\gamma} \right] \right) \left( 0.75M(\eta\beta_j)^2\sqrt{d} \right)\\
        \le & \left( \prod_{k=0}^{K-1} e^{-\frac{\eta m \beta_k}{2\gamma}} \right) A_{\beta,0}  + \sum_{j=0}^{K-1} \left( \prod_{k=j+1}^{K-1} e^{-\frac{\eta m \beta_k}{2\gamma}} \right) \left( 0.75M(\eta\beta_j)^2\sqrt{d} \right)\\
        = & e^{-\frac{\eta m}{2\gamma}\sum_{k=0}^{K-1}\beta_k}A_{\beta,0}+\left(0.75M\sqrt d\sum_{j=0}^{K-1}e^{-\frac{\eta m}{2\gamma}\sum_{k=j+1}^{K-1}\beta_k}\beta_j^2\right)\eta^2.
    \end{align*}
    As a result, for $\eta\le\frac{2\gamma}{m\beta_{\min}}$, we have $\frac{\eta m\beta_{\min}}{2\gamma}-\frac12\left( \frac{\eta m\beta_{\min}}{2\gamma} \right)^2\geq\frac{\eta m\beta_{\min}}{4\gamma}$, and then
    \begin{align*}
        A_{\beta,K}^2
        \le &2e^{-\frac{\eta m}{\gamma}\sum_{k=0}^{K-1}\beta_k}A_{\beta,0}^2
        +2\left(0.75M\sqrt d\sum_{j=0}^{K-1}e^{-\frac{\eta m}{2\gamma}\sum_{k=j+1}^{K-1}\beta_k}\beta_j^2\right)^2\eta^4\\
        \le &2e^{-\frac{\eta m}{\gamma}\sum_{k=0}^{K-1}\beta_k}A_{\beta,0}^2
        +2\cdot 0.75^2\cdot \frac{M^2\beta_{\max}^4d}{\left(1-e^{-\frac{\eta m\beta_{\min}}{2\gamma}}\right)^2}\eta^4\\
        \le &2e^{-\frac{\eta m}{\gamma}\sum_{k=0}^{K-1}\beta_k}A_{\beta,0}^2
        +2\cdot 0.75^2\cdot \frac{M^2\beta_{\max}^4d}{\left(\frac{\eta m\beta_{\min}}{2\gamma}-\frac12\left( \frac{\eta m\beta_{\min}}{2\gamma} \right)^2\right)^2}\eta^4\\
        \le &2e^{-\frac{\eta m}{\gamma}\sum_{k=0}^{K-1}\beta_k}A_{\beta,0}^2
        +2\cdot 0.75^2\cdot \frac{M^2\beta_{\max}^4d}{\left(\frac{\eta m\beta_{\min}}{4\gamma}\right)^2}\eta^4\\
        =&2e^{-\frac{\eta m}{\gamma}\sum_{k=0}^{K-1}\beta_k}A_{\beta,0}^2
        +18\cdot \frac{M^2\beta_{\max}^4d\gamma^2}{m^2\beta_{\min}^2}\eta^2.
    \end{align*}
    Taking expectation on both sides w.r.t. $(\beta_{k})_{k=0}^{K-1}$, we can reuse the results on the 
    $\mathbb E\left[e^{\cdot\sum_{k=0}^{K-1}\beta_k}\right]$
    in the Step 2 in the proof of Proposition \ref{prop:recursive_error_bound} and we obtain for 
    $\eta\le\min\left(\frac{m}{4\beta_{\max}\gamma M},\frac{m\gamma}{(m^2+1.5M\gamma^2)\beta_{\max}},\frac{2\gamma}{m\beta_{\min}}\right)$,
    \[
        A_K^2\le 2\left(1-\frac{\alpha}{2}\eta\right)^KA_0^2+C\eta^2,
    \]
    where
    \[
        \alpha = -\max_{1\leq i\leq N}\left\{ \operatorname{Re}\left(\lambda_i\left(\mathbf{Q} - \frac{m}{\gamma}\Lambda\right)\right) \right\}, \qquad
        C=18\cdot \frac{M^2\beta_{\max}^4d\gamma^2}{m^2\beta_{\min}^2}.
    \]
    Finally, we can use the relationship
    \[
        \mathcal W_2(\nu_K,\pi)
        \le\|x_K-X(K\eta)\|_2
        \le \gamma^{-1}\sqrt 2A_K,
    \]
    given in \cite[p.1973]{dalalyan2018kinetic}, and obtain
    \[
        \mathcal W_2^2(\nu_K,\pi)
        \le \frac2{\gamma^2}\left( 2\left(1-\frac{\alpha}2\eta\right)^KA_0^2+C\eta^2 \right)
        \le4\left(1-\frac{\alpha}2\eta\right)^K\mathcal W_2^2(\nu_0,\pi)+\frac{2C}{\gamma^2}\eta^2,
    \]
    where we use the equality $A_0=\gamma \mathcal W_2(\nu_0,\pi)$ by assuming the initial velocities are drawn from the stationary distribution, i.e. $v_0=V(0)$.
    The proof is complete.
\end{proof}

\subsection{Proof of Theorem~\ref{thm:non:asymptotic:klmc}}

\begin{proof}
The proof is a direct consequence of the recursive error bound for the squared 2-Wasserstein distance established in Proposition~\ref{prop:recursive_bound_RS_KLMC}. Let $w_k^2 := \mathcal{W}_2^2(\nu_k, \pi)$ denote the squared 2-Wasserstein distance at step $k$. From the proposition, we have the final bound after unrolling the recursion and taking the expectation:
\[
    \mathcal{W}_2^2(\nu_K, \pi) \le 4\left(1-\frac{\alpha}{2}\eta\right)^K \mathcal{W}_2^2(\nu_0, \pi) + \frac{2C}{\gamma^2}\eta^2.
\]
To obtain a bound on $\mathcal{W}_2(\nu_K, \pi)$, we take the square root of both sides of the inequality. By applying the elementary inequality $\sqrt{a+b} \le \sqrt{a} + \sqrt{b}$ for non-negative $a,b$, we get:
\begin{align*}
    \mathcal{W}_2(\nu_K, \pi) &\le \sqrt{4\left(1-\frac{\alpha}{2}\eta\right)^K \mathcal{W}_2^2(\nu_0, \pi)} + \sqrt{\frac{2C}{\gamma^2}\eta^2} \\
    &= 2\left(1-\frac{\alpha}{2}\eta\right)^{K/2} \mathcal{W}_2(\nu_0, \pi) + \sqrt{\frac{2C}{\gamma^2}}\eta.
\end{align*}
This completes the proof.
\end{proof}

\subsection{Proof of Corollary~\ref{cor:iteration:complexity:klmc}}

\begin{proof}
The proof follows from the non-asymptotic error bound established in Theorem~\ref{thm:non:asymptotic:klmc}. Our goal is to find conditions on the stepsize $\eta$ and the number of iterations $K$ such that the total error is bounded by a given accuracy level $\epsilon > 0$.
\[
    2\left(1-\frac{\alpha}{2}\eta\right)^{K/2} \mathcal{W}_2(\nu_0, \pi) + \sqrt{\frac{2C}{\gamma^2}}\eta \le \epsilon.
\]
We achieve this by ensuring each of the two terms on the left-hand side is bounded by $\epsilon/2$.

First, we choose the stepsize $\eta$ small enough to control the bias term:
\[
    \sqrt{\frac{2C}{\gamma^2}}\eta \le \frac{\epsilon}{2}.
\]
Solving for $\eta$, we get the condition on the stepsize:
\[
    \eta \le \frac{\epsilon}{2\sqrt{\frac{2C}{\gamma^2}}}.
\]

Next, with the stepsize $\eta$ chosen, we find the number of iterations $K$ required to shrink the initial error term sufficiently:
\[
    2\left(1-\frac{\alpha}{2}\eta\right)^{K/2} \mathcal{W}_2(\nu_0, \pi) \le \frac{\epsilon}{2}.
\]
Rearranging the terms, we have:
\[
    \left(1-\frac{\alpha}{2}\eta\right)^{K/2} \le \frac{\epsilon}{4\mathcal{W}_2(\nu_0, \pi)}.
\]
Using the inequality $1-x \le e^{-x}$ for $x \ge 0$, we can establish a sufficient condition. We can bound the left-hand side from above:
\[
    \left(1-\frac{\alpha}{2}\eta\right)^{K/2} \le \exp\left(-\frac{\alpha\eta}{2} \cdot \frac{K}{2}\right) = \exp\left(-\frac{\alpha\eta K}{4}\right).
\]
Therefore, it is sufficient to choose $K$ such that this upper bound satisfies the requirement:
\[
    \exp\left(-\frac{\alpha\eta K}{4}\right) \le \frac{\epsilon}{4\mathcal{W}_2(\nu_0, \pi)}.
\]
Taking the natural logarithm of both sides and solving for $K$, we get the condition on the number of iterations:
\[
    K \ge \frac{4}{\alpha\eta}\log\left(\frac{4\mathcal{W}_2(\nu_0, \pi)}{\epsilon}\right).
\]

By choosing the stepsize $\eta$ to be at its upper bound, $\eta = \frac{\epsilon\gamma}{2\sqrt{2C}} = \mathcal{O}(\epsilon)$, the required number of iterations $K$ becomes:
\[
    K \ge \frac{4}{\alpha} \cdot \frac{2\sqrt{\frac{2C}{\gamma^2}}}{\epsilon} \log\left(\frac{4\mathcal{W}_2(\nu_0, \pi)}{\epsilon}\right) = \mathcal{O}\left(\frac{1}{\epsilon}\log\left(\frac{1}{\epsilon}\right)\right).
\]
This completes the proof.
\end{proof}

\subsection{Proof of Theorem~\ref{thm:invariant:underdamped:1}}
\label{appendix:proof:invariant:underdamped:1}
\begin{proof}
Recall from \eqref{gamma:infinitesimal} that the infinitesimal generator of the $\gamma(t)$ is given by
\begin{align}
\mathcal{L}_{2}g(\bar{\gamma}_{i})=\sum_{j\neq i}q_{ij}\left[g(\bar{\gamma}_{j})-g(\bar{\gamma}_{i})\right],
\end{align}
for any $i=1,2,\ldots,N$. One can compute that its adjoint operator is given by:
\begin{align}
\mathcal{L}^{\ast}_{2}g(\bar{\gamma}_{i})=\sum_{j\neq i}\left[q_{ji}g(\bar{\gamma}_{j})-q_{ij}g(\bar{\gamma}_{i})\right],
\end{align}
for any $i=1,2,\ldots,N$.
Since $\psi=(\psi_{1},\psi_{2},\ldots,\psi_{N})$ is the invariant distribution of $\gamma(t)$, 
by abusing the notation and defining $\psi(\gamma):=\psi_{i}$ for any $\gamma=\gamma_{i}$, we have
\begin{align}
\mathcal{L}^{\ast}_{2}\psi(\bar{\gamma}_{i})=\sum_{j\neq i}\left[q_{ji}\psi(\bar{\gamma}_{j})-q_{ij}\psi(\bar{\gamma}_{i})\right]
=\sum_{j\neq i}\left[q_{ji}\psi_{j}-q_{ij}\psi_{i}\right]=0,
\end{align}
for any $i=1,2,\ldots,N$.
Next, one can compute that the adjoint operator of the infinitesimal generator of
the joint process $(\gamma(t),V(t),X(t))$ is given by:
\begin{align}
\mathcal{L}^{\ast}g(\bar{\gamma}_{i},v,x)
&=\bar{\gamma}_{i}\sum_{j=1}^{d}\frac{\partial}{\partial v_{j}}\left[v_{j}g\right]
+\sum_{j=1}^{d}\frac{\partial f}{\partial x_{j}}\frac{\partial g}{\partial v_{j}}
+\bar{\gamma}_{i}\sum_{j=1}^{d}\frac{\partial^{2}g}{\partial v_{j}^{2}}
-\sum_{j=1}^{d}v_{j}\frac{\partial g}{\partial x_{j}}
\nonumber
\\
&\qquad\qquad\qquad
+\sum_{j\neq i}\left[q_{ji}g(\bar{\gamma}_{j},v,x)-q_{ij}g(\bar{\gamma}_{i},v,x)\right],
\end{align}
for any $i=1,2,\ldots,N$ and $x\in\mathbb{R}^{d}$ and finally, we can compute that
\begin{align}
&\mathcal{L}^{\ast}\psi(\bar{\gamma}_{i})e^{-f(x)-\frac{1}{2}\Vert v\Vert^{2}}
\nonumber
\\
&=\bar{\gamma}_{i}\sum_{j=1}^{d}\frac{\partial}{\partial v_{j}}\left[v_{j}\psi(\bar{\gamma}_{i})e^{-f(x)-\frac{1}{2}\Vert v\Vert^{2}}\right]
+\sum_{j=1}^{d}\frac{\partial f}{\partial x_{j}}\frac{\partial \psi(\bar{\gamma}_{i})e^{-f(x)-\frac{1}{2}\Vert v\Vert^{2}}}{\partial v_{j}}
+\bar{\gamma}_{i}\sum_{j=1}^{d}\frac{\partial^{2}\psi(\bar{\gamma}_{i})e^{-f(x)-\frac{1}{2}\Vert v\Vert^{2}}}{\partial v_{j}^{2}}
\nonumber
\\
&\qquad\qquad\qquad\qquad\qquad\qquad
-\sum_{j=1}^{d}v_{j}\frac{\partial \psi(\bar{\gamma}_{i})e^{-f(x)-\frac{1}{2}\Vert v\Vert^{2}}}{\partial x_{j}}
\nonumber
\\
&\qquad\qquad\qquad
+\sum_{j\neq i}\left[q_{ji}\psi(\bar{\gamma}_{j})e^{-f(x)-\frac{1}{2}\Vert v\Vert^{2}}-q_{ij}\psi(\bar{\gamma}_{i})e^{-f(x)-\frac{1}{2}\Vert v\Vert^{2}}\right].
\end{align}
We can compute that
\begin{align}
&\bar{\gamma}_{i}\sum_{j=1}^{d}\frac{\partial}{\partial v_{j}}\left[v_{j}\psi(\bar{\gamma}_{i})e^{-f(x)-\frac{1}{2}\Vert v\Vert^{2}}\right]
+\bar{\gamma}_{i}\sum_{j=1}^{d}\frac{\partial^{2}\psi(\bar{\gamma}_{i})e^{-f(x)-\frac{1}{2}\Vert v\Vert^{2}}}{\partial v_{j}^{2}}
\nonumber
\\
&=\bar{\gamma}_{i}\psi(\bar{\gamma}_{i})e^{-f(x)}
\sum_{j=1}^{d}\left(\frac{\partial}{\partial v_{j}}\left[v_{j}e^{-\frac{1}{2}\Vert v\Vert^{2}}\right]
+\frac{\partial^{2}e^{-\frac{1}{2}\Vert v\Vert^{2}}}{\partial v_{j}^{2}}\right)
=0,
\end{align}
and moreover
\begin{align}
&\sum_{j=1}^{d}\frac{\partial f}{\partial x_{j}}\frac{\partial \psi(\bar{\gamma}_{i})e^{-f(x)-\frac{1}{2}\Vert v\Vert^{2}}}{\partial v_{j}}
-\sum_{j=1}^{d}v_{j}\frac{\partial \psi(\bar{\gamma}_{i})e^{-f(x)-\frac{1}{2}\Vert v\Vert^{2}}}{\partial x_{j}}\nonumber
\\
&=\psi(\bar{\gamma}_{i})\sum_{j=1}^{d}\left(\frac{\partial f}{\partial x_{j}}\frac{\partial e^{-f(x)-\frac{1}{2}\Vert v\Vert^{2}}}{\partial v_{j}}
-v_{j}\frac{\partial e^{-f(x)-\frac{1}{2}\Vert v\Vert^{2}}}{\partial x_{j}}\right)=0,
\end{align}
and finally
\begin{align}
\sum_{j\neq i}\left[q_{ji}\psi(\bar{\gamma}_{j})e^{-f(x)-\frac{1}{2}\Vert v\Vert^{2}}-q_{ij}\psi(\bar{\gamma}_{i})e^{-f(x)-\frac{1}{2}\Vert v\Vert^{2}}\right]
=e^{-f(x)-\frac{1}{2}\Vert v\Vert^{2}}\sum_{j\neq i}\left[q_{ji}\psi_{j}-q_{ij}\psi_{i}\right]=0,
\end{align}
for any $i=1,2,\ldots,N$ and $v,x\in\mathbb{R}^{d}$.
Hence, we conclude that
\begin{equation}
\mathcal{L}^{\ast}\psi(\bar{\gamma}_{i})e^{-f(x)-\frac{1}{2}\Vert v\Vert^{2}}=0,
\end{equation}
for any $i=1,2,\ldots,N$ and $v,x\in\mathbb{R}^{d}$, 
and therefore $\psi\otimes\mathcal{N}(0,I_{d})\otimes\pi$
is an invariant distribution of the joint process $(\gamma(t),V(t),X(t))$.
In particular, the Gibbs distribution $\pi\propto e^{-f(x)}$ is an invariant distribution for 
the regime-switching kinetic Langevin dynamics $X(t)$ in \eqref{underdamped:regime:switching}.
This completes the proof.
\end{proof}

\subsection{Proof of Theorem~\ref{thm:underdamped:continuous}}

\begin{proof}
Consider the classical kinetic Langevin dynamics with constant friction coefficient $\gamma$:
\begin{align}
&dV(t)=-\gamma V(t)dt-\nabla f(X(t))dt+\sqrt{2\gamma}dB_{t},
\\
&dX(t)=V(t)dt.
\end{align}
Let $P_{t}^{X}$ denote the Markov kernal of $(X(t))_{t\geq 0}$. 
That is, $P_{t}^{X}((x,v),A)=\mathbb{P}(X(t)\in A|V(0)=v,X(0)=x)$
for any Borel set $A\subset\mathbb{R}^{d}$. 
We denote $\mu P_{t}^{X}$ the unconditional distribution of the random variable
$X(t)$ when the starting distribution of the process $(V,X)$ is $\mu$, i.e. $(V(0),X(0))\sim\mu$.
According to Theorem~1 in \cite{dalalyan2018kinetic}, 
for any measures $\mu,\mu'$, and every $\gamma>0$, $t\geq 0$,
\begin{equation}\label{eq:dalalyan_frictional_result}
\mathcal{W}_{2}(\mu P_{t}^{X},\mu'P_{t}^{X})\leq\frac{\sqrt{2}}{\gamma}e^{-\frac{m\wedge(\gamma^{2}-M)}{\gamma}t}\mathcal{W}_{2}(\mu,\mu').
\end{equation}
If $\gamma\geq\max(\sqrt{2},\sqrt{m+M})$, then we have
\begin{equation}\label{eq:dalalyan_frictional_result_1}
\mathcal{W}_{2}(\mu P_{t}^{X},\mu'P_{t}^{X})\leq e^{-\frac{m}{\gamma}t}\mathcal{W}_{2}(\mu,\mu').
\end{equation}
By letting $\mu\sim\mathcal{N}(0,I_{d})\otimes\nu_{0}$, 
where $\nu_{0}$ is the law of $X(0)$ 
and $\mu'\sim\mathcal{N}(0,I_{d})\otimes\pi$,
we have
\begin{equation}
\mathcal{W}_{2}(\nu_{t},\pi)\leq e^{-\frac{m}{\gamma}t}\mathcal{W}_{2}(\nu_{0},\pi),
\end{equation}
where $\nu_{t}$ is the law of $X(t)$.
Next, consider frictional-regime-switching Langevin dynamics:
\begin{align}
&dV(t)=-\gamma(t)V(t)dt-\nabla f(X(t))dt+\sqrt{2\gamma(t)}dB_{t},
\\
&dX(t)=V(t)dt.
\end{align}
Under our assumption $\min_{1\leq i\leq N}\bar{\gamma}_{i}\geq\max(\sqrt{2},\sqrt{m+M})$,
we have $\gamma(t)\geq\max(\sqrt{2},\sqrt{m+M})$ for every $t$. 
Conditional on the CTMC process $(\gamma(t))_{t\geq 0}$, we have
\begin{equation}
\mathcal{W}_{2}(\nu_{\gamma,t},\pi)\leq e^{-\int_{0}^{t}\frac{m}{\gamma(s)}ds}\mathcal{W}_{2}(\nu_{\gamma,0},\pi),
\end{equation}
where $\nu_{\gamma,t}$ is the law of $X(t)$ conditional on $(\gamma(t))_{t\geq 0}$.
By taking the expectations over $(\gamma(t))_{t\geq 0}$
and letting $\gamma(0)\sim\psi$,
we get
\begin{align}
\mathcal{W}_{2}^{2}(\nu_{t},\pi)
\leq
\mathbb{E}_{\gamma(0)\sim\psi}\left[\mathcal{W}_{2}^{2}(\nu_{\gamma,t},\pi)\right]
\leq
\mathbb{E}_{\gamma(0)\sim\psi}\left[e^{-2\int_{0}^{t}\frac{m}{\gamma(s)}ds}\right]
\mathcal{W}_{2}^{2}(\nu_{0},\pi),
\end{align}
where $\nu_{t}$ is the unconditional law of $X(t)$, 
and we used the fact that $\nu_{\gamma,0}=\nu_{0}$ in distribution, 
that is independent of $(\gamma(t))_{t\geq 0}$.

Let $u(t):=(u_{1}(t),\ldots,u_{N}(t))$, where $u_{i}(t):=\mathbb{E}_{\gamma(0)=\bar{\gamma}_{i}}\left[e^{-2m\int_{0}^{t}\frac{1}{\gamma(s)}ds}\right]$. 
By Feynman-Kac formula, 
\begin{equation}
\frac{\partial u}{\partial t}=\mathbf{Q}u-2m\Lambda_{\gamma}^{-1} u,
\end{equation}
where $\Lambda_{\gamma}^{-1}$ is the diagonal matrix with diagonal entries $1/\bar{\gamma}_{i}$, which implies that
\begin{equation}
u(t)=e^{(\mathbf{Q}-2m\Lambda_{\gamma}^{-1})t}\mathbf{1},
\end{equation}
where $\mathbf{1}$ is an all-one vector.
This implies that 
\begin{equation}
\mathbb{E}_{\gamma(0)\sim\psi}\left[e^{-2m\int_{0}^{t}\frac{1}{\gamma(s)}ds}\right]=\left\langle e^{(\mathbf{Q}-2m\Lambda_{\gamma}^{-1})t}\mathbf{1},\psi\right\rangle.
\end{equation}
This completes the proof.
\end{proof}


\subsection{Proof of Proposition~\ref{prop:frs_klmc_final}}
\label{appendix:proof:frs_klmc_final}
\begin{proof}
    Let $(x_{\gamma,k}, v_{\gamma,k})$ be the state of the algorithm at step $k$ and $\mathcal F_{\gamma,k}$ be the $\sigma$-algebra generated by $\{(x_{\gamma,n}, v_{\gamma,n})\}_{0\le n\le k}$.
    To analyze the error at step $k+1$, we introduce an \textbf{auxiliary continuous process} $\{(X'_\gamma(t), V'_\gamma(t))\}_{t \in [k\eta, (k+1)\eta]}$. This process follows the same SDE as $(X_\gamma(t), V_\gamma(t))$ with a constant friction $\gamma_k$, but it is initialized at the algorithm's current state: $(X'_\gamma(k\eta), V'_\gamma(k\eta)) = (x_{\gamma,k}, v_{\gamma,k})$.

    Conditioning on $\mathcal F_{\gamma,k}$, the total error at step $k+1$ can then be bounded using the triangle inequality:
    \begin{align*}
        &\underbrace{\left\| x_{\gamma,k+1} - X_\gamma((k+1)\eta) \right\|_2^2}_{\text{Total Error at step k+1}} \\
        \le &\underbrace{\left\| x_{\gamma,k+1} - X'_\gamma((k+1)\eta-0) \right\|_2^2}_{\text{Discretization Error}}+ \underbrace{\left\| X'_\gamma((k+1)\eta-0)- X_\gamma((k+1)\eta) \right\|_2^2}_{\text{Process Error}}.
    \end{align*}

    Let us analyze each term separately.
    \paragraph{Discretization Error:} 
    The difference between the algorithm's velocity update and the true SDE evolution over one step $t \in [k\eta, (k+1)\eta]$ with friction $\gamma_k$ is given by:
    \[
        v_{\gamma,k+1} - V'_\gamma((k+1)\eta-0) = -\int_{k\eta}^{(k+1)\eta} e^{-\gamma_k((k+1)\eta - s)} \left( \nabla f(X'_\gamma(s)) - \nabla f(x_{\gamma,k}) \right) ds.
    \]
    By taking the $L^2$-norm and applying Minkowski's inequality, the Lipschitz property of the gradient, and the relation $X'_\gamma(s) - x_{\gamma,k} = \int_{k\eta}^s V'_\gamma(u)du$, we obtain the bound for the velocity error (see \cite[p. 1971]{dalalyan2018kinetic})
    \begin{equation}\label{eq:frictional_v}
        \left\| v_{\gamma,k+1} - V'_\gamma((k+1)\eta-0) \right\|_2 \le \frac{M\eta^2}{2} \max_{u \in [k\eta, (k+1)\eta]} \|V'_\gamma(u)\|_2,
    \end{equation}
    and the position error 
    \begin{equation}\label{eq:frictional_x}
        \left\| x_{\gamma,k+1} - X'_\gamma((k+1)\eta-0) \right\|_2 \le \frac{M\eta^3}{6} \max_{u \in [k\eta, (k+1)\eta]} \|V'_\gamma(u)\|_2.
    \end{equation}
    Following the argument in \cite{dalalyan2018kinetic}, from $[k\eta,(k+1)\eta)$, we define the transformation matrix $\mathbf P_{\gamma,k}$ and its inverse $\mathbf P_{\gamma,k}^{-1}$ as:
    \begin{equation}
        \mathbf{P}_{\gamma,k} = \frac{1}{\gamma_k}\begin{pmatrix} 0 & -\gamma_k I_d \\ I_d & I_d \end{pmatrix}, \qquad
        \mathbf{P}_{\gamma,k}^{-1} = \begin{pmatrix} I_d & \gamma_k I_d \\ -I_d & 0 \end{pmatrix}.
    \end{equation}
    Given the regime chain $(\gamma_n)_{n\ge0}$, the maximum velocity of the auxiliary process can be bounded in terms of the transformed error at the beginning of the step, $A_{\gamma,k}$, which is defined as
    \[
        A_{\gamma,k}:=\left\| \mathbf P_{\gamma,k}^{-1}\begin{pmatrix}
            v_{\gamma,k} - V'_\gamma(k\eta-0) \\
            x_{\gamma,k} - X'_\gamma(k\eta-0)
        \end{pmatrix}  \right\|_2,
    \]
    where $V'_\gamma(\cdot-0)$ and $X'_\gamma(\cdot-0)$ denote the left limit of $V'_\gamma(\cdot)$ and $X'_\gamma(\cdot)$, respectively.
    As shown in Lemma~2 of \cite{dalalyan2018kinetic},
\begin{equation}\label{eq:frictional_v_prime_upper_bound}
        \max_{u \in [k\eta, (k+1)\eta]} \|V'_\gamma(u)\|_2 \le \sqrt{d} + A_{\gamma,k}.
    \end{equation}
    Now, let us bound $A_{\gamma,k}$. Like the strategy we have used in the RS-KLMC case, we use \eqref{eq:dalalyan_result_kinetic}. In our case, we have
    \[
        A_{\gamma,k+1}\le 0.75M\eta^2\sqrt{d}+\left(e^{-\eta m/\gamma_k}+0.75M\eta^2\right)A_{\gamma,k}.
    \]
    By iterating, for $\eta\le\frac{m\gamma_{\min}}{m^2+1.5M\gamma_{\max}^2}$, which guarantees $1-\frac{\eta m}{\gamma_{k}}+\frac12\left( \frac{\eta m}{\gamma_{k}} \right)^2 + 0.75M\eta^2\le1-\frac{\eta m}{2\gamma_{k}}$ for all $k=1,\ldots,N$, we have
    \begin{align*}
        A_{\gamma,K} 
        \le & \left( \prod_{k=0}^{K-1} \left[ e^{-\eta m/\gamma_k} + 0.75M\eta^2 \right] \right) A_{\gamma,0} + \left( 0.75M\eta^2\sqrt{d} \right) \cdot \sum_{j=0}^{K-1} \left( \prod_{k=j+1}^{K-1} \left[ e^{-\eta m/\gamma_k} + 0.75M\eta^2 \right] \right)\\
        \le & \prod_{k=0}^{K-1}\left(1-\frac{\eta m}{\gamma_k}+\frac12\left( \frac{\eta m}{\gamma_k} \right)^2 + 0.75M\eta^2\right) A_{\gamma,0} \\
        &+ \left( 0.75M\eta^2\sqrt{d} \right) \cdot \sum_{j=0}^{K-1} \left( 1-\frac{\eta m}{\gamma_{\max}}+\frac12\left( \frac{\eta m}{\gamma_{\max}} \right)^2 + 0.75M\eta^2 \right)^{K-j-1}\\
        \le&\prod_{k=0}^{K-1}\left( 1-\frac{\eta m}{2\gamma_k} \right)A_{\gamma,0}+1.5\frac{M\sqrt d\gamma_{\max}}{m}\eta^2\\
        \le &A_{\gamma,0}+1.5\frac{M\sqrt d\gamma_{\max}}{m}\eta^2\\
        = &\sqrt{\sum_{i=1}^N\psi_i\bar\gamma_i^2}\mathcal W_2(\nu_0,\pi)+1.5\frac{M\sqrt d\gamma_{\max}}{m}\eta^2,
    \end{align*}
    where we use the equality $A_{\gamma,0}=\|\gamma_0\|_2\mathcal W_2(\nu_0,\pi)=\sqrt{\sum_{i=1}^N\psi_i\bar\gamma_i^2}\mathcal W_2(\nu_0,\pi)$.
    Plugging into \eqref{eq:frictional_v_prime_upper_bound}, and then \eqref{eq:frictional_x}, we obtain
    \[
        \left\| x_{\gamma,k+1} - X'_\gamma((k+1)\eta) \right\|_2 \le \frac{M\eta^3}{6} \left(\sqrt d+\sqrt{\sum_{i=1}^N\psi_i\bar\gamma_i^2}\mathcal W_2(\nu_0,\pi)+1.5\frac{M\sqrt d\gamma_{\max}}{m}\eta^2\right).
    \]
    Let $1.5\frac{M\sqrt d\gamma_{\max}}{m}\eta^2\le \sqrt d$, i.e. $\eta\le \sqrt{\frac{m}{1.5M\gamma_{\max}}}$, we have
    \[
        \left\| x_{\gamma,k+1} - X'_\gamma((k+1)\eta) \right\|_2 \le \frac{M\eta^3}{6} \left(2\sqrt d+\sqrt{\sum_{i=1}^N\psi_i\bar\gamma_i^2}\mathcal W_2(\nu_0,\pi)\right).
    \]

    \paragraph{Process Error:} Let $\{(X_\gamma(t), V_\gamma(t))\}_{t\ge 0}$ be the stationary continuous RS-KLD process defined as
            \begin{align*}
                &dV_\gamma(t)=-\gamma_{\lfloor t/\eta \rfloor}V_\gamma(t)dt-\nabla f(X_\gamma(t))dt+\sqrt{2\gamma_{\lfloor t/\eta \rfloor}}dB_t,\\
                &dX_\gamma(t)=V_\gamma(t)dt,
            \end{align*}
            where $(B_t)_{t\ge0}$ is a standard $d$-dimensional Brownian motion.

        Assume the constant friction coefficient $\gamma\ge\max(\sqrt2,\sqrt{M+m})$, where $M$ and $m$ are the convexity and smoothness of the potential $f$, respectively. Let $\mu=\mu_1\otimes \mu_2$ and $\mu'=\mu_1\otimes \mu'_2$, where $\mu_1$ and $\mu'_1$ are the distributions of the initial position $X(0)$, and $\mu_2$ and $\mu'_2$ are the distributions of the initial velocity $V(0)$. Recall the equation \eqref{eq:dalalyan_frictional_result_1} in Appendix~\ref{appendix:proof:invariant:underdamped:1}, we have mentioned that for any $t\ge 0$, we have
        \begin{equation}
            \mathcal W_2\left(\mu P_t^{X},\mu' P_t^{X}\right)
            \le  e^{-\frac{m}{\gamma}\eta}\mathcal W_2(\mu,\mu'),
        \end{equation}
        where $P_t^{X}$ is the transition probability of the process $(X_s)_{s\ge0}$, and $\mathcal W_2$ is the 2-Wasserstein distance.

        Denote 
        $\gamma_{\max}=\max(\bar\gamma_1,\ldots,\bar\gamma_N)$ and $\gamma_{\min}=\min(\bar\gamma_1,\ldots,\bar\gamma_N)$. If $\gamma_{\min}\geq\max(\sqrt2,\sqrt{M+m})$, since $\gamma(t)$ remains constant $\gamma_{\lfloor t/\eta \rfloor}$ during the time interval $[k\eta, (k+1)\eta)$, applying \eqref{eq:dalalyan_frictional_result} to the process $(X_\gamma(t), V_\gamma(t))$ gives:
        \[
            \mathcal W_2\left(\mu P_{(k+1)\eta}^{X},\mu' P_{(k+1)\eta}^{X}\right)
            \le e^{-\frac{m}{\gamma_k}t}\mathcal W_2(\mu P_{k\eta}^X, \mu'P_{k\eta}^X),
            \qquad k\ge0.
        \]
        Let $\mu=\mathcal{N}(0,I_{d})\otimes\nu_0$ and $\mu'=\mathcal{N}(0,I_{d})\otimes\pi$, then $(X_\gamma, V_\gamma)((k+1)\eta)\sim \mathcal{N}(0,I_{d})\otimes\pi$.
        Since $(X'_\gamma,V'_\gamma)(K\eta)\sim \mu P_{K\eta}^X$, 
        we have
        \begin{align*}
            \left\| X'_\gamma((K+1)\eta) - X_\gamma((K+1)\eta) \right\|_2
            \le e^{-\frac{m\eta}{\gamma_K}}\mathcal W_2(\mu P_{K\eta}^X,\mu' P_{K\eta}^X)=e^{-\frac{m\eta}{\gamma_K}}\mathcal W_2(\nu_K,\pi).
        \end{align*}
Combining the Discretization Error and Process Error together, we obtain
\begin{align*}
    \mathcal W_2(\nu_{K+1},\pi)
    \le&\left\| x_{\gamma,K+1} - X_\gamma((K+1)\eta) \right\|_2\\
    \le &e^{-\frac{m\eta}{\gamma_{K}}}\mathcal W_2(\nu_K,\pi)+\frac{M\eta^3}{6} \left(2\sqrt d+\sqrt{\sum_{i=1}^N\psi_i\bar\gamma_i^2}\mathcal W_2(\nu_0,\pi)\right).
\end{align*}

By iterating, we have
\begin{align*}
    \mathcal W_2(\nu_{K},\pi) 
     &\le  e^{-m\eta\sum_{k=0}^{K-1}\frac1{\gamma_k}} \mathcal W_2(\nu_0,\pi) 
     \\
     &\qquad\qquad\qquad+ \frac{1}{\frac{m}{\gamma_{\max}}\left(1-\frac{m\eta}{2\gamma_{\max}}\right)} \cdot \frac{M\eta^2}{6} \left(2\sqrt d+\sqrt{\sum_{i=1}^N\psi_i\bar\gamma_i^2}\mathcal W_2(\nu_0,\pi)\right).
\end{align*}
For $\eta\leq\frac{\gamma_{\max}}{m}$, which guarantees $1-\frac{m\eta}{2\gamma_{\max}}\geq\frac12$, we have
\begin{align*}
    \mathcal W_2(\nu_{K},\pi) 
     \le & e^{-m\eta\sum_{k=0}^{K-1}\frac1{\gamma_k}} \mathcal W_2(\nu_0,\pi) 
     + \frac{\gamma_{\max}M\eta^2}{3m}\left(2\sqrt d+\sqrt{\sum_{i=1}^N\psi_i\bar\gamma_i^2}\mathcal W_2(\nu_0,\pi)\right),
\end{align*}
and then
\begin{align*}
    \mathcal W_2^2(\nu_{K},\pi) 
     \le & 2e^{-2m\eta\sum_{k=0}^{K-1}\frac1{\gamma_k}} \mathcal W_2^2(\nu_0,\pi) 
     + \frac{2\gamma_{\max}^2M^2\eta^4}{9m^2}\left(2\sqrt d+\sqrt{\sum_{i=1}^N\psi_i\bar\gamma_i^2}\mathcal W_2(\nu_0,\pi)\right)^2.
\end{align*}

Taking expectations on both sides w.r.t. $(\gamma_{k})_{k=0}^{K-1}$, we can reuse the results on the $\mathbb E\left[e^{\cdot\sum_{k=0}^{K-1}\beta_k}\right]$
in the Step 2 in Appendix~\ref{appendix:overdamped_recursive_error_bound} and we obtain
\begin{align*}
    \mathcal W_2^2(\nu_{K},\pi) 
     \le & 2\left(1-\frac\alpha2\eta\right)^K \mathcal W_2^2(\nu_0,\pi) 
     + \frac{2\gamma_{\max}^2M^2\eta^4}{9m^2}\left(2\sqrt d+\sqrt{\sum_{i=1}^N\psi_i\bar\gamma_i^2}\mathcal W_2(\nu_0,\pi)\right)^2,
\end{align*}
where
\[
    \alpha = -\max_{1\leq i\leq N} \left\{ \operatorname{Re}\left(\lambda_i\left(\mathbf{Q} - 2m\Lambda_{\gamma}^{-1}\right)\right) \right\},\quad      \Lambda_{\gamma}^{-1}=\text{diag}\left(\frac1{\bar\gamma_i},\ldots,\frac1{\bar\gamma_N}\right).
\]
The proof is complete.
\end{proof}

\subsection{Proof of Theorem~\ref{thm:non:asymptotic:frs-klmc}}
\begin{proof}
The result is obtained by taking the square root of both sides of the inequality in Proposition~\ref{prop:frs_klmc_final} and applying the elementary inequality $\sqrt{a+b} \le \sqrt{a} + \sqrt{b}$ for non-negative $a, b$.
\end{proof}

\subsection{Proof of Corollary~\ref{cor:iteration:complexity:frs-klmc}}

\begin{proof}
The proof follows from the non-asymptotic error bound established in Theorem~\ref{thm:non:asymptotic:frs-klmc}. Our goal is to find conditions on the stepsize $\eta$ and the number of iterations $K$ such that the total error is bounded by a given accuracy level $\epsilon>0$.

From Theorem~\ref{thm:non:asymptotic:frs-klmc}, we have the bound:
\[
\mathcal{W}_2(\nu_K, \pi) \le \sqrt{2} \left(1-\frac{\alpha}{2}\eta\right)^{K/2} \mathcal{W}_2(\nu_0,\pi) + C_B \eta^2.
\]
We want to ensure that the right-hand side is less than or equal to $\epsilon$. We can achieve this by ensuring each of the two terms is bounded by $\epsilon/2$.

First, we choose the stepsize $\eta$ small enough to control the second term:
\[
    C_B\eta^2\le \frac{\epsilon}2.
\]
Solving for $\eta$, we get the condition on the stepsize:
$\eta^2\le \frac{\epsilon}{2C_B}$, which is equivalent to $\eta \le \sqrt{\frac{\epsilon}{2C_B}}$.
This matches the first condition stated in the corollary.

Next, with the stepsize $\eta$ chosen, we find the number of iterations $K$ required to shrink the initial error term sufficiently:
\[
\sqrt{2} \left(1-\frac{\alpha}{2}\eta\right)^{K/2} \mathcal{W}_2(\nu_0,\pi) \le \frac{\epsilon}{2}.
\]
Rearranging the terms, we have:
\[
    \left(1-\frac\alpha2\eta\right)^{K/2} \le \left(e^{-\frac{\alpha\eta}{2}}\right)^{K/2}=e^{-\frac{\alpha\eta K}{4}}.
\]
Therefore, it is suffcient to choose $K$ such that this upper bound satisfies the requirement:
\[
    e^{-\frac{\alpha\eta K}{4}}\le \frac{\epsilon}{2\sqrt2}\mathcal W_2(\nu_0,\pi).
\]
Taking the logarithm on both sides and solving for $K$, we have
\begin{align*}
    K \ge \frac{4}{\alpha\eta}\log\left(\frac{2\sqrt{2}\mathcal{W}_2(\nu_0, \pi)}{\epsilon}\right).
\end{align*}
This gives the required number of iterations. To find the overall iteration complexity, we can choose the stepsize $\eta$ to be proportional to its upper bound i.e., $\eta=\mathcal O(\sqrt\epsilon)$. Substituting this into the expression for $K$:
\[
    K \ge \frac{4}{\alpha \cdot \mathcal{O}(\sqrt{\epsilon})} \log\left(\frac{2\sqrt{2}\mathcal{W}_2(\nu_0, \pi)}{\epsilon}\right).
\]
Thus, the complexity is:
\[
    K = \mathcal{O}\left(\frac{1}{\sqrt{\epsilon}}\log\left(\frac{1}{\epsilon}\right)\right).
\]
This completes the proof.
\end{proof}
\end{document}